\documentclass{jfm}
\usepackage{graphicx}
\usepackage{natbib}
\usepackage{epstopdf, epsfig}
\usepackage{amsmath}
\usepackage{amssymb}
\usepackage{latexsym}
\usepackage{wasysym}
\usepackage{multirow}
\usepackage{color}
\usepackage{subfigure}
\usepackage{pdflscape}
\usepackage{multirow}
\usepackage{dashrule}
\usepackage{tikz}
\usepackage{cancel}
\usetikzlibrary{arrows}
\usepackage{mathtools}
\usepackage[utf8]{inputenc}
\usepackage{float}
\usepackage[version=4]{mhchem}
\usepackage{siunitx}
\usepackage{longtable,tabularx}
\usepackage{gensymb}
\DeclareGraphicsExtensions{.png}

\usepackage [english]{babel}
\usepackage [autostyle, english = american]{csquotes}
\MakeOuterQuote{"}

\newcommand{\bea}{\begin{equation}\begin{aligned}}
\newcommand{\eea}{\end{aligned}\end{equation}}

\shorttitle{Guidelines for authors}
\shortauthor{R. Ma and K. Mahesh}

\title{Boundary layer transition due to distributed roughness: Effect of roughness spacing}
\author{Rong Ma\aff{1} 
 \and Krishnan Mahesh\aff{2}
  \corresp{\email{krmahesh@umich.edu}}}

\affiliation{\aff{1}Department of Aerospace Engineering and Mechanics, University of Minnesota,
Minneapolis, MN 55455, USA
\aff{2}Department of Naval Architecture and Marine Engineering, University of Michigan,
Ann Arbor, MI 48109, USA}

\begin{document}

\maketitle

\begin{abstract}
The influence of roughness spacing on boundary layer transition over distributed roughness elements is studied using direct numerical simulation (DNS) and global stability analysis, and compared to isolated roughness elements at the same $Re_h$. Small spanwise spacing ($\lambda_z=2.5h$) inhibits the formation of counter-rotating vortices (CVP) and as a result, hairpin vortices are not generated and the downstream shear layer is steady. For $\lambda_z=5h$, the CVP and hairpin vortices are induced by the first row of roughness, perturbing the downstream shear layer and causing transition. The temporal periodicity of the primary hairpin vortices appears to be independent of the streamwise spacing. Distributed roughness leads to a lower critical $Re_h$ for instability to occur and a more significant breakdown of the boundary layer compared to isolated roughness. When the streamwise spacing is comparable to the region of flow separation ($\lambda_x=5h$), the high-momentum fluid barely moves downward into the cavities and the wake flow has little impact on the following roughness elements. The leading unstable varicose mode is associated with the central low-speed streaks along the aligned roughness elements, and its frequency is close to the shedding frequency of hairpin vortices in the isolated case. For larger streamwise spacing ($\lambda_x=10h$), two distinct modes are obtained from global stability analysis. The first mode shows varicose symmetry, corresponding to the primary hairpin vortex shedding induced by the first-row roughness. %It is also found to be associated with the high-speed streaks form in the longitudinal grooves.
The high-speed streaks formed in the longitudinal grooves are also found to be unstable and interacting with the varicose mode. The second mode is a sinuous mode with lower frequency, induced as the wake flow of the first-row roughness runs into the second row. It extracts most energy from the spanwise shear between the high- and low-speed streaks.
%For larger streamwise spacing ($\lambda_x=10h$), high-momentum fluid penetrates into the cavities and impinges onto the following roughness, the high-speed streaks with large amplitude form in the longitudinal grooves farther downstream. Two leading eigenvalues are obtained from global stability analysis. One corresponding to the primary hairpin vortex shedding is also associated with the high-speed streaks in the grooves. The other unstable sinuous mode with lower frequency is induced as the wake flow of the first-row roughness runs onto the second row, and extracts most of energy from the spanwise shear induced between the high- and low-speed streaks.
\end{abstract}


\section{Introduction}\label{sec:intro}
When laminar boundary layers encounter rough surfaces, the flow field can be greatly modified by surface roughness, and transition to turbulence can occur. Understanding roughness-induced transition is important since it leads to increase in friction drag and affects the performance of aeronautical and naval applications. Three-dimensional (3-D) surface roughness can be generally categorized into isolated and distributed roughness elements, both of which are commonly involved in engineering applications. While an isolated roughness element represents a single protuberance or a trip on the surface, it can also be considered as a foundational model to be extended to distributed surface roughness. 

The effects of isolated roughness elements on boundary layers have been studied in past literature \citep{baker1979laminar}. The streamwise vortices induced by an isolated roughness element create longitudinal streaks downstream whose unstable nature plays a crucial role in the transition process \citep{reshotko2001transient,fransson2004experimental,fransson2005experimental}. The linear stability properties of isolated-roughness-induced transition have been investigated both computationally and experimentally \citep{de2013laminar,loiseau2014investigation,citro2015global,bucci2021influence,ma2022global}. They are found to depend on the combined effects of parameters such as the ratio of roughness height $h$ to the displacement boundary layer thickness $\delta^*$, aspect ratio $\eta=d/h$ ($d$ is roughness width) and $Re_h=U_eh/\nu$ ($U_e$ is the boundary layer edge velocity and $\nu$ is the kinetic viscosity of the fluid). %Past work has largely focused on boundary layer transition induced by an isolated roughness element. 
Compared to the isolated roughness element, distributed roughness elements display phenomena not present in the isolated case. Fewer studies have been performed on how 3-D distributed roughness elements affect the stability properties and flow structures in transitional boundary layers.

Global linear stability analysis \citep{theofilis2011global} provides insight into the temporal disturbance growth during the early stages of transition, and is useful for non-parallel flows such as roughness wakes. It has been used to capture and understand the leading unstable modes in boundary layer transition due to isolated roughness \citep{loiseau2014investigation,citro2015global,bucci2021influence,ma2022global}. The potential of global stability analysis to detect the leading unstable modes induced by distributed surface roughness remains relatively unexplored. %It could also be useful to detect the leading unstable modes induced by distributed surface roughness. However, this has not been explored and evaluated. 
In this work, we therefore combine DNS and global stability analysis to investigate transition due to distributed surface roughness with varying streamwise and spanwise spacing. 

%literature review on distributed roughness in transitional BL, compared to isolated:
%distributed roughness with large height --> bypass transition, transient growth~streamwise vortices:
%3-D distributed surface roughness can also modify boundary layer flows and destabilize the shear layer to trigger transition.
Past studies of transition over distributed surface roughness have mainly focused on the effects of roughness height. \cite{corke1986experiments} studied the effects of distributed roughness on transition and suggested that transition is most likely to be triggered by the few highest peaks. They also found that the low-inertia fluid in the valleys between roughness elements %and for roughness with high grain density 
could be influenced more easily by free-stream disturbances. For roughness with small amplitudes, transition is induced through a linear amplification of temporal disturbance growth followed by secondary instabilities and breakdown to turbulence \citep{reshotko2001transient}. In contrast, large-amplitude roughness creates local separations, leading to strong 3-D disturbances that can develop into turbulence directly. This is the so-called "bypass" transition which means that the linear instability processes such as Tollmien-Schlichting waves are bypassed. \cite{vadlamani2018distributed} conducted numerical investigations on transition of a subsonic boundary layer over sinusoidal roughness elements with different height. They suggested that secondary instabilities on the streaks promote transition to turbulence for roughness elements that are inside the boundary layer, and that the instability wavelengths appear to be governed by the fixed streamwise and spanwise spacings between roughness elements. For roughness elements that are higher than the boundary layer, they found that the scale of instability is shorter and the shedding from the obstacles leads to transition. \cite{von2020direct} investigated the influence of distributed roughness and freestream turbulence on bypass transition. They found that the bypass transition occurs as a result of secondary instability of low-speed streaks and suggested that the streak spacing does not vary with different roughness density at constant roughness height.

While roughness height is an important parameter to affect transition, roughness spacing can also potentially modify the transitional flow behavior and associated instability mechanisms. Investigation of spanwise proximity of roughness elements has commonly involved spanwise arrays of roughness elements. \cite{ergin2006unsteady} studied transitional flow behind a spanwise array of cylindrical roughness elements and suggested that transition results from a competition between unsteady disturbance growth and steady disturbance decay. \cite{choudhari2006roughness} performed numerical simulations of a flat plate boundary layer past a spanwise array of cylindrical roughness elements (spaced three diameters away from each other). They observed self-sustained and unsteady vortical structures that resemble the flow structures during natural transition for roughness with large height.  \cite{muppidi2012direct} investigated supersonic boundary layer flow over distributed surface roughness and showed that the counter-rotating vortex pairs induced by roughness perturb the shear layer and triggers transition. With closely packed roughness elements, there is little upstream spacing to generate a strong horseshoe vortex and little spanwise spacing to produce counter-rotating vortices, which can prevent transition from occurring. %The spanwise spacing of distributed roughness elements has hardly been varied to quantify its influence on instability and transition process in past literature. 
The influence of spanwise spacing on instability and transition has been less-explored in past work. Our present study aims to quantitatively evaluate the effects of spanwise spacing of roughness elements on the vortical structures and associated instability characteristics in transition. %The isolated cube with $\eta=1$ and $h/\delta^*=2.86$ in \cite{ma2022global} serves as the baseline and is compared to the distributed roughness cases in the present work. %, with a comparison to an isolated roughness element. 
% Although the influence of roughness density %on secondary instability 
% have been indicated in past literature, %the roughness spacing has not been varied systematically %quantitatively 
% the effects of roughness spacing on instability and transition process need further understanding.

% %literature review distributed -- spanwise array:
% %A spanwise array of roughness elements has been investigated. 
% %Transition due to spanwise arrays of distributed roughness elements have been investigated in past literature. 
% Past literature has mainly focused on the effects of different roughness height for a spanwise row of roughness elements with fixed spanwise spacing. We thus vary spanwise spacing of distributed surface roughness, aiming to quantitatively investigate how the spanwise proximity of roughness elements would affect the vortical structures and associated instability characteristics in transition, with a comparison to an isolated roughness element. %Less is known about how the varying spanwise spacings would affect transition process with a comparison to an isolated roughness element.  

% streamwise spacing: (turbulent rough regime - d,k type roughness
The role of streamwise proximity of roughness elements is also a crucial factor of transition due to distributed roughness. \cite{choudhari2010llaminar} performed simulations of a high-speed boundary layer past an isolated diamond trip as well as two trips aligned with the flow direction (spaced three times the trip width). They found that the introduction of an additional trip could amplify the streak amplitude and cause transition even at a smaller trip height, while the case with larger trip height presents a weaker augmentation of the streak amplitude since the flow has not recovered as much as for smaller trip height. %The region of flow separation behind a cube is known to be approximately $4$-$5h$ in a turbulent boundary layer according to \cite{castro1977flow}. 
In a turbulent boundary layer, the classification of d-type and k-type behaviors \citep{jimenez2004turbulent} is related to the streamwise roughness spacing and might be made based on the ratio of roughness pitch to height being equal to $3$ \citep{tani1987turbulent}. \cite{perry1969rough} and \cite{ikeda2007direct} further claimed that the difference between k-type and d-type roughness is related to the state of vortex shedding: for d-type roughness, stable vortices form within the grooves and no eddy sheds into the flow above the crests; for k-type roughness, eddies with length scale of order $h$ shed into the flow above the crests.  \cite{vanderwel2015effects} suggested that distributed roughness with streamwise gaps less than $4$-$5h$ would act like continuous strips in turbulent boundary layers, while more than $5h$ would act like 3-D distributed roughness. Although the dependence of flow behaviors on different streamwise spacing has been discussed extensively in turbulent boundary layers, less is studied on how streamwise spacing affects transition. This work therefore investigates transition due to distributed surface roughness with streamwise spacing $5h$ and $10h$, and compares the results to the isolated roughness case.
%A proper distinction between d-type and k-type can not be based on the state of vortex shedding. For d-type roughness, the roughness function is not dependent on k+, the frictional drag is much larger than the pressure drag. (Leonardi et al., 2007)

The paper is organized as follows. The numerical methodology is introduced in \S \ref{sec:numerical}.
The flow configuration and simulation details are discussed in \S \ref{sec:simulation}. In \S \ref{sec:results}, the influence of distributed roughness on the transitional boundary layer is studied, the formation mechanisms of hairpin vortices in different roughness distributions are examined, and the base flow and global stability results are analyzed and compared to the isolated case. Finally, the paper is summarized in \S \ref{sec:summary}.

\section{Numerical methodology}\label{sec:numerical}
The governing equations and numerical method are briefly summarized. An overview of linear stability and details regarding the iterative eigenvalue solver are provided. 

\subsection{Direct numerical simulation}
The incompressible Navier-Stokes (N-S) equations are solved using the finite volume algorithm developed by \cite{mahesh2004numerical}:%. The governing equations for the momentum and continuity equations are given by the Navier-Stokes equations:
\begin{equation}
\frac{\partial u_i}{\partial t} + \frac{\partial}{\partial x_j}(u_iu_j)=-\frac{\partial p}{\partial x_i} + \nu\frac{\partial^2 u_i}{\partial x_j x_j}+K_i,~~\frac{\partial u_i}{\partial x_i} = 0,
\label{eqn:nsme}
\end{equation}
where $u_i$ and $x_i$ are the $i$-th component of the velocity and position vectors respectively, $p$ denotes pressure divided by density, $\nu$ is the kinematic viscosity of the fluid and $K_i$ is a constant pressure gradient (divided by density). Note that the density is absorbed in the pressure and $K_i$.
The algorithm is robust and emphasizes discrete kinetic energy conservation in the inviscid limit which enables it to simulate high-Re flows without adding numerical dissipation. A predictor-corrector methodology is used where the velocities are first predicted using the momentum equation, and then corrected using the pressure gradient obtained from the Poisson equation yielded by the continuity equation. The Poisson equation is solved using a multigrid pre-conditioned conjugate gradient method (CGM) using the Trilinos libraries (Sandia National Labs).

The DNS solver has been validated for a variety of problems on related wall-bounded flows, including: realistically rough superhydrophobic surfaces \citep{alame2019wall}, random rough surfaces in turbulent channel flow \citep{ma2021direct} and boundary layer transition with an isolated roughness element \citep{ma2022global}. %response of a plate in turbulent channel flow \citep{anantharamu2021response}.
%The implicit time advancement uses the second-order Crank-Nicolson discretization:
%\begin{equation}
%     \frac{\hat{u}_i-u_i^n}{\Delta t}=\frac{1}{2}[(NL+VISC)^{n+1}+(NL+VISC)^{n}] \, ,
% \end{equation}
% where the face normal velocities $V^{n+1}_N$ are linearized in time (time-lagged) such that $V^{n}_N$ is used instead; the linearization in time yields an error of $O(\Delta t^2)$, which is the same order as that of the overall scheme. All the terms expressed as $\hat{u}_i$ are taken to the left hand side and a system of equations is solved using SOR until convergence.
\subsection{Linear stability analysis}
%Linear stability analysis enables the investigation of the linearized dynamics of infinitesimal perturbations evolving on a base state. 
Modal linear stability analysis is the study of the dynamic response of a base state subject to infinitesimal perturbations \citep{theofilis2011global}. In the present work, the incompressible Navier-Stokes equations (\ref{eqn:nsme}) are linearized about a base state, $\overline{u}_i$ and $\overline{p}$. The flow can be decomposed into a base state subject to a small $O(\epsilon)$ perturbation $\Tilde{u}_i$ and $\Tilde{p}$. The linearized Navier-Stokes (LNS) equations are obtained by subtracting the base state from equations (\ref{eqn:nsme}) and can be written as follows:
\begin{equation}
\frac{\partial \Tilde{u}_i}{\partial t} + \frac{\partial}{\partial x_j}\Tilde{u}_i\overline{u}_j + \frac{\partial}{\partial x_j}\overline{u}_i\Tilde{u}_j=-\frac{\partial \Tilde{p}}{\partial x_i} + \nu\frac{\partial^2 \Tilde{u}_i}{\partial x_j x_j},~~\frac{\partial \Tilde{u}_i}{\partial x_i} = 0.
\label{eqn:lnsme}
\end{equation}
The same numerical schemes as the N-S equations are used to solve the LNS equations. The LNS equations are subject to the following %initial and
boundary conditions:
\begin{equation}
    %\Tilde{u}_i(x,y,z,t=0)=\Tilde{u}_{i}|_{t=0}\neq 0, ~~ 
    \Tilde{u}_i(S,t)=0,
\end{equation}
where $S$ is the boundary of the spatial domain.

The LNS equations can be rewritten as a system of linear equations,
\begin{equation}
\frac{\partial \Tilde{u}_i}{\partial t} = A \Tilde{u}_i,
\label{eqn:linear_sys}
\end{equation}
where $A$ is the LNS operator and $\Tilde{u}_i$ is the velocity perturbation. The solutions to the linear system of equations (\ref{eqn:linear_sys}) are:
\begin{equation}
\Tilde{u}_i(x,y,z,t)=\sum_{\omega}\hat{u}_i(x,y,z)e^{\omega t} 
\label{eqn:linear_sol}
\end{equation}
where $\hat{u_i}$ is the velocity coefficient, and both $\hat{u_i}$ and $\omega$ can be complex. %The complex conjugate in equation ($\ref{eqn:linear_sol}$) is omitted for the sake of brevity. 
This defines $\sigma=Re(\omega)$ as the growth/damping rate and $\omega_a=Im(\omega)$ as the temporal frequency of $\hat{u_i}$. The linear system of equations can then be transformed into a linear eigenvalue problem:
\begin{equation}
\Omega \hat{U}_i = A \hat{U}_i,
\label{eqn:eigen_direct}
\end{equation}
where $\omega_j = diag(\Omega)_j$ is the $j$-th eigenvalue and $\hat{u}_i^j = U_i[j,:]$ is the $j$-th eigenvector. For global stability analysis, solving the eigenvalue problem using direct methods is very computationally expensive. Instead, a matrix-free method, the implicitly restarted Arnoldi method (IRAM) implemented in the PARPACK library is used to efficiently solve for the leading eigenvalues and eigenmodes in the present work.

\section{Simulation details}\label{sec:simulation}
\begin{figure}
\centering
\includegraphics[width=130mm]{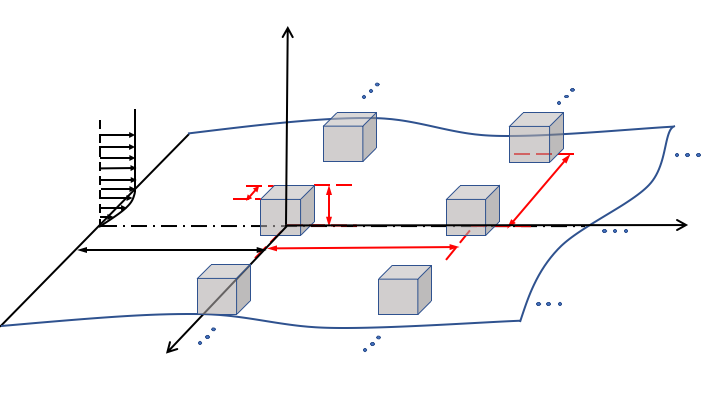}
% \put(-115,20){$L_x$}
% \put(-265,45){{$L_z$}}
% \put(-295,125){$L_y$}
\put(-193,70){\color{red}$\lambda_x$}
\put(-88,105){\color{red}$\lambda_z$}
\put(-290,70){$l$}
\put(-320,155){$U_e$}
\put(-196,100){\color{red}$h$}
\put(-256,108){\color{red}$d$}
\put(-250,83){$O$}
\put(-34,98){$x$}
\put(-295,28){{$z$}}
\put(-235,185){$y$}
\caption{Schematic of the flow configuration and roughness distribution. $x$, $y$ and $z$ coordinates are the streamwise, wall-normal and spanwise directions. The distance from the inlet of the computational domain to the first row of roughness elements remains constant as $l=15h$. The streamwise and spanwise spacing are denoted by $\lambda_x$ and $\lambda_z$ respectively.} 
\label{fig:config}
\end{figure}
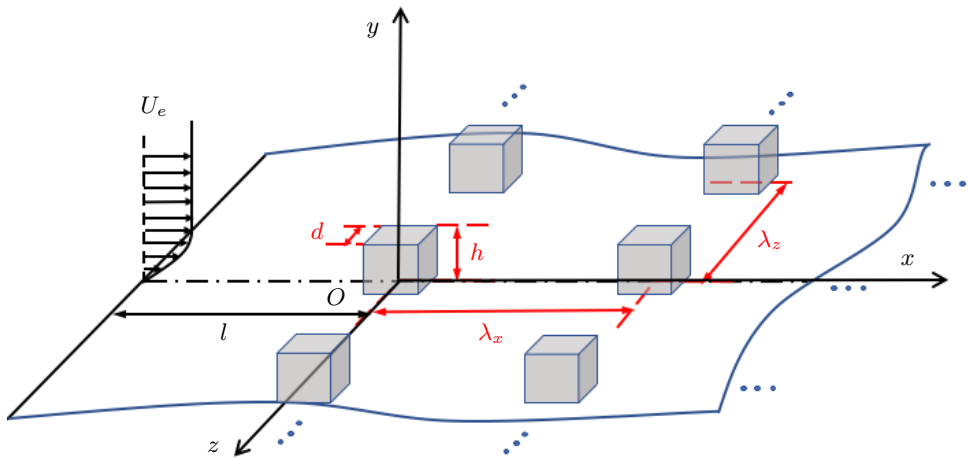

\begin{table}
\begin{center}
\def~{\hphantom{0}}
    \begin{tabular}{lccccccc}
    Case &  $\lambda_x$ & $\lambda_z$ & $N_x \times N_y \times N_z$ & $L_x \times L_y \times L_z$ & $\Delta x^+, \Delta z^+$ & $\Delta y_{wall}^+$ & $\Delta y_{top}^+$ \\
    isolated   & - & - & $1080 \times 240 \times 240$ & $45h \times 15h \times 10h$ & $0.72-2$ & $0.24-0.36$ & $12.6$ \\
    (5h,2.5h) & 5h & 2.5h & $1860 \times 240 \times 240$ & $77.5h \times 15h \times 10h$ & $0.72-2$ & $0.24-0.36$ & $12.6$ \\
    (5h,5h) & 5h & 5h & $1860 \times 240 \times 240$ & $77.5h \times 15h \times 10h$ & $0.72-2$ & $0.24-0.36$ & $12.6$ \\
    (10h,5h)   & 10h & 5h & $2160 \times 240 \times 240$ & $90h \times 15h \times 10h$ & $0.72-2$ & $0.24-0.36$ & $12.6$
    \end{tabular}
    \caption{\label{tab:grid} Simulation parameters for the isolated and distributed roughness cases at $Re_h=600$. }
    \end{center}
\end{table}

Figure \ref{fig:config} depicts the flow configuration and roughness distribution. At the inflow, a laminar Blasius boundary layer profile is prescribed. Cuboids with aspect ratio of width to height $\eta=d/h=1$ are aligned in both the streamwise and spanwise directions. The ratio of the first-row roughness height to the displacement boundary layer thickness $h/\delta^*$ is $2.86$, consistent with the isolated cube in \cite{ma2022global}. The roughness height is $h=1$, the reference length in the simulations. The Blasius laminar boundary layer solution is specified at the inflow boundary, and convective boundary conditions are used at the outflow boundary. Periodic boundary conditions are used in the spanwise direction. No-slip boundary conditions are imposed on the flat plate and the roughness surfaces. The boundary conditions $U_e=1$, $\partial v/\partial y=0$ and $\partial w/\partial y=0$ are used at the upper boundary. Uniform grids are used in the streamwise and spanwise directions, and the grid in the wall-normal direction is clustered near the flat plate. Three distributed roughness cases with different $\lambda_x$ and $\lambda_z$ are investigated in the present work. The grid size and domain length are determined based on the grid convergence and domain length sensitivity studies in \cite{ma2022global} for an isolated cube with the similar flow configuration. They are also comparable to past literature on flow simulations over cube roughness \citep{coceal2006mean,leonardi2010channel,xu2021flow}. Details of domain length and grid information are shown in table \ref{tab:grid}.

% Leonardi2010 shows density lambda=1/25=0.04 with 10h*10h domain is good.

\section{Results}\label{sec:results}

\subsection{Influence of roughness spacing on boundary layer}%Influence of roughness spacing on boundary layers}breakdown of boundary layer
\label{transition}
\subsubsection{Downstream flow separation}
\begin{figure}
\includegraphics[width=130mm,trim={0.5cm 1.0cm 0.5cm 0.2cm},clip]{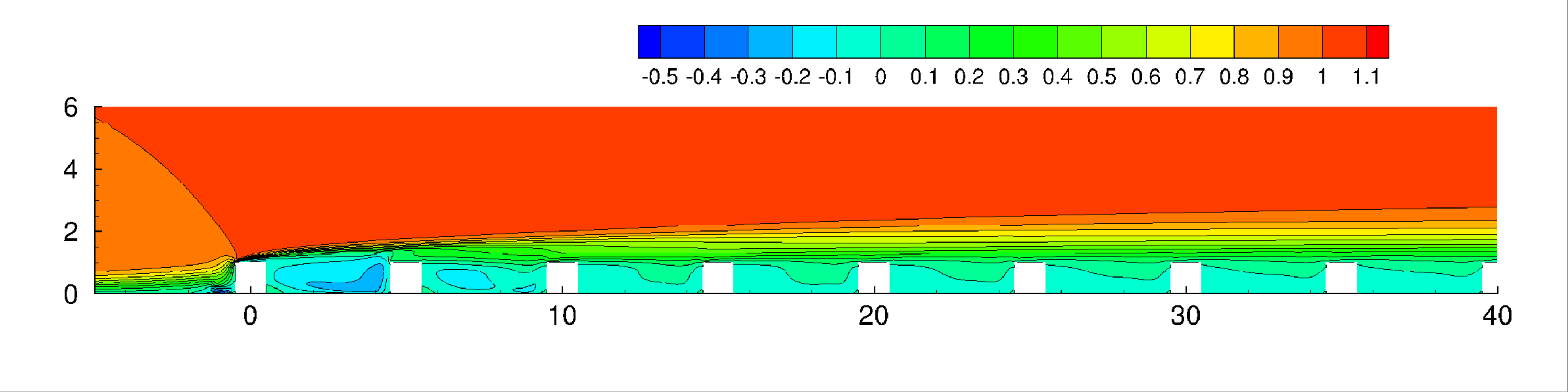}
\put(-375,55){$(a)$}
\put(-375,30){\rotatebox{90}{$y/h$}}
%\put(-183,-3){$x/h$}
\put(-242,70){\scriptsize{$u/U_e$}}
\hspace{3mm}
\includegraphics[width=130mm,trim={0.5cm 1.0cm 0.5cm 1.8cm},clip]{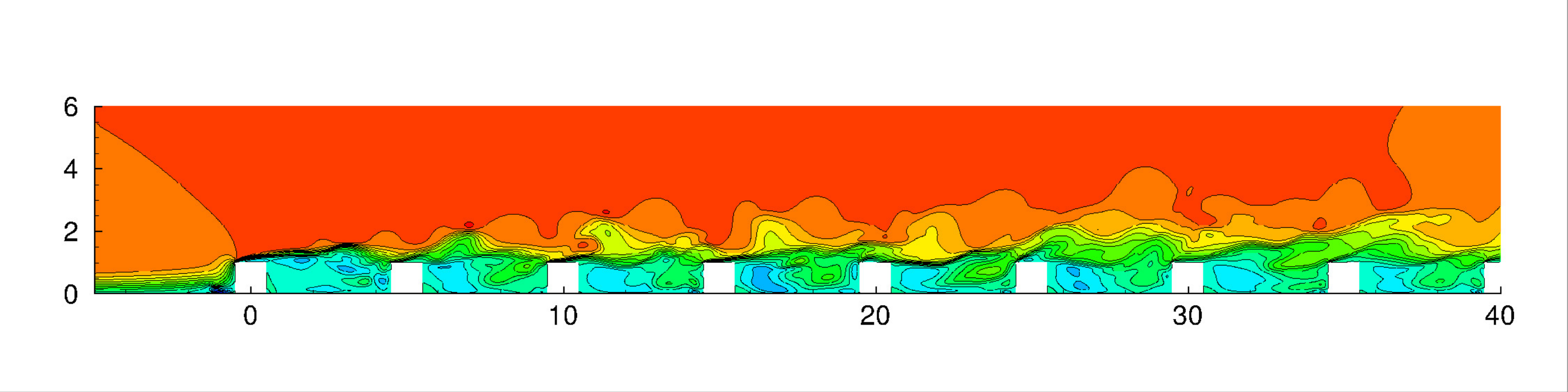}
\put(-375,55){$(b)$}
\put(-375,30){\rotatebox{90}{$y/h$}}
%\put(-183,-3){$x/h$}
%\put(-22,50){\scriptsize{$u/U_e$}}
\hspace{3mm}
% \includegraphics[width=130mm,trim={0.5cm 1.0cm 0.5cm 1.8cm},clip]{images/umean_z0_Lx5.pdf}
% \put(-375,55){$(b)$}
% \put(-375,30){\rotatebox{90}{$z/h$}}
% %\put(-183,-3){$x/h$}
% %\put(-22,50){\scriptsize{$u_d/U_e$}}
% \hspace{3mm}
\includegraphics[width=130mm,trim={0.5cm 1.0cm 0.5cm 1.8cm},clip]{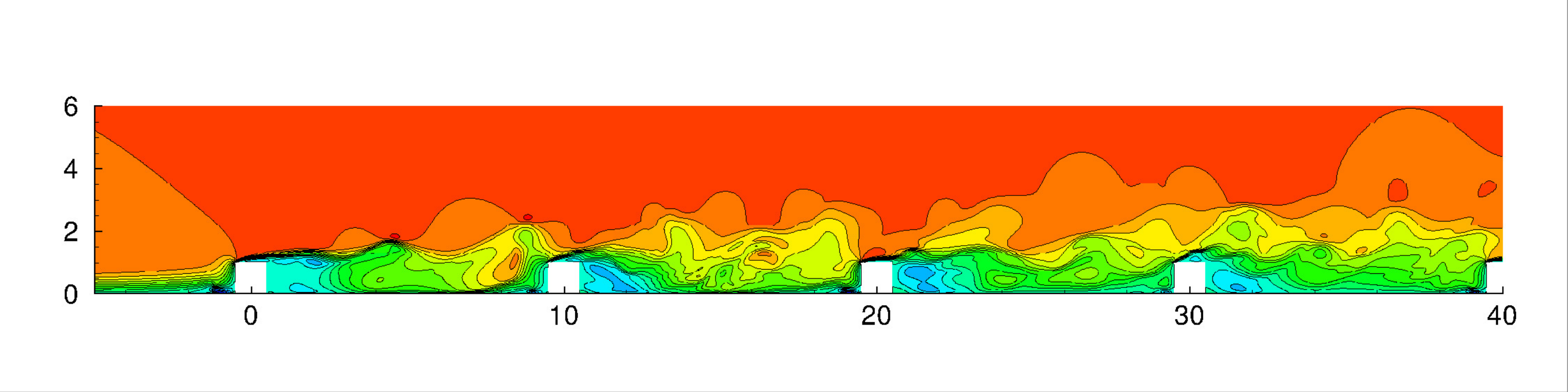}
\put(-375,55){$(c)$}
\put(-375,30){\rotatebox{90}{$y/h$}}
\put(-183,-3){$x/h$}
\hspace{3mm}
% \includegraphics[width=130mm,trim={0.5cm 1.0cm 0.5cm 1.8cm},clip]{images/umean_z0_Lx10.pdf}
% \put(-375,55){$(c)$}
% \put(-375,30){\rotatebox{90}{$z/h$}}
% \put(-183,-3){$x/h$}
% \includegraphics[width=130mm,trim={0.5cm 1.0cm 0.5cm 1.5cm},clip]{images/vorx_y05_Lx10.pdf}
% \put(-370,70){$(c)$}
% \put(-382,30){\rotatebox{90}{$z/h$}}
% \put(-180,0){$x/h$}
% %\put(-22,50){\scriptsize{$u_d/U_e$}}
% %\put(-22,50){\scriptsize{$u_d/U_e$}}
\caption{Instantaneous streamwise velocity $u/U_e$ in the symmetry plane from the DNS at $Re_h=600$ for $(a)$ Case $(5h, 2.5h)$, $(b)$ Case $(5h, 5h)$ and $(c)$ Case $(10h, 5h)$. } 
\label{fig:inst_u_z0}
\end{figure}
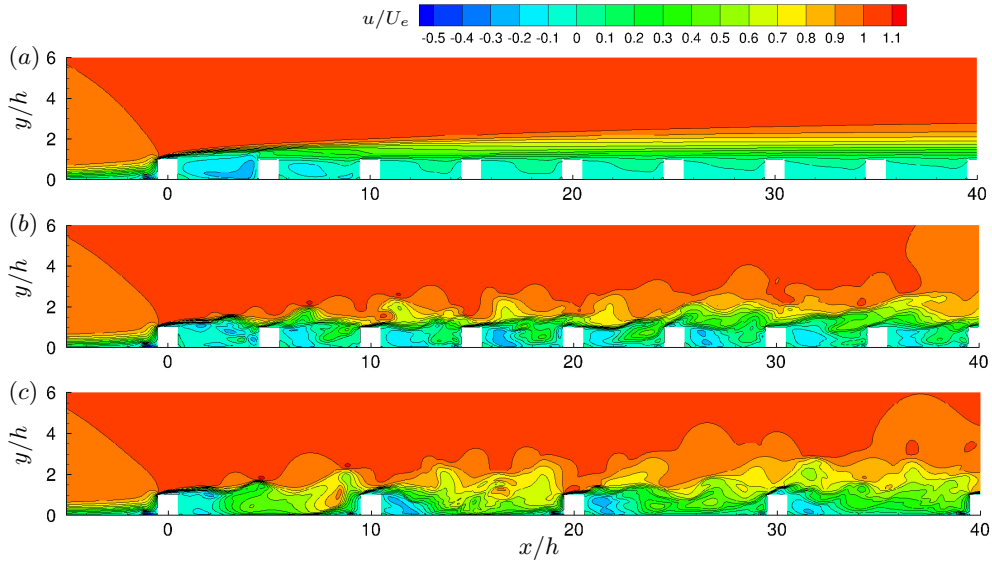

A laminar boundary layer undergoes flow separation downstream of roughness elements. The wake flow downstream of roughness elements is visualized by instantaneous streamwise velocity in the symmetry plane in figure \ref{fig:inst_u_z0}. Figure \ref{fig:inst_u_z0}$(a)$ shows that for Case $(5h, 2.5h)$, the reverse flow is strong in the first cavity and becomes weak from the second cavity onwards. Wall-normal momentum transfer hardly occurs and the boundary layer above the roughness layer remains laminar. %The reversed flow is filled within the cavities between roughness elements and the flow within the cavities turns quiescent from $x=10h$.
For Case $(5h, 5h)$, the unsteadiness of the reverse flow is observed in the first cavity in figure \ref{fig:inst_u_z0}$(b)$. %The shear layer induced by each line of roughness can be identified and its length scale is equivalent with the streamwise spacing $\lambda_x=5h$. 
The length scale of flow separation is comparable to the streamwise spacing $\lambda_x=5h$. %The region of flow separation behind a cube is known to be approximately $4$-$5h$ in a turbulent boundary layer according to \cite{castro1977flow}. 
The high-momentum fluid above roughness tips hardly penetrates into the cavities. %, and the streamwise velocity gradient remains small in the cavities. 
For a larger streamwise spacing $\lambda_x=10h$, as shown in figure \ref{fig:inst_u_z0}$(c)$, the high-momentum fluid transfers downwards into the first cavity, and impinges on the second-row roughness, which could possibly induce different instability modes. % and could induce another unstable mode. 
The penetration of high-momentum fluid into the cavities becomes weaker with increasing downstream distance. 

\subsubsection{Boundary layer integral parameters}
%To investigate the breakdown of boundary layers for 3-D distributed roughness, the boundary layer integral parameters are examined for Case $(\lambda_x,\lambda_z)=(5h, 5h)$ and Case $(\lambda_x,\lambda_z)=(10h, 5h)$. 

\begin{figure}
\includegraphics[height=40mm,trim={0.2cm 0.2cm 0.2cm 0.2cm},clip]{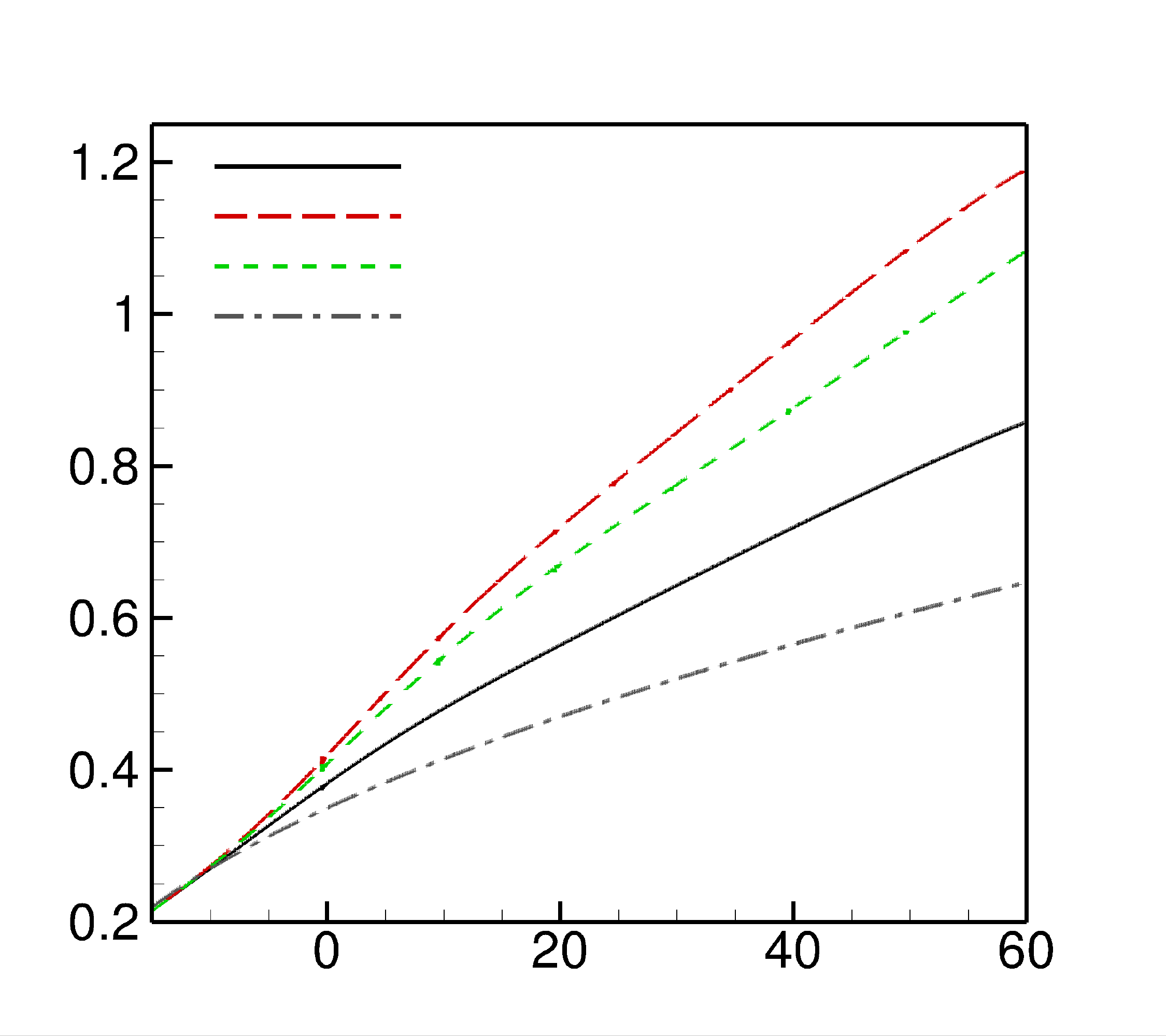}
\put(-135,100){$(a)$}
\put(-132,52){\rotatebox{90}{$\delta^*/h$}}
\put(-75,-3){$x/h$}
\put(-80,95){\scriptsize{Isolated}}
\put(-80,89){\scriptsize{$\lambda_x=5h$}}
\put(-80,83){\scriptsize{$\lambda_x=10h$}}
\put(-80,76){\scriptsize{Blasius}}
\includegraphics[height=40mm,trim={0.2cm 0.2cm 0.2cm 0.2cm},clip]{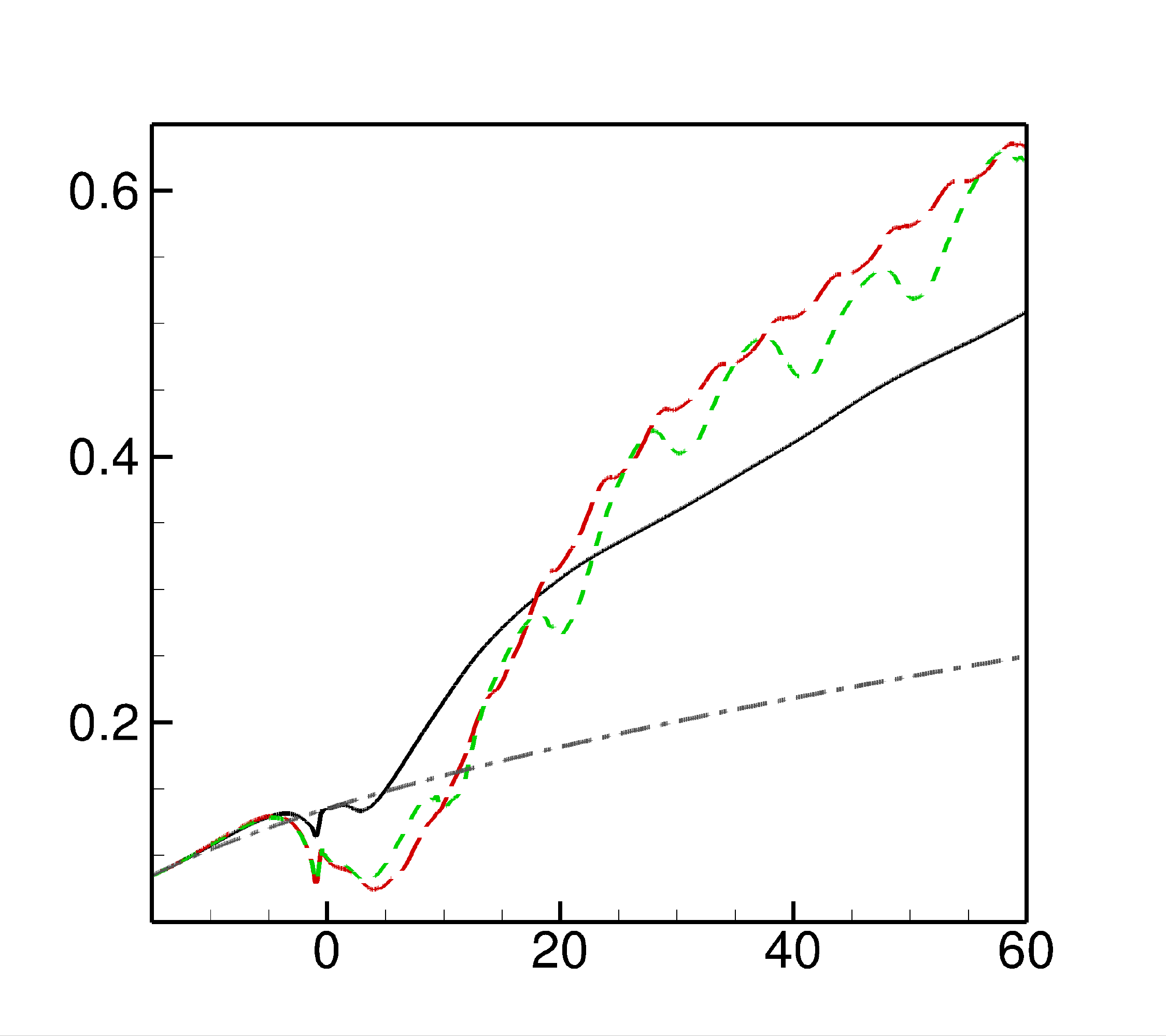}
\put(-135,100){$(b)$}
\put(-132,52){\rotatebox{90}{$\theta/h$}}
\put(-70,-3){$x/h$}
%\put(-56,136){Original}
%\put(-56,126){Processed}
\includegraphics[height=40mm,trim={0.2cm 0.2cm 0.2cm 0.2cm},clip]{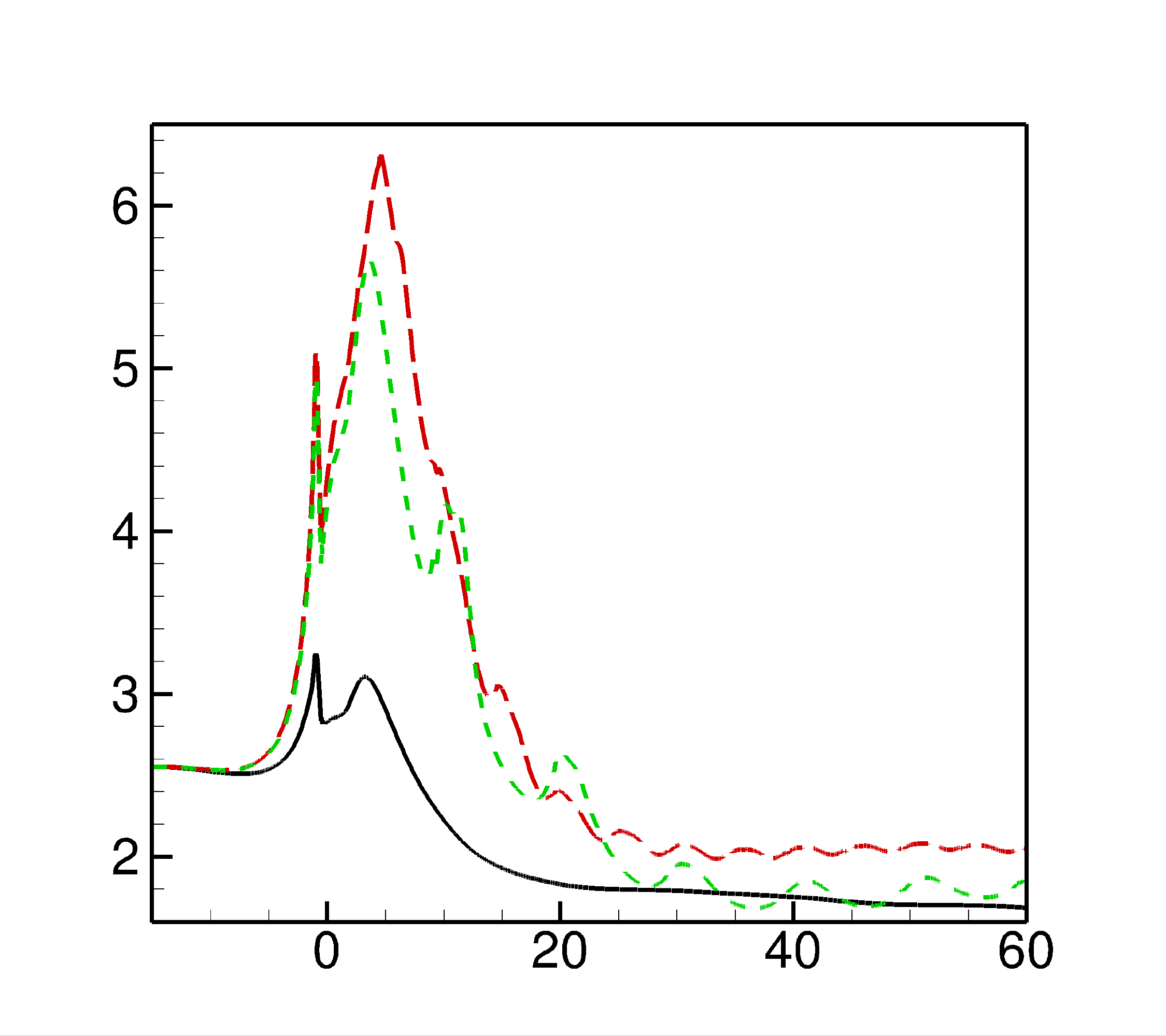}
\put(-135,100){$(c)$}
\put(-132,48){\rotatebox{90}{$H$}}
\put(-70,-3){$x/h$}
\caption{Streamwise variations of $(a)$ displacement boundary layer thickness, $(b)$ momentum boundary layer thickness and $(c)$ shape factor of Cases $(5h, 5h)$ and $(10h, 5h)$ at $Re_h=600$, with comparison to the isolated roughness and laminar Blasius solution.}
\label{fig:integral}
\end{figure}  

\begin{figure}
\includegraphics[width=130mm,trim={0.5cm 2.5cm 0.5cm 0.2cm},clip]{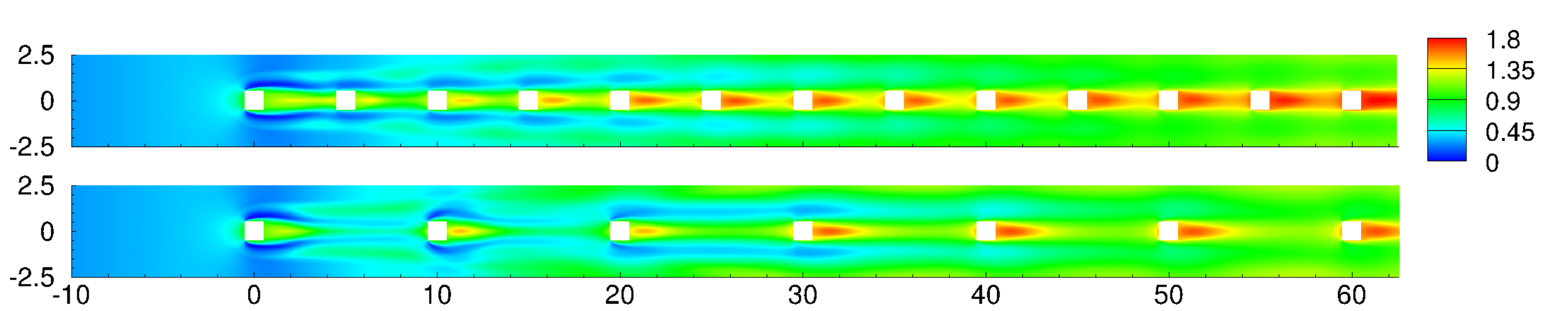}
\put(-380,63){$(a)$}
\put(-380,42){\rotatebox{90}{$z/h$}}
\put(-380,12){\rotatebox{90}{$z/h$}}
\put(-20,70){\scriptsize{$\delta^*/h$}}
%\put(-22,50){\scriptsize{$u_d/U_e$}}
\hspace{3mm}
\includegraphics[width=130mm,trim={0.5cm 2.5cm 0.5cm 0.2cm},clip]{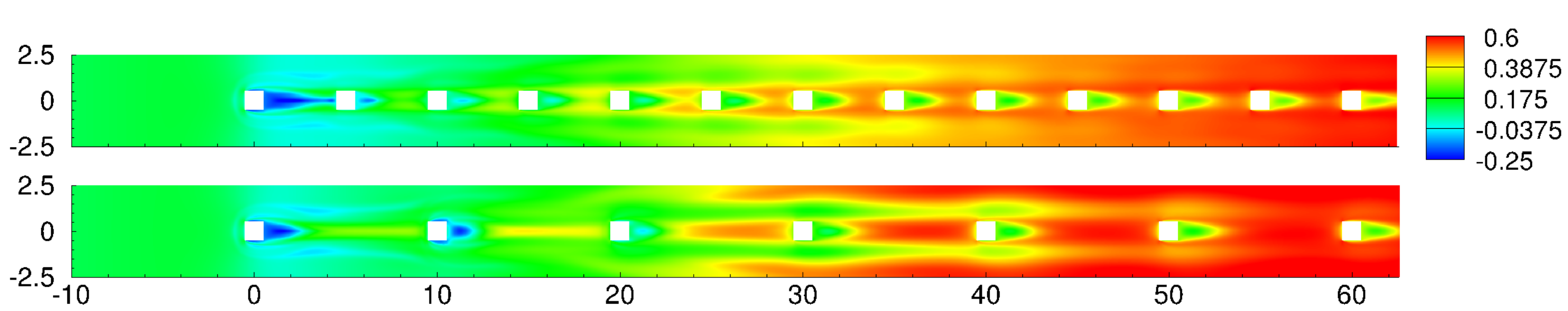}
\put(-380,63){$(b)$}
\put(-380,42){\rotatebox{90}{$z/h$}}
\put(-380,12){\rotatebox{90}{$z/h$}}
\put(-20,70){\scriptsize{$\theta/h$}}
%\put(-200,-3){$x/h$}
%\put(-22,50){\scriptsize{$u_d/U_e$}}
\hspace{3mm}
\includegraphics[width=130mm,trim={0.5cm 2.5cm 0.5cm 0.2cm},clip]{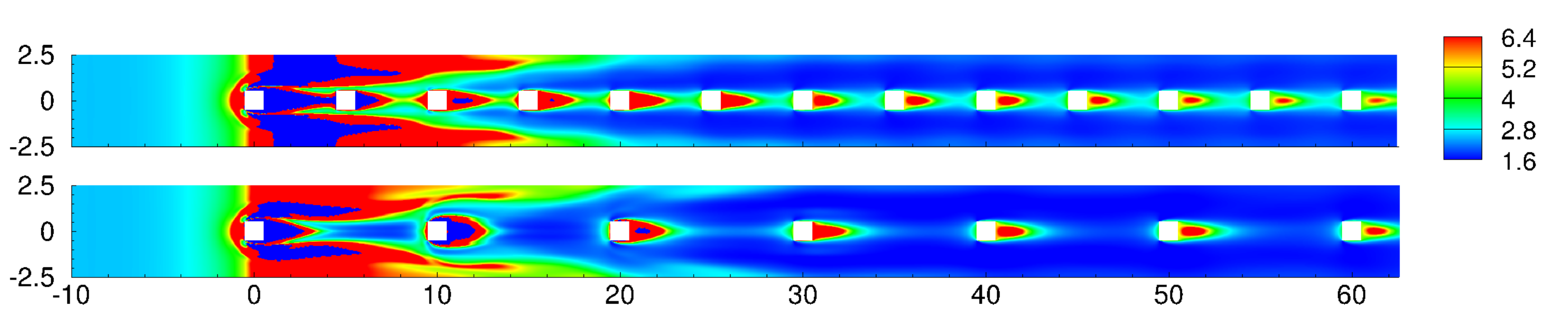}
\put(-380,63){$(c)$}
\put(-380,42){\rotatebox{90}{$z/h$}}
\put(-380,12){\rotatebox{90}{$z/h$}}
\put(-20,70){\scriptsize{$H$}}
\put(-200,-5){$x/h$}
%\put(-22,50){\scriptsize{$u_d/U_e$}}
\caption{Contour plots of the local integral boundary layer parameters for Case $(5h, 5h)$ and Case $(10h, 5h)$ at $Re_h=600$. } 
\label{fig:local_integral}
\end{figure}
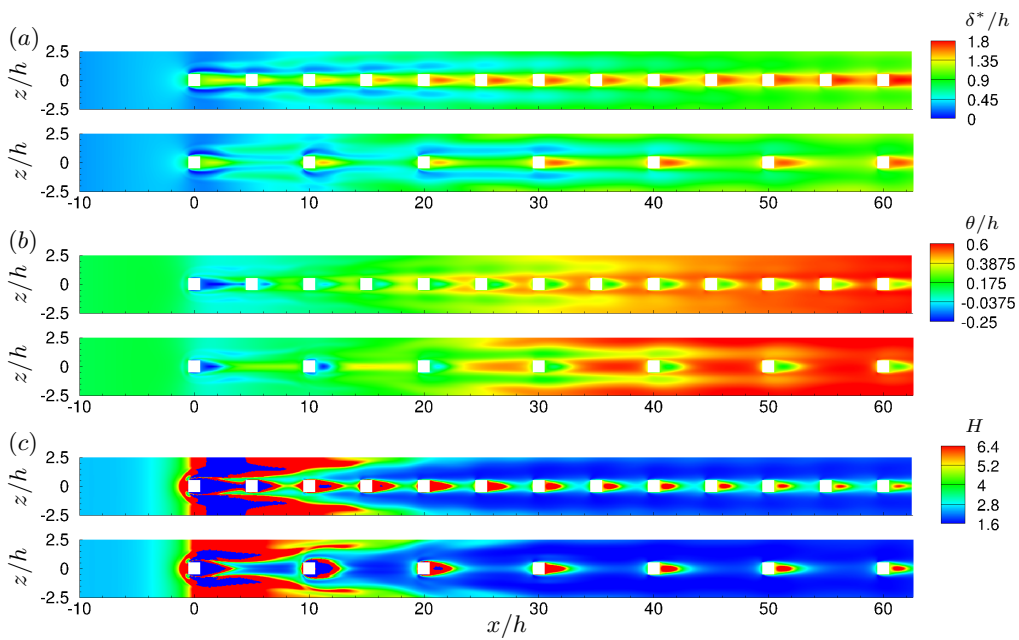

The effect of streamwise roughness spacing on transitional boundary layers is examined through boundary layer integral parameters for Cases $(5h, 5h)$ and $(10h, 5h)$, with a comparison to the isolated cube. The streamwise variation of the displacement boundary layer thickness ($\delta^*$), momentum boundary layer thickness ($\theta$) and shape factor ($H$) is presented in figure \ref{fig:integral}$(a)$-$(c)$. The parameters are computed from a time-averaged flow field and spanwise averaging is performed across the span. Comprehensive spatial averaging where the averaging volume includes both the fluid and solid area is used to ensure continuity in the profiles \citep{xie2018note}.
% \begin{equation}
%     \langle \phi \rangle_c (z) = \frac{1}{A_c}\int_{A_f} \phi(x,y,z) dxdy
% \end{equation}
% where $\phi$ is a flow quantity, $\langle  \rangle$ denotes spatial averaging, $A_c$ is the total area including that occupied by solid and $A_f$ is the fluid area. 
To understand the spatial inhomogeneity of the flow field caused by distributed roughness, the local integral boundary layer parameters are also shown in figure \ref{fig:local_integral}.

In figure \ref{fig:integral}$(a)$, an increase in $\delta^*$ is seen due to the presence of roughness elements compared to the laminar Blasius solution. A more significant increase is observed for distributed roughness than the isolated roughness. The increase is more pronounced for distributed roughness with smaller streamwise spacing. Figure \ref{fig:local_integral}$(a)$ shows that the region with lower values of local $\delta^*$ corresponds to the location of lateral high-speed streaks induced by roughness. Figure \ref{fig:integral}$(b)$ shows negative deviation from the Blasius solution in the profiles of $\theta$ for the distributed roughness cases. This is related to the strong reverse flow that occurs closely behind the first two rows of roughness elements, as demonstrated by figure \ref{fig:local_integral}$(b)$. A steeper increase in the profiles of $\theta$ indicates that the breakdown of boundary layers is more significant for distributed surface roughness compared to the isolated roughness. 

The shape factor in figure \ref{fig:integral}$(c)$ shows significant increase compared to the isolated roughness. The high values of $H$ are associated with inflection points and reveals locally the instability of the streaks induced by roughness elements, as shown in figure \ref{fig:local_integral}$(c)$. The steep drop in $H$ begins at approximately $x=5h$ for the three cases, suggesting that the streaks with high amplitudes break down and transition happens at similar downstream positions for three different roughness distributions. The shape factor converges farther downstream after the breakdown. The values are higher for Case $(5h, 5h)$ when compared to Case $(10h, 5h)$ and the isolated case, resulting from a stronger blockage effect of a denser roughness distribution.

%breakdown does not depend on roughness distribution, more likely to depend on roughness height.

\subsection{Formation of hairpin-shaped vortices}\label{hairpin}
Packets of hairpin-shaped vortices are key structures in roughness-induced transition. They are associated with global instability as known for isolated roughness.  \cite{cohen2014minimal} proposed a model consisting of three key elements required for the formation of hairpin vortices: simple shear, counter-rotating vortices and two-dimensional vortex sheet, and highlights that this combination is unstable. The influence of roughness spacing on the generation of hairpin vortices is therefore important to understand and is investigated in this section for distributed surface roughness. %Recall that the key flow elements, counter-rotating vortices and simple shear, for the generation of hairpin vortices are examined in \S \ref{cvp} and \S \ref{perturb}. 

\subsubsection{Counter-rotating vortex pairs (CVP)}\label{cvp}
% Base flow (streamwise deviation or streamwise velocity with streamlines) in cross-flow planes
% Mean streamwise vorticity field

\begin{figure}
\includegraphics[width=68mm,trim={1.8cm 0.1cm 0.1cm 0cm},clip]{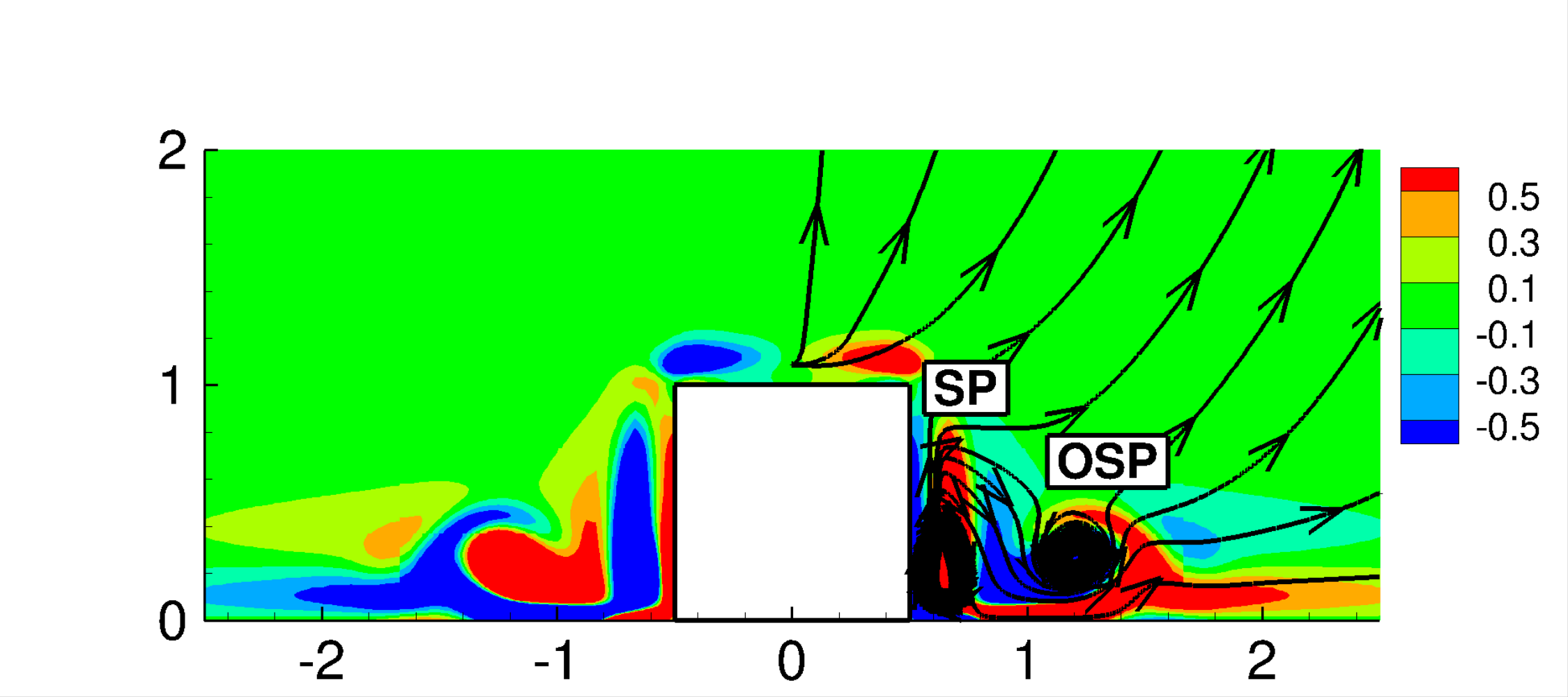}
\put(-206,35){\rotatebox{90}{$y/h$}}
\put(-110,-7){$z/h$}
\put(-22,75){\scriptsize{$\overline{\omega}_x$}}
 \put(-170,60){$x=0$}
\includegraphics[width=68mm,trim={1.8cm 0.1cm 0.1cm 0cm},clip]{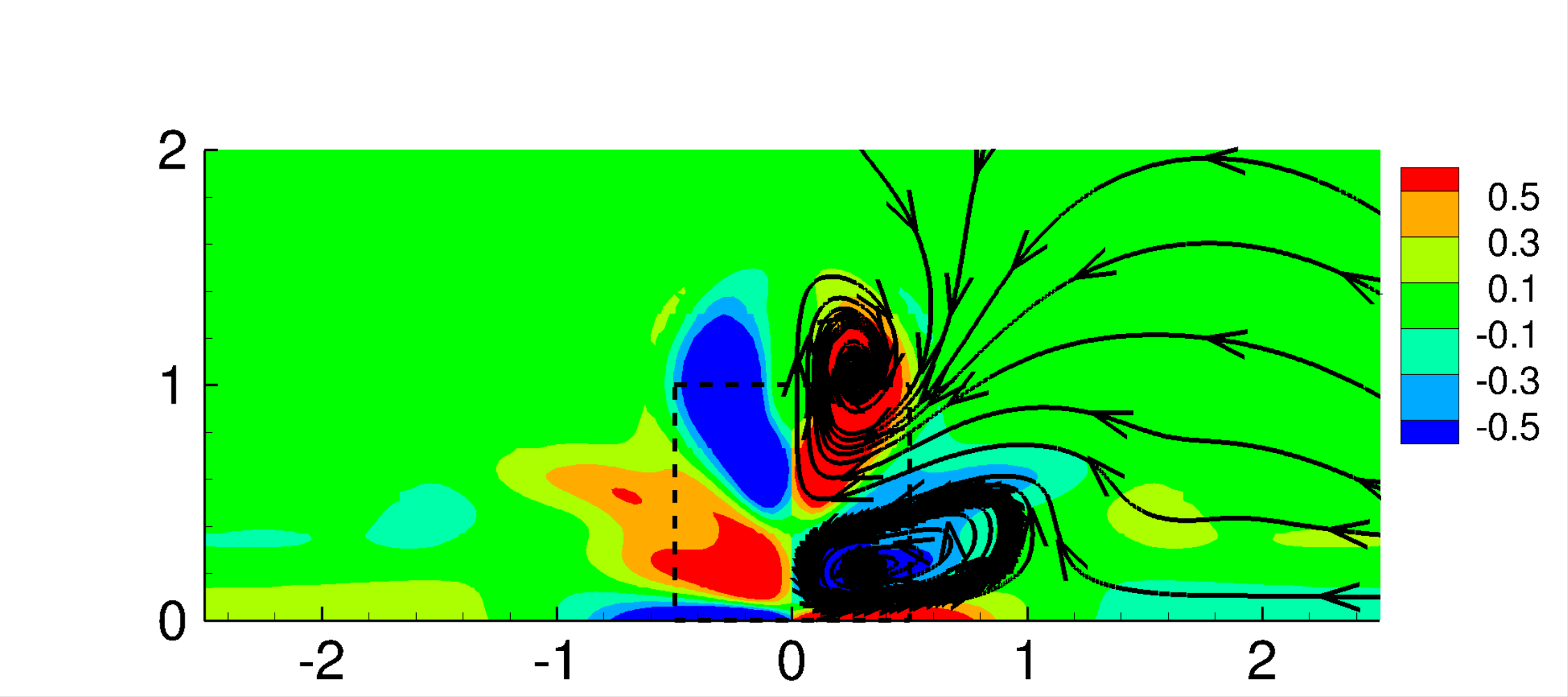}
\put(-110,-7){$z/h$}
\put(-22,75){\scriptsize{$\overline{\omega}_x$}}
 \put(-170,60){$x=4h$}
 \hspace{3mm}
\includegraphics[width=130mm,trim={0.5cm 1.0cm 0.5cm 0.5cm},clip]{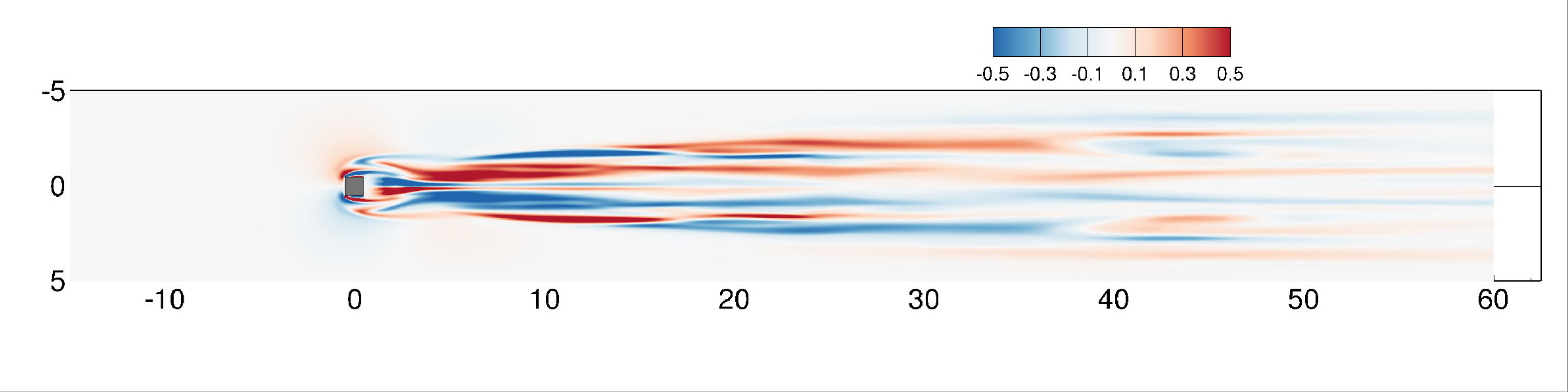}
%\put(-380,60){$(a)$}
\put(-382,30){\rotatebox{90}{$z/h$}}
\put(-180,0){$x/h$}
\put(-150,70){\scriptsize{$\overline{\omega}_x$}}
\caption{Mean streamwise vorticity $\overline{\omega}_x=\pm 0.5$ with mean streamlines in the cross-flow planes (top) and the plane of $y=0.5h$ (bottom) from the DNS at $Re_h=600$ for the isolated roughness. } 
\label{fig:vorx_iso}
\end{figure}
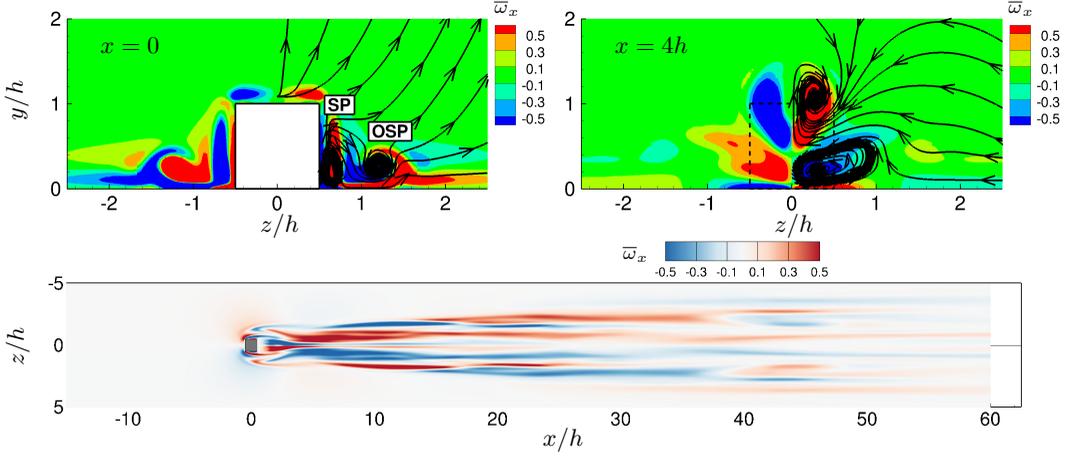

As known for isolated roughness, the spanwise vortices formed upstream wrap around the roughness element and give birth to the streamwise vortices downstream of the roughness, and the streamwise counter-rotating vortices are known to play a critical role in the generation of hairpin vortices \citep{iyer2013high,bucci2021influence}. As the baseline, figure \ref{fig:vorx_iso} depicts the characteristics of streamwise vortices induced by an isolated cube. The symmetry plane vortices (SP) %(also referred to as rear pair vortices)
located very close to the roughness have an upwash between them. They lift up and move towards each other with increasing downstream distance, and are dissipated at $x=25h$. The off-symmetry plane vortices (OSP) located farther from the symmetry plane are the continuation of the vortex tubes from upstream. They have a central downwash which keeps them away from each other with increasing downstream distance.

%msi/laminar_BL_Diaz_Re600_DNS_medium_domain/data_files/re600/

\begin{figure}
\includegraphics[width=140mm,trim={0.2cm 0.1cm 0.1cm 0.5cm},clip]{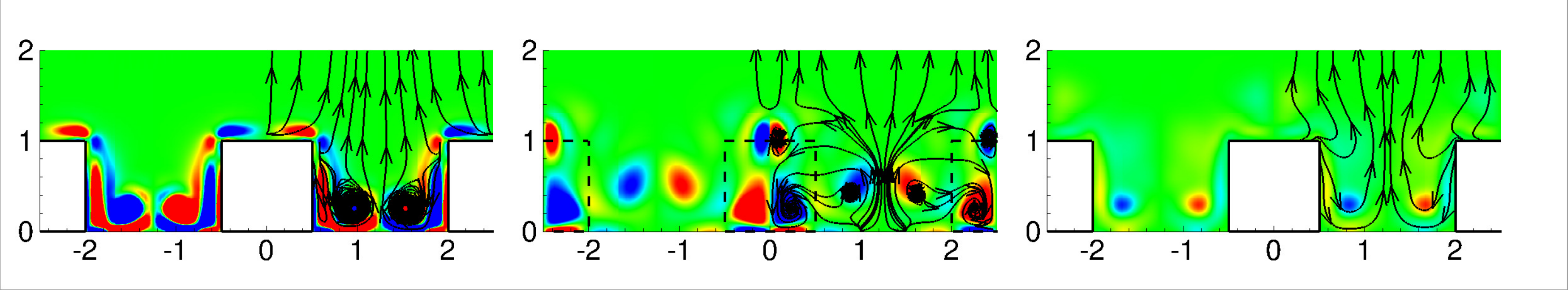}
\put(-413,30){\rotatebox{90}{$y/h$}}
\put(-80,-3){$z/h$}
\put(-210,-3){$z/h$}
\put(-340,-3){$z/h$}
\put(-340,65){$x=0$}
\put(-215,65){$x=4h$}
\put(-90,65){$x=10h$}
\put(-413,60){$(a)$}
\hspace{3mm}
\centering
\includegraphics[width=90mm,trim={0.5cm 2.0cm 0.5cm 4cm},clip]{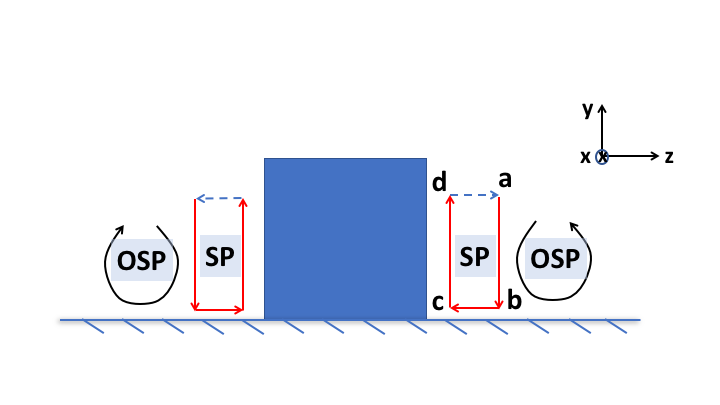}
\put(-250,80){$(b)$}
 \hspace{3mm}
\includegraphics[width=130mm,trim={0.5cm 1.0cm 0.5cm 1.5cm},clip]{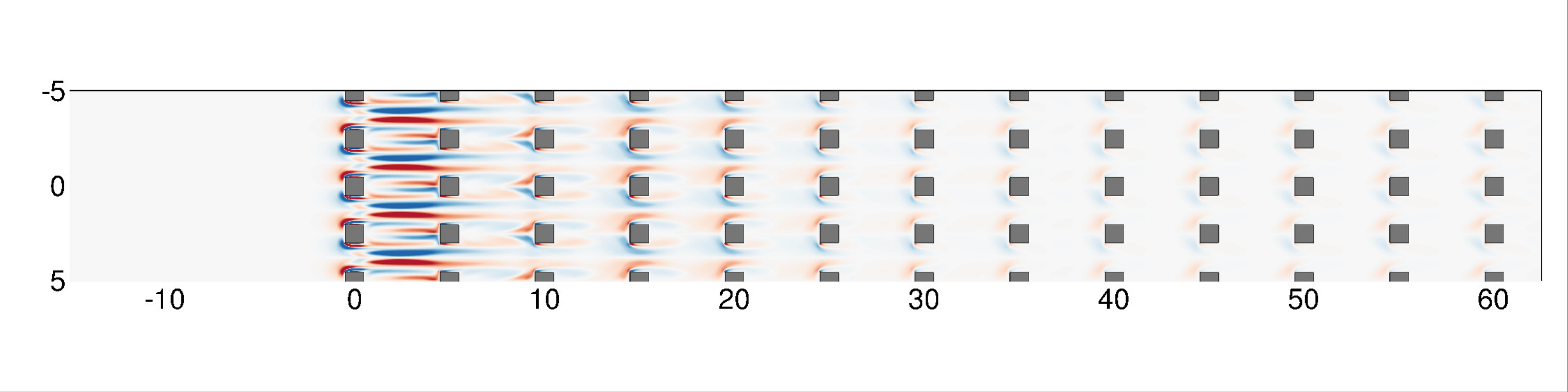}
\put(-383,60){$(c)$}
\put(-382,30){\rotatebox{90}{$z/h$}}
\put(-180,0){$x/h$}
%\put(-22,50){\scriptsize{$u_d/U_e$}}
\caption{$(a)$ Mean streamwise vorticity $\overline{\omega}_x=\pm 0.5$ with mean streamlines in the cross-flow planes, $(b)$ illustration of the formation mechanisms of SP vortices and $(c)$ streamwise evolution of $\overline{\omega}_x$ at the plane of $y=0.5h$ from the DNS of Case $(5h, 2.5h)$ at $Re_h=600$. The contour levels are the same as figure \ref{fig:vorx_iso}.} 
\label{fig:vorx_Lz25}
\end{figure}
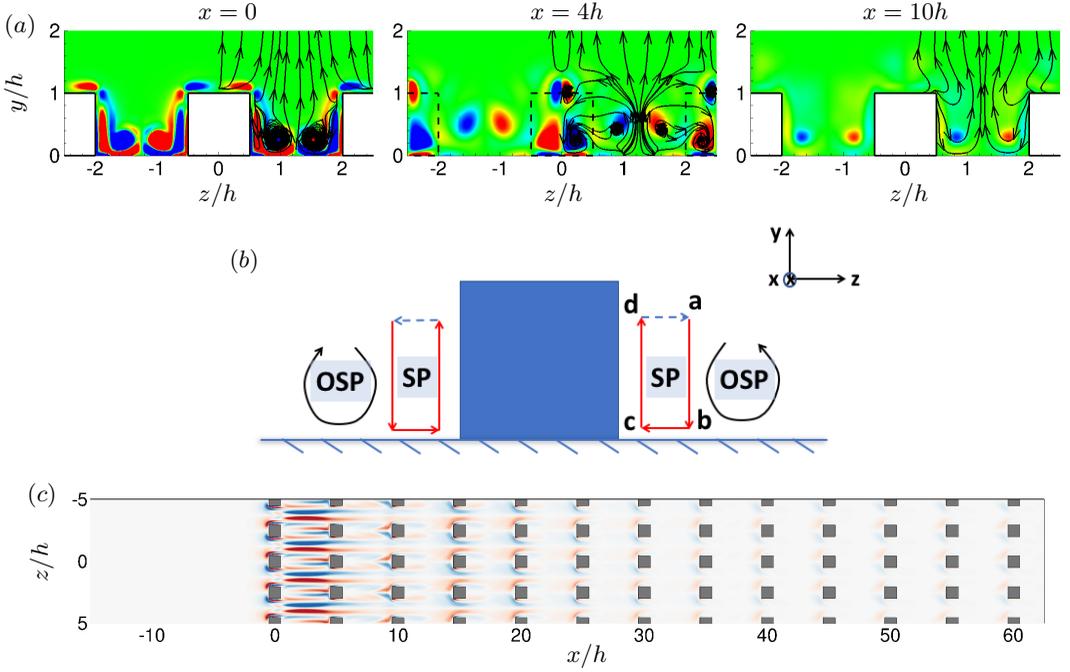

The effects of spanwise spacing on CVP are investigated for Case $(5h, 2.5h)$ in figure \ref{fig:vorx_Lz25}. %Figure \ref{fig:vorx_Lz25} shows the contours of mean streamwise vorticity with mean streamlines in the cross-flow planes and the streamwise evolution of mean streamwise vorticity in the plane of $y=0.5h$. 
At $x=0$, the OSP vortices are observed in the groove between two adjacent cubes, similar as the isolated case. However, the SP vortices do not appear on either side of obstacles, in contrast to the isolated case. The generation mechanism of the SP vortices is examined using the mean streamlines in figure \ref{fig:vorx_Lz25}$(a)$ and illustrated in figure \ref{fig:vorx_Lz25}$(b)$. The secondary flow close to the cube sides moves downward due to the motion of the OSP vortices (from $a$ to $b$), then moves toward the cube due to a positive spanwise pressure gradient (from $b$ to $c$), and moves upward for mass conservation (from $c$ to $d$). With small spanwise spacing, the OSP vortices in the groove are located closer toward each other, which strengthens the upward fluid motion in the groove and weakens the centrifugal forces. The last step under the effects of centrifugal forces for the generation of SP vortices (from $d$ to $a$) is therefore missing. As a result, a weak CVP is observed at the roughness tip location at $x=4h$, and is dissipated within a short downstream distance. %and disappears at $x=10h$.
The OSP vortices in the groove are also diminished with increasing downstream distance and mostly disappears beyond the second row of roughness elements, as shown in figure \ref{fig:vorx_Lz25}$(c)$. %A pair of downwash streamwise vortices is observed upstream of each cube in figure \ref{fig:vorx_Lz25}$(b)$, which is produced by the impulsive upward fluid motion induced by the obstacles. %They are also observed in front of each line of roughness elements in figure \ref{fig:vorx_Lz25}$(b)$, which have a less significant impact on the generation of CVP.

\begin{figure}
\includegraphics[width=140mm,trim={0.2cm 0.1cm 0.1cm 0.5cm},clip]{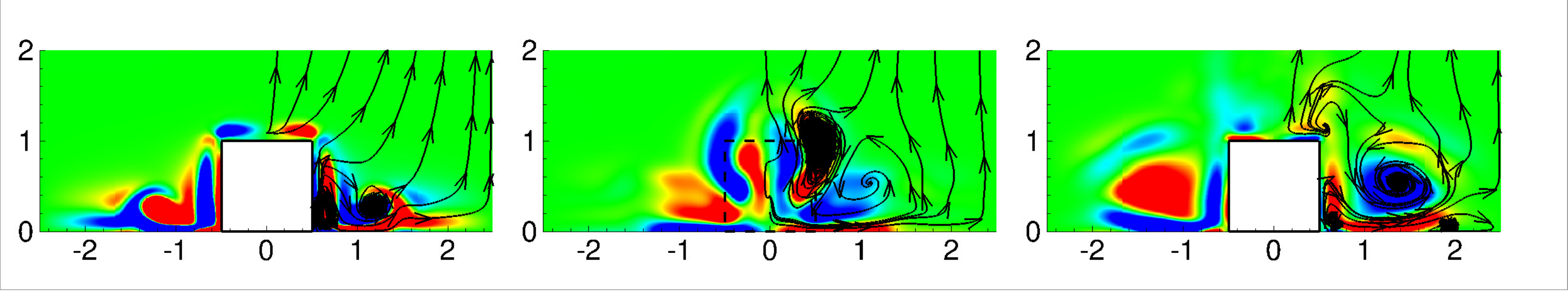}
\put(-413,30){\rotatebox{90}{$y/h$}}
\put(-340,65){$x=0$}
\put(-215,65){$x=4h$}
\put(-90,65){$x=10h$}
\put(-413,60){$(a)$}
 \hspace{3mm}
 \includegraphics[width=140mm,trim={0.2cm 0.1cm 0.1cm 0.5cm},clip]{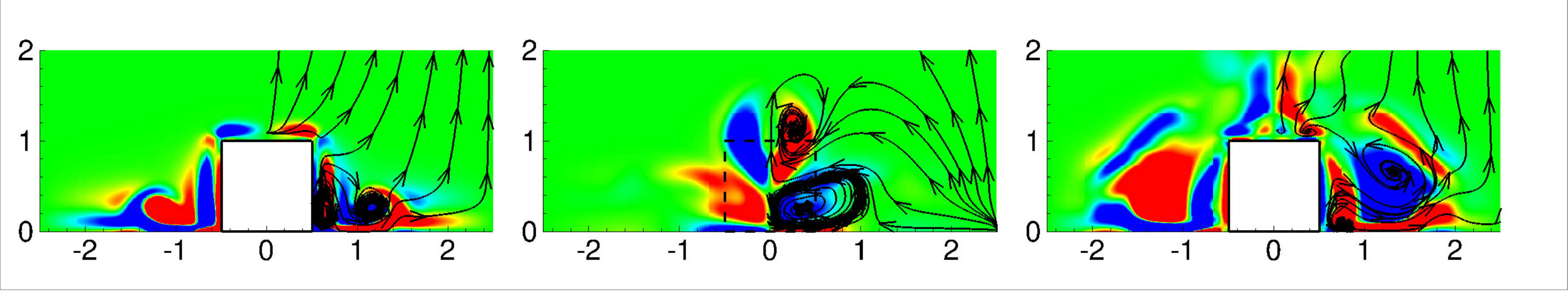}
\put(-413,30){\rotatebox{90}{$y/h$}}
\put(-80,-3){$z/h$}
\put(-210,-3){$z/h$}
\put(-340,-3){$z/h$}
\put(-413,60){$(b)$}
 \hspace{3mm}
\includegraphics[width=130mm,trim={0.5cm 1.0cm 0.5cm 1.5cm},clip]{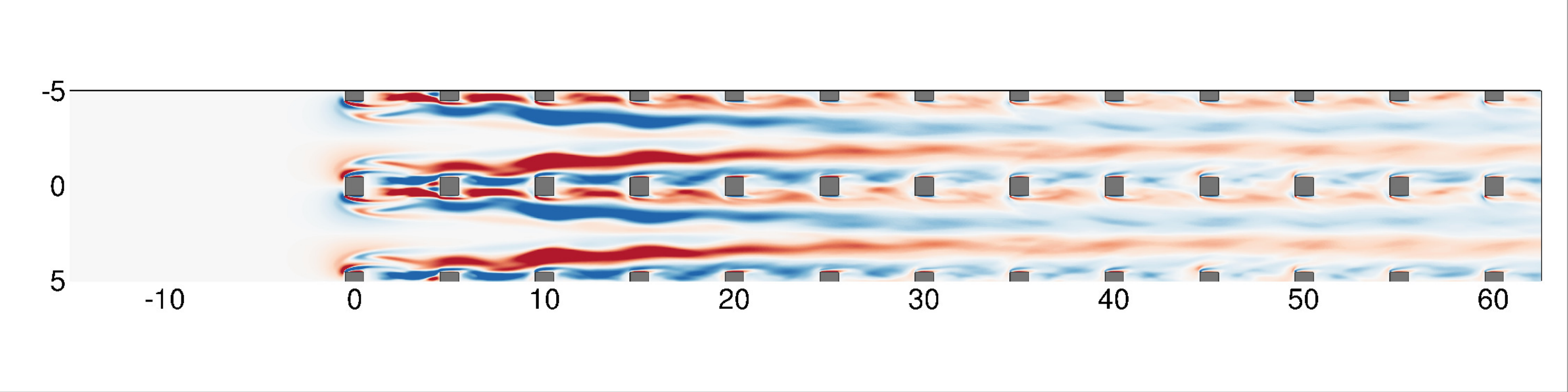}
\put(-383,60){$(c)$}
\put(-382,30){\rotatebox{90}{$z/h$}}
% \put(-180,0){$x/h$}
%\put(-22,50){\scriptsize{$u_d/U_e$}}
 \hspace{3mm}
\includegraphics[width=130mm,trim={0.2cm 1.0cm 4.35cm 1.5cm},clip]{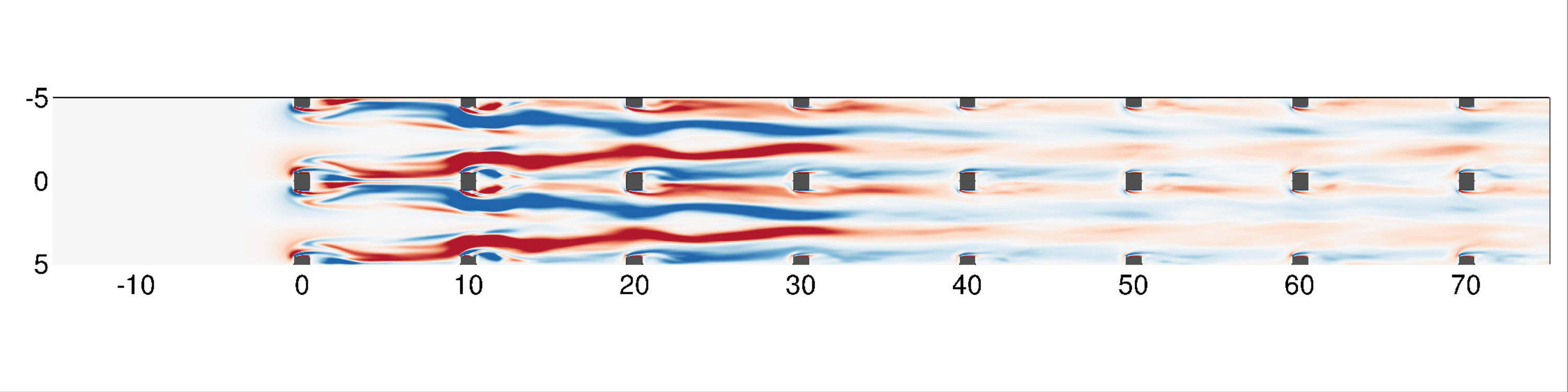}
\put(-383,65){$(d)$}
\put(-382,30){\rotatebox{90}{$z/h$}}
\put(-180,5){$x/h$}
\caption{Mean streamwise vorticity $\overline{\omega}_x=\pm 0.5$ with mean streamlines in the cross-flow planes and the plane of $y=0.5h$ from the DNS at $Re_h=600$ for $(a,c)$ Case $(5h, 5h)$ and $(b,d)$ Case $(10h, 5h)$. The contour levels are the same as figure \ref{fig:vorx_iso}.} 
\label{fig:vorx_Lz5}
\end{figure}
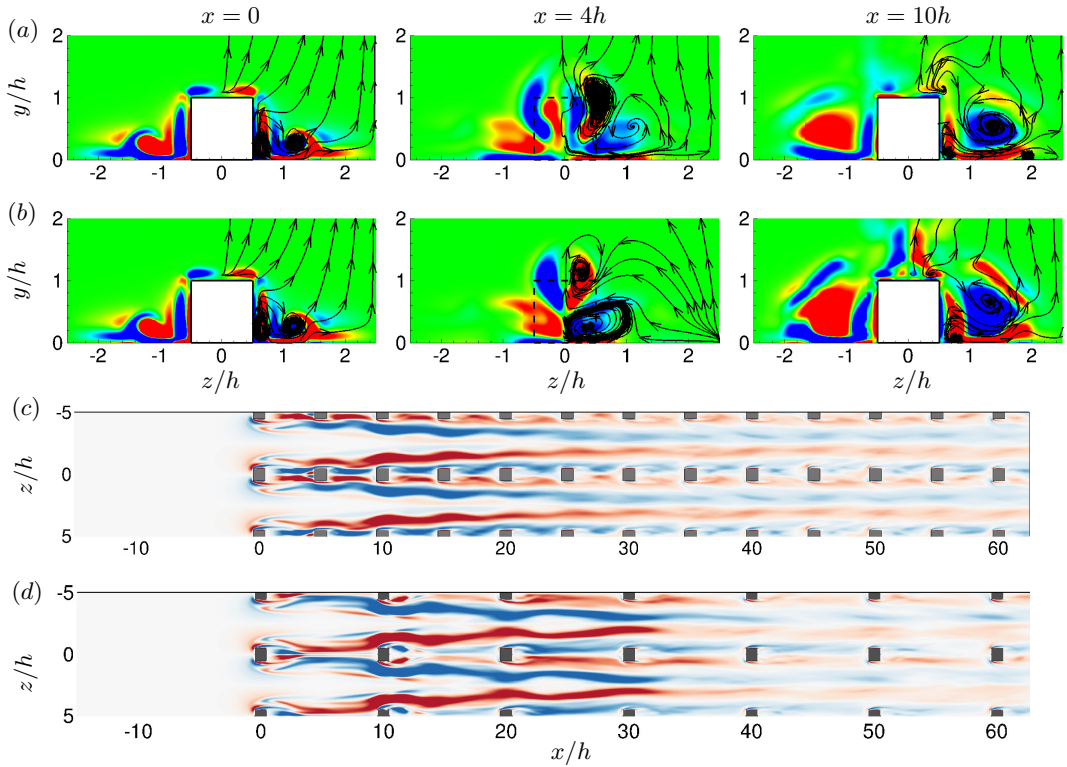

The effects of streamwise spacing on CVP are investigated in figure \ref{fig:vorx_Lz5} for Cases $(5h, 5h)$ and $(10h, 5h)$. For both cases, the SP and OSP vortices are observed and behave similarly as the isolated roughness at $x=0$. Figure \ref{fig:vorx_Lz5}$(a)$ shows that for Case $(5h, 5h)$, the CVP grows with increasing downstream distance, and is distorted and pushed away from each other by the following roughness as observed at $x=4h$. At $x=10h$, the SP vortices developed from the front obstacles are weakened due to the stagnation effects of the following obstacles, but a new pair of SP vortices is induced again on either side of the cube, and the OSP vortices on the lateral sides are strengthened. Figure \ref{fig:vorx_Lz5}$(c)$ shows that the CVP is dissipated beyond $x=30h$, similarly as the isolated roughness case. %Figure \ref{fig:vorx_Lz5}$(c)$ depicts a clear view that even though the roughness elements have stagnation effects on the CVP, the SP vortices are induced again and the OSP vortices are enhanced by the following roughness elements. The CVP are dissipated at $x=30h$, similarly as the isolated roughness case. %\cite{vanderwel2015effects} suggested that rough surfaces with a streamwise spacing less than $4$-$5h$ would act like continuous strips since the separation region behind a cube is known to be $4$-$5h$. 

For Case $(10h, 5h)$, the streamwise spacing is much larger than the streamwise length scale of the separation region. The behavior of CVP is similar as that for an isolated roughness at $x=0$ and $4h$, as shown in figure \ref{fig:vorx_Lz5}$(b)$. Instead of being distorted by the roughness as Case $(5h, 5h)$, both the SP and OSP vortices move closer towards the symmetry plane, enhancing the momentum exchange within the cavities. At $x=10h$, they impinge onto the second row of roughness elements, and break down into small vortical structures. A new pair of SP vortices is generated and the OSP vortices are strengthened on the lateral sides. Figure \ref{fig:vorx_Lz5}$(d)$ shows that after the second-row cubes, the SP vortices maintain in the cavities while the OSP vortices develop in the longitudinal grooves. The CVP persists farther than Case $(5h, 5h)$, and is dissipated beyond $x=40h$. %This indicates that as the wake flow of the first-row roughness impinges on the second-row roughness, another unstable mode is possibly associated with the wake induced by the second-line roughness.

% \begin{figure}
% % \includegraphics[width=130mm,trim={0.5cm 1.0cm 0.5cm 1.5cm},clip]{images/vorx_y05_Lz25.pdf}
% % \put(-380,60){$(b)$}
% % \put(-382,30){\rotatebox{90}{$z/h$}}
% % %\put(-180,0){$x/h$}
% % %\put(-22,50){\scriptsize{$u_d/U_e$}}
% % \hspace{3mm}
% \includegraphics[width=130mm,trim={0.5cm 1.0cm 0.5cm 1.5cm},clip]{images/vorx_y05_Lx5.pdf}
% \put(-380,60){$(c)$}
% \put(-382,30){\rotatebox{90}{$z/h$}}
% %\put(-180,0){$x/h$}
% %\put(-22,50){\scriptsize{$u_d/U_e$}}
% \hspace{3mm}
% \includegraphics[width=130mm,trim={0.2cm 1.0cm 4.35cm 1.5cm},clip]{images/vorx_y05_Lx10.pdf}
% \put(-380,65){$(d)$}
% \put(-382,30){\rotatebox{90}{$z/h$}}
% \put(-180,5){$x/h$}
% %\put(-22,50){\scriptsize{$u_d/U_e$}}
% %\put(-22,50){\scriptsize{$u_d/U_e$}}
% \caption{Mean streamwise vorticity $\overline{\omega}_x=\pm 0.5$ in the plane of $y=0.5h$ from the DNS at $Re_h=600$ for $(a)$ Case $(\lambda_x,\lambda_z)=(5h, 2.5h)$, $(b)$ Case $(\lambda_x,\lambda_z)=(5h, 5h)$ and $(c)$ Case $(\lambda_x,\lambda_z)=(10h, 5h)$. } 
% \label{fig:vorx_yslice}
% \end{figure}

\subsubsection{Perturbation to the shear layer}\label{perturb}

\begin{figure}
\includegraphics[width=130mm,trim={0.5cm 1.0cm 0.5cm 0.5cm},clip]{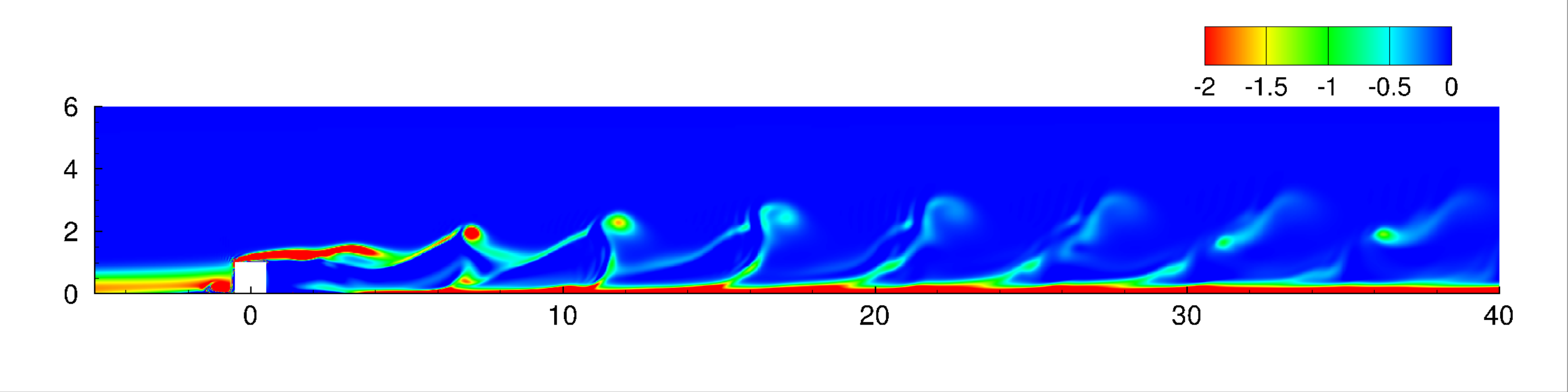}
\put(-375,55){$(a)$}
\put(-375,30){\rotatebox{90}{$y/h$}}
%\put(-183,-3){$x/h$}
\put(-96,70){\scriptsize{$\omega_z$}}
\hspace{3mm}
\includegraphics[width=130mm,trim={0.5cm 1.0cm 0.5cm 1.8cm},clip]{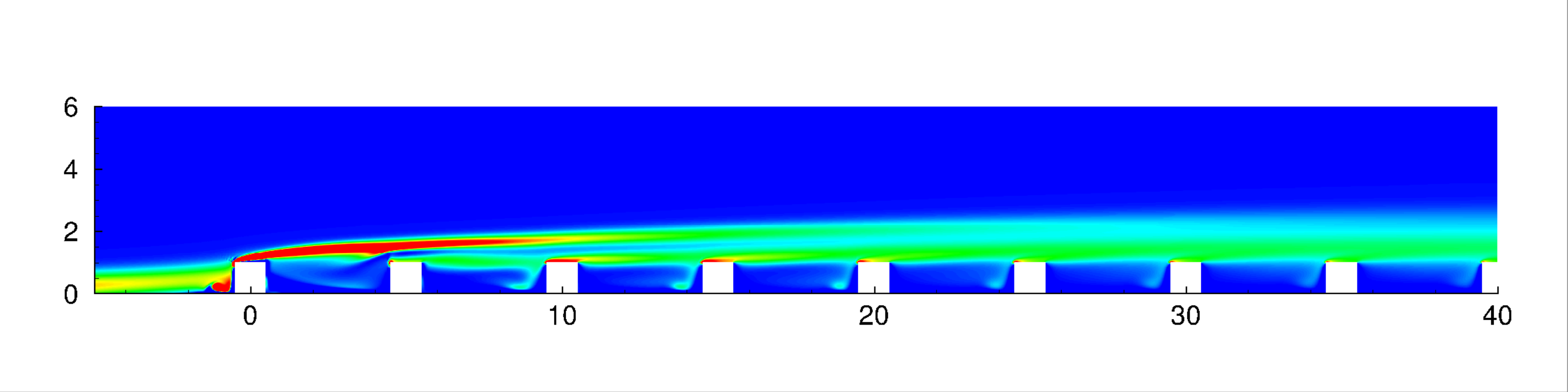}
\put(-375,55){$(b)$}
\put(-375,30){\rotatebox{90}{$y/h$}}
%\put(-183,-3){$x/h$}
%\put(-22,50){\scriptsize{$u_d/U_e$}}
\hspace{3mm}
\includegraphics[width=130mm,trim={0.5cm 1.0cm 0.5cm 1.8cm},clip]{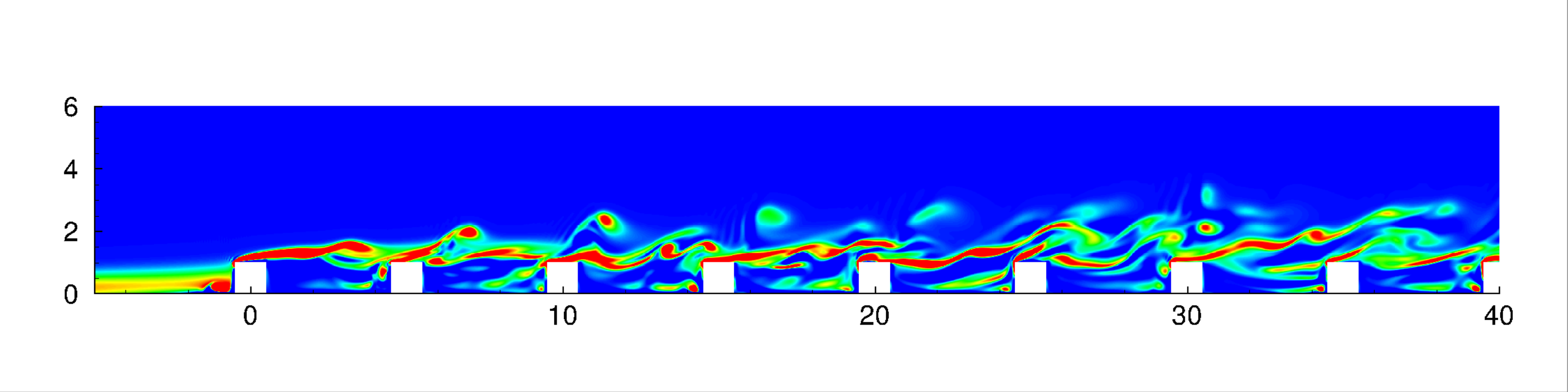}
\put(-375,55){$(c)$}
\put(-375,30){\rotatebox{90}{$y/h$}}
%\put(-183,-3){$x/h$}
%\put(-22,50){\scriptsize{$u_d/U_e$}}
\hspace{3mm}
\includegraphics[width=130mm,trim={0.5cm 1.0cm 0.5cm 1.8cm},clip]{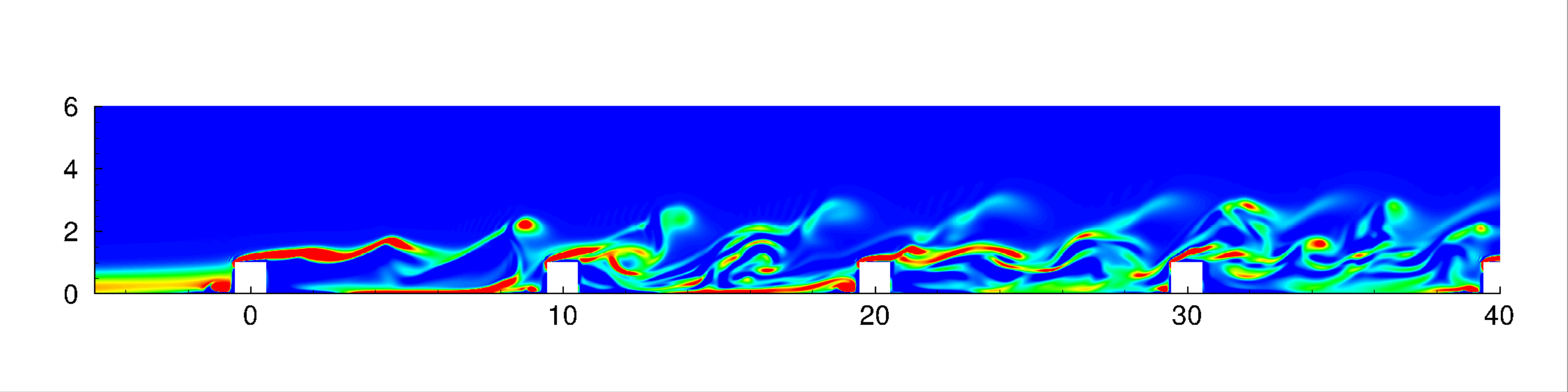}
\put(-375,55){$(d)$}
\put(-375,30){\rotatebox{90}{$y/h$}}
\put(-183,-3){$x/h$}
%\put(-22,50){\scriptsize{$u_d/U_e$}}
% \hspace{3mm}
% \includegraphics[width=130mm,trim={0.5cm 1.0cm 0.5cm 1.5cm},clip]{images/vorx_y05_Lx10.pdf}
% \put(-370,70){$(c)$}
% \put(-382,30){\rotatebox{90}{$z/h$}}
% \put(-180,0){$x/h$}
% %\put(-22,50){\scriptsize{$u_d/U_e$}}
% %\put(-22,50){\scriptsize{$u_d/U_e$}}
\caption{Instantaneous spanwise vorticity in the symmetry plane from the DNS at $Re_h=600$ for $(a)$ isolated roughness, $(b)$ Case $(5h, 2.5h)$, $(c)$ Case $(5h, 5h)$ and $(d)$ Case $(10h, 5h)$. } 
\label{fig:vorz_zslice}
\end{figure}
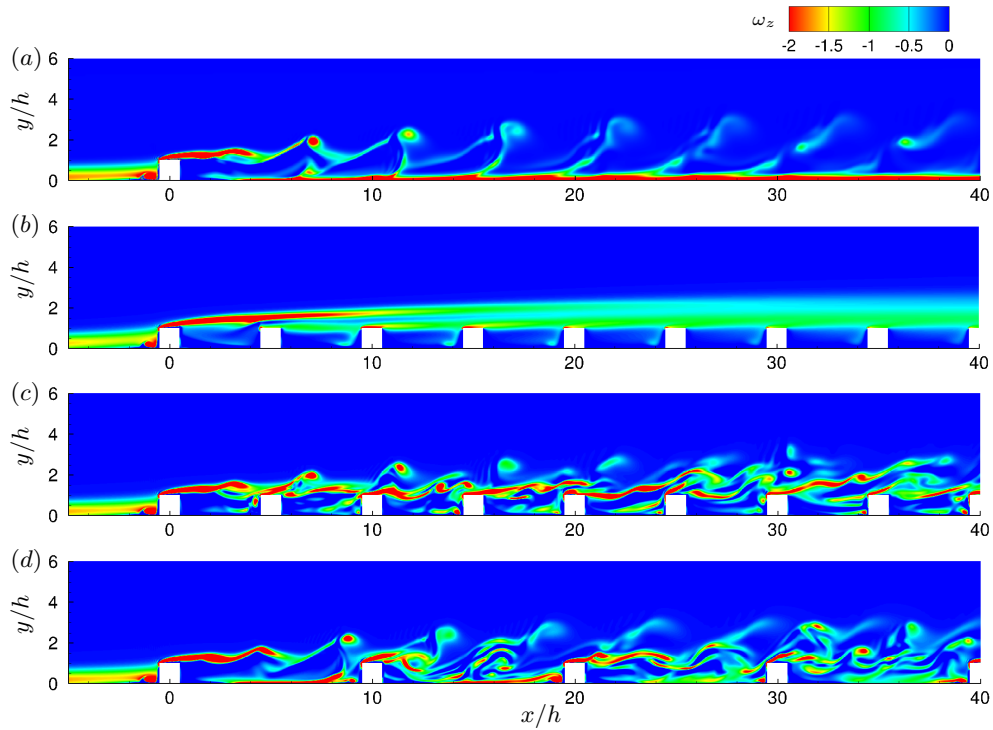

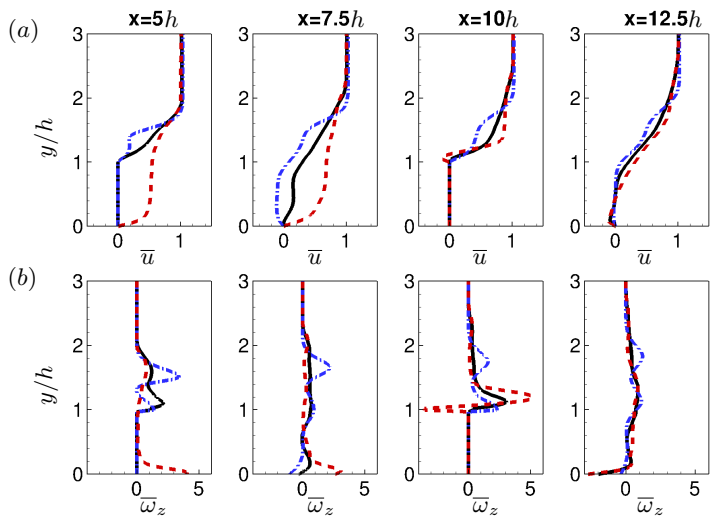
\begin{figure}
\centering
% \includegraphics[width=135mm,trim={0.1cm 0.2cm 0.5cm 0.5cm},clip]{images/inflec_Lx5.pdf}
% \put(-395,65){\rotatebox{90}{$y$}}
% \put(-195,0){$\overline{u}$}
% % \put(-180,160){$(a)$}
% \hspace{3mm}
% \includegraphics[width=135mm,trim={0.1cm 0.2cm 0.5cm 0.5cm},clip]{images/inflec_Lz25.pdf}
% \put(-395,65){\rotatebox{90}{$y$}}
% \put(-195,0){$\overline{u}$}
% % \put(-180,160){$(a)$}
% \includegraphics[width=135mm,trim={0.1cm 0.2cm 0.5cm 0.5cm},clip]{images/umean_combine.pdf}
% \put(-395,65){\rotatebox{90}{$y/h$}}
% \put(-195,0){$\overline{u}$}
% \put(-395,120){$(a)$}
%  \hspace{3mm}
% \includegraphics[width=135mm,trim={0.1cm 0.2cm 0.5cm 0.5cm},clip]{images/inflec_combine.pdf}
% \put(-395,65){\rotatebox{90}{$y/h$}}
% \put(-195,0){$\overline{\omega}_z$}
% \put(-395,120){$(b)$}
\includegraphics[width=85mm,trim={0.1cm 0.1cm 0.5cm 0cm},clip]{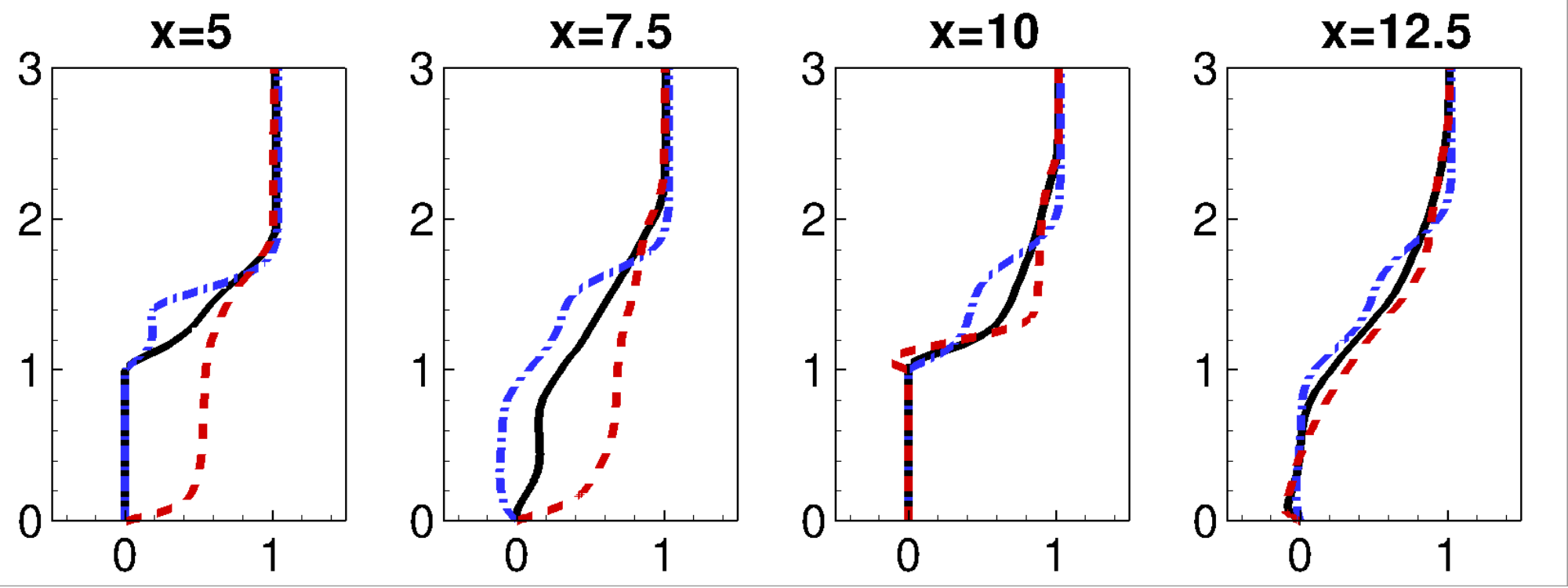}
\put(-255,35){\rotatebox{90}{$y/h$}}
\put(-215,-5){$\overline{u}$}
\put(-152,-5){$\overline{u}$}
\put(-90,-5){$\overline{u}$}
\put(-28,-5){$\overline{u}$}
\put(-265,80){$(a)$}
\put(-206,84){$h$}
\put(-136,84){$h$}
\put(-78,84){$h$}
\put(-9,84){$h$}
 \hspace{3mm}
\includegraphics[width=85mm,trim={0.1cm 0.1cm 0.5cm 0.5cm},clip]{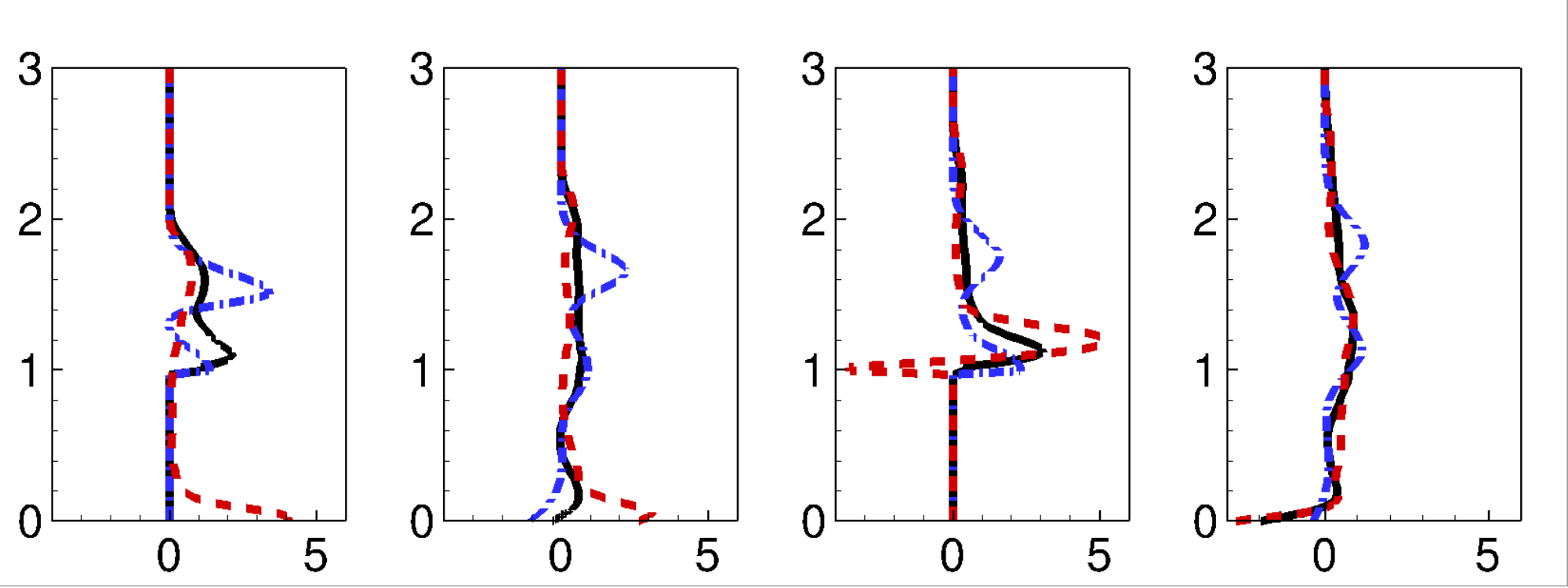}
\put(-255,35){\rotatebox{90}{$y/h$}}
\put(-215,-5){$\overline{\omega}_z$}
\put(-152,-5){$\overline{\omega}_z$}
\put(-90,-5){$\overline{\omega}_z$}
\put(-28,-5){$\overline{\omega}_z$}
\put(-265,80){$(b)$}
\caption{Streamwise variation of $(a)$ $\overline{u}$ and $(b)$ $\overline{\omega}_z$ profiles in the symmetry plane for Case $(5h, 5h)$ (black solid), Case $(5h, 2.5h)$ (blue dash-dot) and Case $(10h, 5h)$ (red dashed) at $Re_h=600$.}
\label{fig:vorz_profiles}
\end{figure}
%msi/ns/laminar_BL_collocated_Re600_DNS_medium_domain

% Mean velocity and spanwise vorticity profiles (inflection points)
% instantaneous / mean spanwise vorticity field
% mean streamwise velocity field
The CVP examined in \S \ref{cvp} perturbs the shear layer in their vicinity, and their size and strength determine the nature of perturbation to the shear layer. %The perturbation of shear layer is inhomogeneous in the spanwise direction and therefore the global linear stability analysis could lead to a better understanding of the instability mechanisms associated with the flows.
The perturbation to the shear layer is visualized by instantaneous spanwise vorticity at the symmetry plane in figure \ref{fig:vorz_zslice}. The shear layer downstream of the isolated cube is shown as the baseline in figure \ref{fig:vorz_zslice}$(a)$. The breakdown of the shear layer is observed downstream of the cube due to the perturbation of vortex shedding, and the vortex heads are dissipated at $x=20h$. 

% %The simple shear is also a key flow element for the generation of hairpin vortices \citep{cohen2014minimal}.
% \cite{suponitsky2005generation} suggested that the interaction between simple shear and a localized vortical disturbance can lead to the formation of a hairpin vortex if the initial magnitude is sufficiently large. The shear in the base flow and the strength of vorticity thus determine the nature of perturbation to the shear layer. To quantitatively identify how roughness distribution modifies the local shear, % and localized vorticity magnitude, %and how this would affect the formation of hairpin vortices, 
% the profiles of mean streamwise velocity and spanwise vorticity at the symmetry plane are examined in figure \ref{fig:vorz_profiles}. The inflection point in the mean velocity profiles is a necessary condition for the instability in shear flows, which can be identified at the peak location of the spanwise vorticity. Although the wall-normal shear is induced by the first-row roughness and is lifted up with increasing downstream distance, the wake flow is still steady due to the absence of CVP in Case $(5h, 2.5h)$. Case $(10h, 5h)$ demonstrates the strongest localized mean shear at the roughness location.

For distributed surface roughness, %inhibits the generation of CVP, consequently, no vortex shedding is observed in 
figure \ref{fig:vorz_zslice}$(b)$ shows that for small spanwise spacing $\lambda_z=2.5h$, the downstream shear layer formed right above the roughness tips remains steady and no hairpin vortex shedding is produced in the wake flow. %This is due to the fact that the counter-rotating vortices (rear pair vortices) is absent and the streamwise vortices are dissipated in a short downstream distance when the spanwise spacing is sufficiently small, as analyzed in \S \ref{cvp}, they thus are unable to grow and provide strong disturbances to the shear layer. 
%The vortical structures visualized using the Q criterion \citep{hunt1988eddies} in figure \ref{fig:Q_Lz25} show that only the OSP vortices are induced by the first-row cubes and no hairpin vortices are produced in the wake flow. 
The inflection point in the mean velocity profiles is examined in figure \ref{fig:vorz_profiles}. It is a necessary condition for instability in shear flows, and is corresponding to the peak location of spanwise vorticity. Although the inflection points formed by wall-normal shear can be identified for Case $(5h, 2.5h)$ in figure \ref{fig:vorz_profiles}, the absence of CVP due to the small spanwise spacing results in a failure of hairpin vortex generation. %The small spanwise roughness spacing can inhibit the generation of hairpin vortices through the weakening of streamwise vortices within the longitudinal grooves. As a consequence, the perturbation to the shear layer is weak, and a stable shear layer is formed above the roughness layer.

% \begin{figure}
%  \includegraphics[width=130mm,trim={0.2cm 0.1cm 0.1cm 0.1cm},clip]{images/Q_tile_Lz25.pdf}
% %\put(-335,115){$u/U_e$}
% %\put(-350,37){\rotatebox{90}{$z/h$}}
% \put(-286,10){u}
% %\put(-75,0){$x/h$}
%  \caption{Instantaneous vortical structures for Case $(5h, 2.5h)$ in perspective and top-down views, visualized by isocontours of $Q=0.1U_e^2/h^2$ and colored with instantaneous streamwise velocity. Plotted in the x-y and z-y planes are the contours of instantaneous spanwise vorticity, in the range of $-1$ (black) and $0$ (white). } 
% \label{fig:Q_Lz25}
% \end{figure}

With larger spanwise spacing $\lambda_z=5h$, the breakdown of the shear layer is seen in figures \ref{fig:vorz_zslice}$(c)$ and \ref{fig:vorz_zslice}$(d)$. Although the localized shear layers are induced by each roughness tip in Cases $(5h, 5h)$ and $(10h, 5h)$, the primary vortex shedding behaves similarly as the isolated case, and the vortex heads are dissipated at around $x=20h$. %In addition to this, for Case $(\lambda_x,\lambda_z)=(5h, 5h)$, the localized shear layer induced by the roughness tips also perturbs the flow, resulting in a more complex vorticity field.
The length scale of localized shear layers is equivalent to $5h$, and few vortices penetrate into cavities for Case $(5h, 5h)$. In contrast, for Case $(10h, 5h)$, more complex vortical structures are evolved into the cavities from the second-row cubes and the strongest localized mean shear is demonstrated at the roughness location in figure \ref{fig:vorz_profiles}. The periodicity of the primary vortex shedding seems to be similar as that for the isolated roughness and independent with different streamwise roughness spacing. The associated instability mechanisms will be further examined and discussed in \S \ref{glsa}.

\begin{figure}
 \includegraphics[width=130mm,trim={0.2cm 0.1cm 0.1cm 0.1cm},clip]{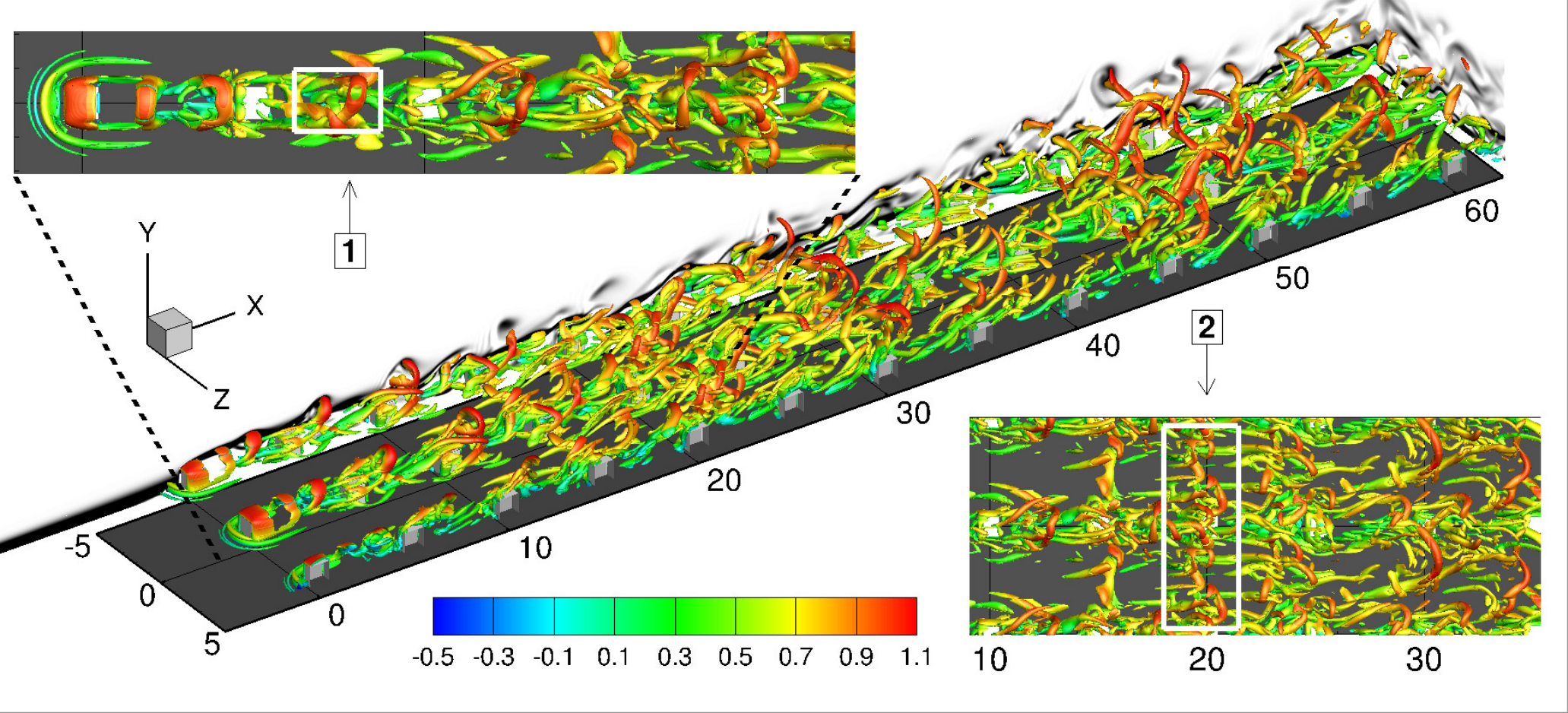}
%\put(-335,115){$u/U_e$}
%\put(-350,37){\rotatebox{90}{$z/h$}}
\put(-298,10){$u/U_e$}
\put(-75,0){$x/h$}
 \caption{Instantaneous vortical structures for Case $(5h, 5h)$ in perspective and top-down views, visualized by isocontours of $Q=0.1U_e^2/h^2$ and colored with instantaneous streamwise velocity. Plotted in the x-y and z-y planes are the contours of instantaneous spanwise vorticity, in the range of $-1$ (black) and $0$ (white).} 
\label{fig:Q_Lx5}
\end{figure}

The evolution of vortical structures is examined using the Q criterion \citep{hunt1988eddies} in figure \ref{fig:Q_Lx5} for Case $(5h, 5h)$. Both the SP and OSP vortices are observed in the vicinity of the first-row cubes. They interact with the shear layer, leading to the hairpin vortex shedding downstream of the first-row roughness elements. The packets of hairpin-type structure with small legs labelled $1$ can be identified in the top-left inset of figure \ref{fig:Q_Lx5}. The shorter streamwise extent of the vortical motions is due to the blockage effects of the closely distributed cubes. As the vortex legs are inclined upward and undergo stretching by a positive wall-normal velocity gradient, they are cut off by the following cubes and break down into smaller vortical structures with low momentum within the roughness layer. The bottom-right inset of figure \ref{fig:Q_Lx5} shows that the interactions between vortical structures in the longitudinal grooves occur at approximately $x=15h$. As the hairpin-type vortices break down into smaller structures at $x=20h$, the small scales are still organized in spanwise coherent structures, labelled $2$. The arch-shaped spanwise coherent structures are observed farther downstream.

\begin{figure}
 \includegraphics[width=130mm,trim={0.2cm 0.1cm 0.1cm 0.1cm},clip]{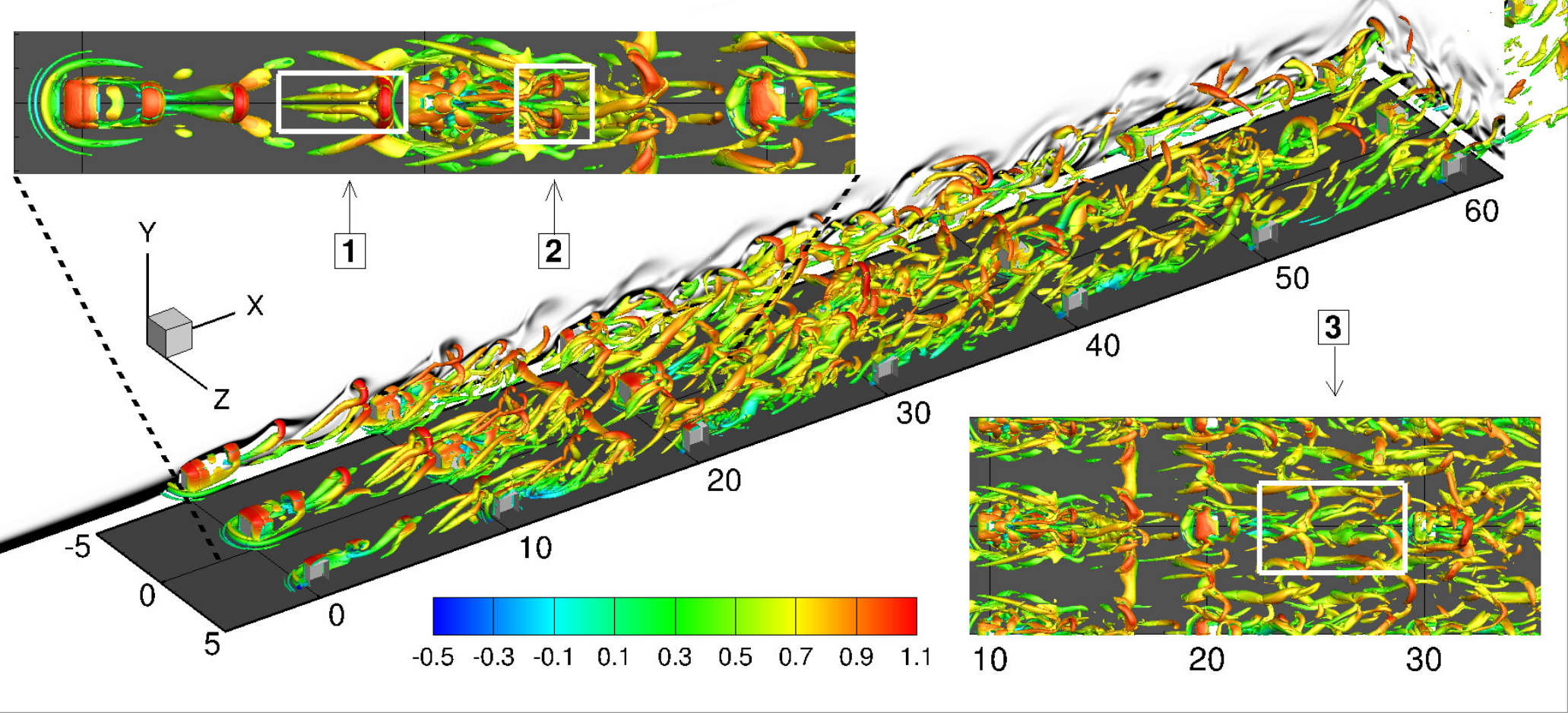}
%\put(-335,115){$u/U_e$}
%\put(-350,37){\rotatebox{90}{$z/h$}}
\put(-298,10){$u/U_e$}
\put(-75,0){$x/h$}
 \caption{Instantaneous vortical structures for Case $(10h, 5h)$ in perspective and top-down views, visualized by isocontours of $Q=0.1U_e^2/h^2$ and colored with instantaneous streamwise velocity. The contour levels are the same as figure \ref{fig:Q_Lx5}.} 
\label{fig:Q_Lx10}
\end{figure}

For Case $(10h, 5h)$, both SP and OSP vortices are observed, and the hairpin vortices are generated from the first-row cubes, as shown in figure \ref{fig:Q_Lx10}. The top-left inset shows that the hairpin vortices labelled $1$ behave similarly as the isolated roughness case. As the high-momentum fluid impinges onto the second row of roughness, another spanwise vortex wraps around the second-row cubes, and the vortical structures break down into small scales downstream of the second-row cubes. %The hairpin vortex labelled $2$ can still be identified downstream of the second-row roughness. 
The CVP sheds from the second-row roughness tips, and a separation of the vortex heads labelled $2$ occurs. This %results from the interaction between the hairpin vortices shedding from the first-line roughness elements and the localized shear layer induced by the second-line roughness elements, and 
indicates that a different unstable mode might be induced by the second row of cubes. % resulting from the interaction between the localized shear layer induced by the second-line cube and  The separation region behind the second-line cube diminishes the development of the %indicating a different unstable mode shape. 
The bottom-right inset shows that the vortical structures in the longitudinal grooves interact with each other from $x=15h$. The spanwise coherent structures are less prominent than Case $(5h, 5h)$. The longitudinal vortical structures labelled $3$ are possibly related to the streak interactions in the grooves.

\subsection{Global stability analysis}\label{glsa}
%\cite{cohen2014minimal} highlighted that the formation of hairpin vortices is inherently unstable as the CVP perturbs the shear layer, forming the inflection point in the base-flow velocity profiles. 
The hairpin vortices induced by roughness elements are inherently unstable as the CVP perturbs the shear layer, forming the inflection point in the base-flow velocity profiles. The perturbation of the shear layer is inhomogeneous in the spanwise direction and therefore the global stability analysis is useful to provide insights into the instability mechanisms associated with the distributed roughness wakes.

\subsubsection{Base flow computation}
The base flow for global instability analysis is computed for the distributed roughness cases. For Cases $(5h, 2.5h)$ and $(5h, 5h)$, the selective frequency damping (SFD) method \citep{aakervik2006steady} is used to artificially settle the flow towards a steady equilibrium. An encapsulated formation of the SFD method developed by \cite{jordi2014encapsulated} is applied in the present work. 

For Case $(10h, 5h)$, it is found that the SFD method is unable to damp the oscillations due to the unsteady part of the solutions, even though careful choices are made for the control coefficient $\chi$ and the filter width $\Delta$. The possible reason is that the SFD method is unable to get the steady state when there are multiple unstable modes, as indicated by \cite{casacuberta2018effectivity}. They discussed the effectivity of SFD for systems with more than one unstable eigenmode where the most unstable eigenvalue is $\mu^c$ and other unstable eigenvalues are denoted by $\mu^k$. They concluded that SFD is unable to drive the system towards the base flow when $\mu^k$ with large values of $Im(\mu^k)/Re(\mu^k)$ is present close to the origin. As discussed in \cite{ma2022global}, using the time-averaged mean flow as the base state for global stability analysis is able to capture the temporal frequency and associated mode shape of the primary vortical structures for roughness-induced transition. The time-averaged mean flow is therefore considered as an alternate base flow in the present work to investigate global instability for Case $(10h, 5h)$.

%The problem is considered to have converged when $||q-\overline{q}||_{inf} \le 10^{-8}$ according to \cite{jordi2014encapsulated}, where $\overline{q}$ is the filtered state. When using SFD, the control coefficient $\chi$ and the filter width $\Delta$ play important roles in the convergence process. The control coefficient $\chi$ should be positive and larger than the growth rate of the desired mode, while the filter cut-off frequency $\omega_c=1/\Delta$ must be lower than all of the flow instabilities to ensure the unstable disturbances are well within the damped region. For example, $\chi=0.5$ and $\Delta=2$ are used for the unstable case ($Re_h,\eta$)=($600,1$), 

\subsubsection{High- and low-speed longitudinal streaks}

%Figure \ref{fig:streaks_Lx5} compares the base flow of isolated roughness, Cases $(\lambda_x,\lambda_z)=(5h, 5h)$ and $(\lambda_x,\lambda_z)=(5h, 2.5h)$ at different $x$ positions. For the isolated roughness element, The roughness elements act like continuous strips when the streamwise spacing $\lambda_x \le 5h$. The strength of the central low-speed streak is amplified due to the continuous roughness arrays in the sreamwise direction. As the spanwise spacing is decreased to $\lambda_z=2.5h$, the lateral low-speed streaks merge with those of the adjacent roughness elements, and the central low-speed streak shrinks due to the spanwise interaction of the roughness elements. the low-momentum fluid is located within the cavities for both two cases which is a typical phenomenon for d-type roughness. The shear layer induced by the roughness elements for Case $\lambda_z=5h$ is stronger and sustains in a longer streamwise extent than Case $\lambda_z=2.5h$. 

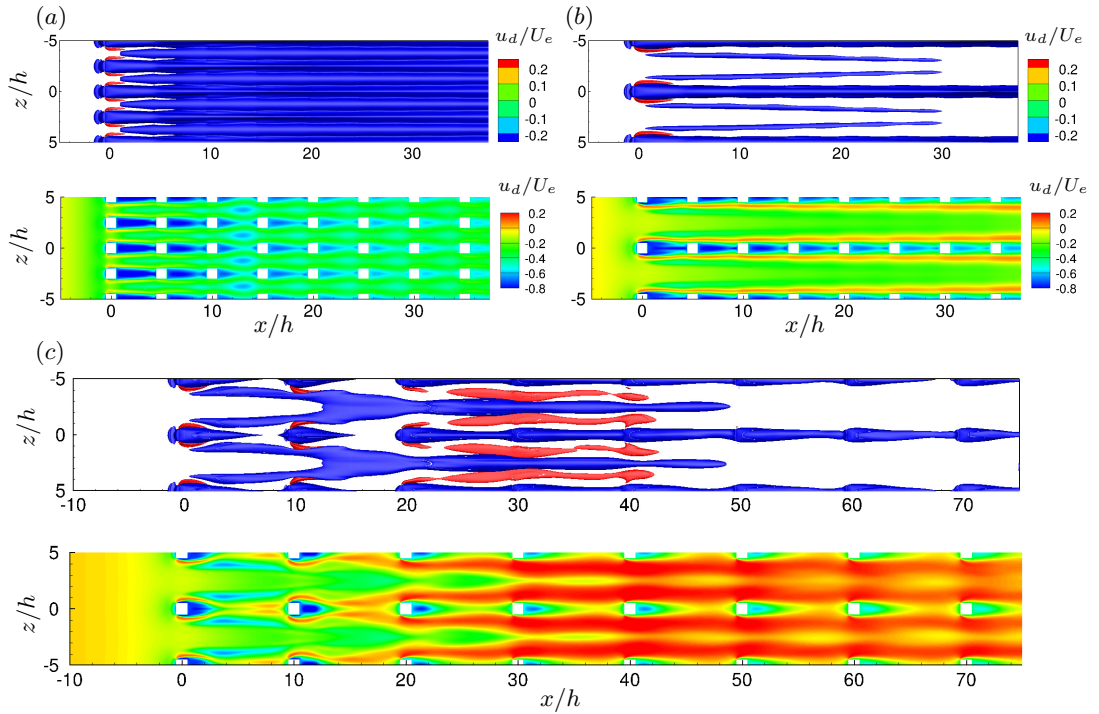
\begin{figure}
% \includegraphics[width=70mm,trim={1.5cm 0.2cm 0.5cm 1cm},clip]{images/ud_z0_Lx5.pdf}
% % \put(-200,60){$(a)$}
% \put(-200,23){\rotatebox{90}{$z/h$}}
% \put(-110,0){$x/h$}
% %\put(-22,50){\scriptsize{$u_d/U_e$}}
% \includegraphics[width=70mm,trim={1.5cm 0.2cm 0.5cm 1cm},clip]{images/ud_z0_Lx25.pdf}
% \put(-200,23){\rotatebox{90}{$z/h$}}
% \put(-110,0){$x/h$}
% %\put(-22,50){\scriptsize{$u_d/U_e$}}
% \hspace{3mm}
\includegraphics[width=70mm,trim={0.2cm 0.8cm 0.5cm 0.2cm},clip]{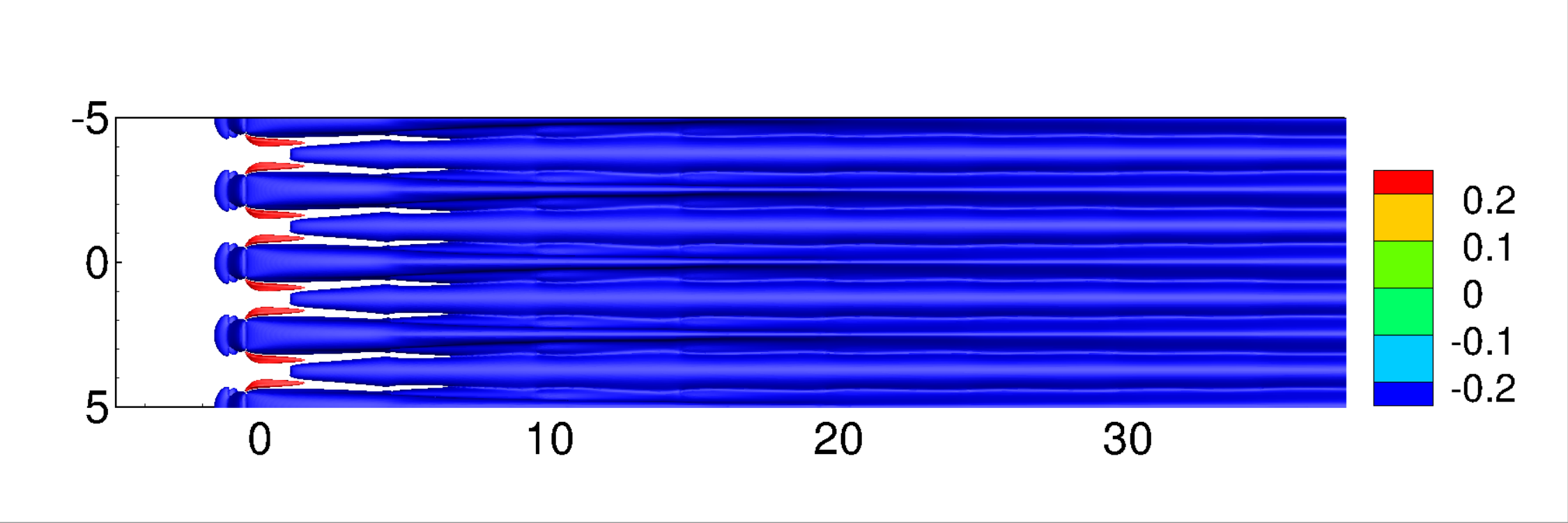}
\put(-195,52){$(a)$}
\put(-205,20){\rotatebox{90}{$z/h$}}
%\put(-110,0){$x/h$}
\put(-22,46){\scriptsize{$u_d/U_e$}}
\includegraphics[width=70mm,trim={0.2cm 0.8cm 0.5cm 0.2cm},clip]{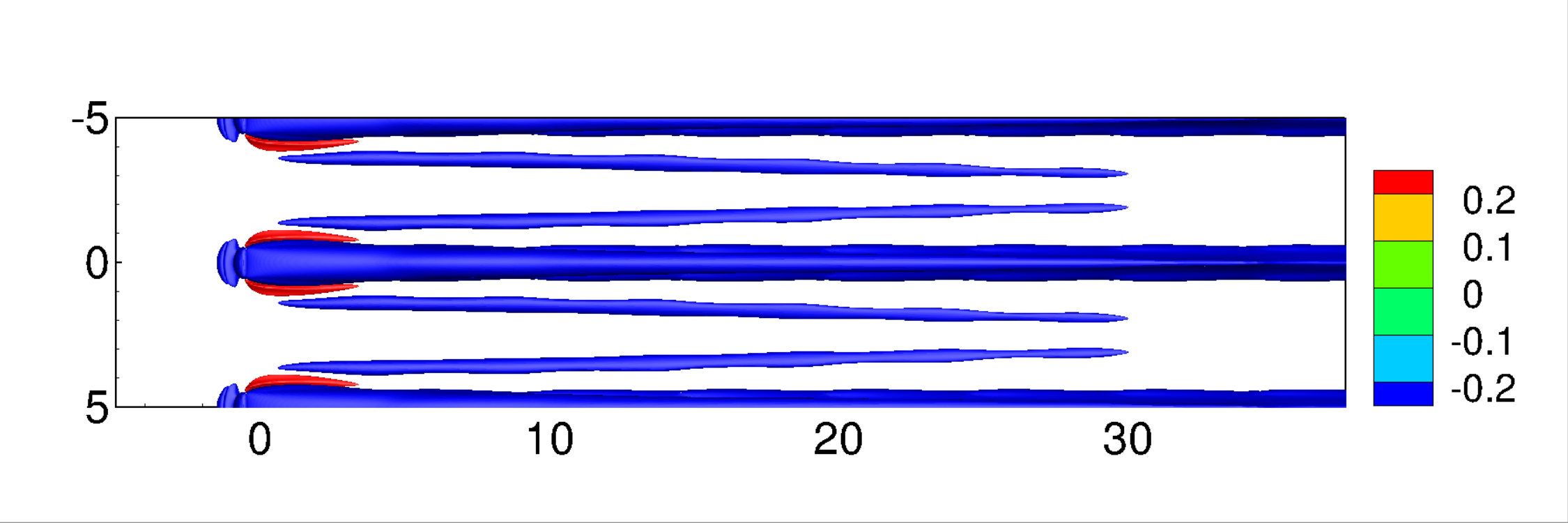}
\put(-195,52){$(b)$}
%\put(-200,28){\rotatebox{90}{$z/h$}}
%\put(-110,0){$x/h$}
\put(-22,46){\scriptsize{$u_d/U_e$}}
\hspace{3mm}
\includegraphics[width=69mm,trim={1.5cm 0.2cm 0.5cm 0.8cm},clip]{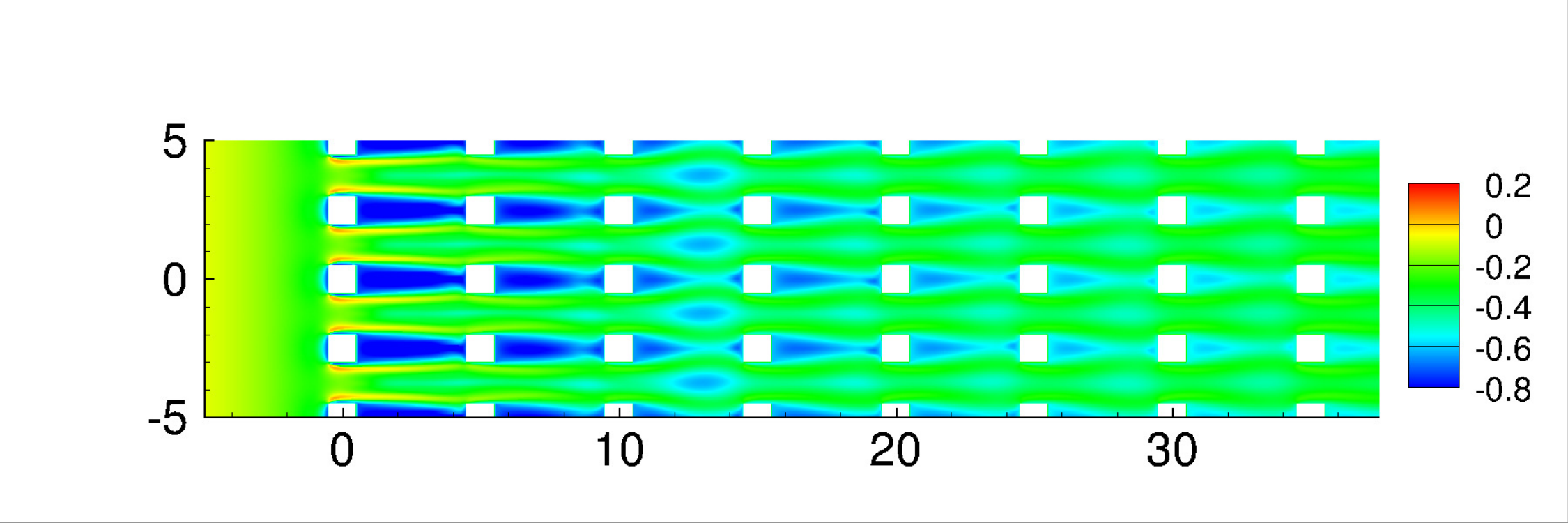}
\put(-202,25){\rotatebox{90}{$z/h$}}
\put(-110,0){$x/h$}
\put(-18,53){\scriptsize{$u_d/U_e$}}
\hspace{0.5mm}
\includegraphics[width=69mm,trim={1.5cm 0.2cm 0.5cm 0.8cm},clip]{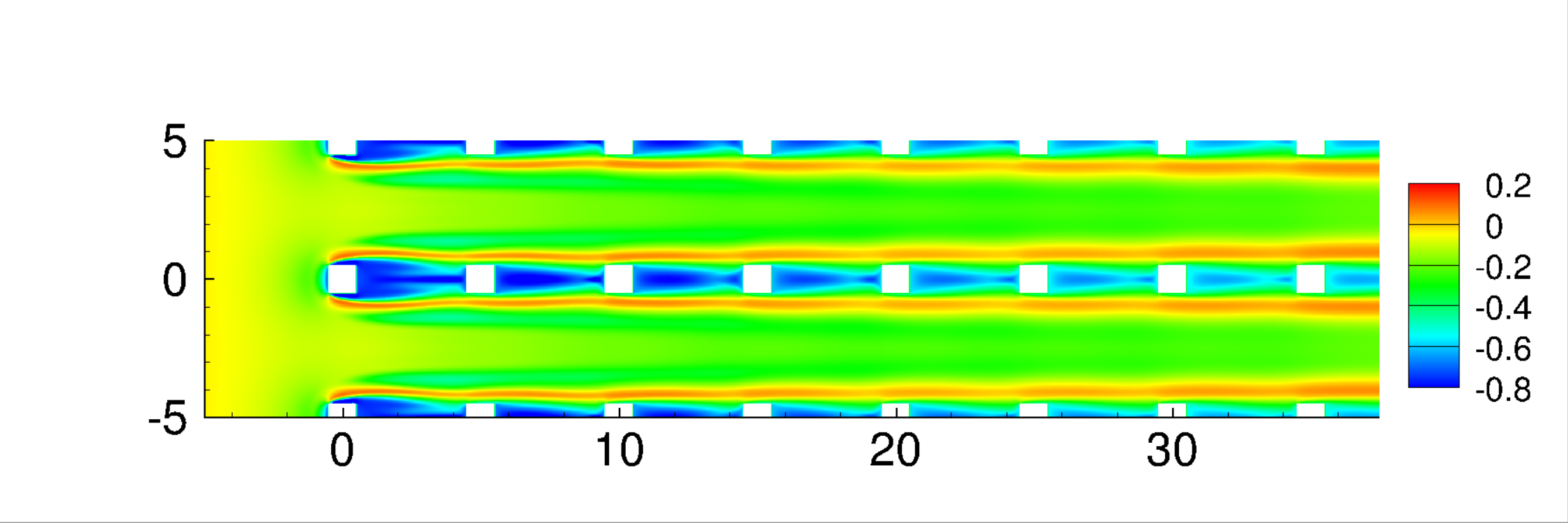}
%\put(-200,23){\rotatebox{90}{$z/h$}}
\put(-110,0){$x/h$}
\put(-18,53){\scriptsize{$u_d/U_e$}}
\hspace{3mm}
\includegraphics[width=135mm,trim={0.7cm 1.7cm 0.5cm 1.0cm},clip]{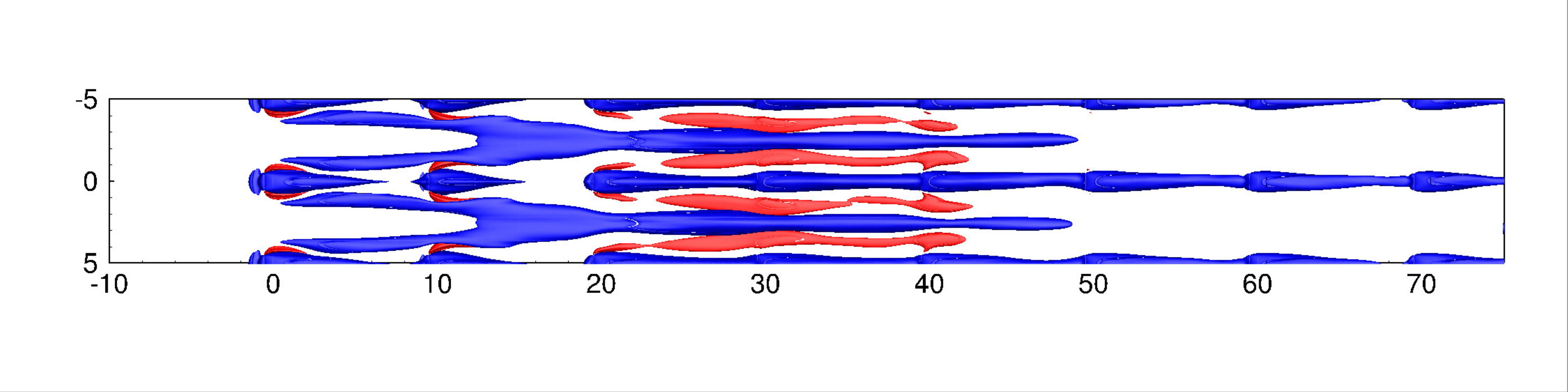}
\put(-380,60){$(c)$}
\put(-387,27){\rotatebox{90}{$z/h$}}
%\put(-190,9){$x/h$}
%\put(-22,50){\scriptsize{$u_d/U_e$}}
\hspace{3mm}
\includegraphics[width=135mm,trim={0.5cm 0.5cm 0.5cm 1.6cm},clip]{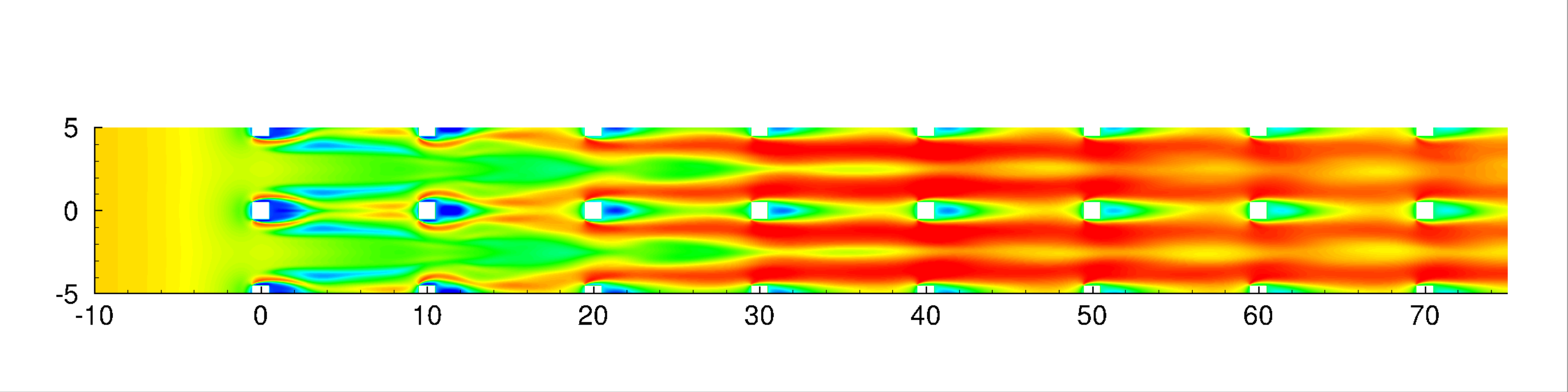}
% \put(-200,60){$(a)$}
\put(-387,30){\rotatebox{90}{$z/h$}}
\put(-190,3){$x/h$}
%\put(-22,50){\scriptsize{$u_d/U_e$}}
\caption{Top-down views of high- and low-speed streaks, visualized by isosurfaces (top) and contour plots at the plane $y=0.5h$ (bottom) of the streamwise velocity deviation of the base flow from the Blasius boundary layer solution, $u_d=U_b-u_{bl}$, for $(a)$ Case $(5h, 2.5h)$, $(b)$ Case $(5h, 5h)$ and $(c)$ Case $(10h, 5h)$. The contour levels in $(c)$ are the same as $(a)$ and $(b)$.} 
\label{fig:streaks_Lx5}
\end{figure}

The high- and low-speed streaks are visualized in figure \ref{fig:streaks_Lx5} by the streamwise deviation of the base flow from the Blasius solution for the three distributed roughness cases. Figure \ref{fig:streaks_Lx5}$(a)$ shows that for Case $(5h, 2.5h)$, the central and lateral low-speed streaks are merged with each other and form a shear layer above the roughness layer.
The high-speed streaks are only induced at the first row of roughness elements and disappear within a short downstream distance due to the dense spanwise roughness distribution. In this case, the instability mechanism might be Kelvin-Helmholtz instability rather than streak instability. For Case $(5h, 5h)$, both the central and lateral streaks are observed in figure \ref{fig:streaks_Lx5}$(b)$, behaving similarly as those in an isolated roughness case. The lateral low-speed streaks develop away from the symmetry plane with increasing downstream distance, due to the presence of the following roughness elements. 

As the streamwise spacing is increased to $\lambda_x=10h$, different behavior of high- and low-speed streaks is seen in figure \ref{fig:streaks_Lx5}$(c)$. In contrast to Case $(5h, 5h)$, the central low-speed streak only forms within the vicinity downstream of the cubes. The high-speed streaks induced by the first-row cubes move close to each other and collide onto the following obstacles, which induces the high-speed streaks from the second-row cubes. The high-speed streaks grow and interact in the longitudinal grooves farther downstream.

\subsubsection{Eigenspectra and eigenmodes}
% eigenspectra; eigenmode; production

\begin{figure}
\centering
\includegraphics[width=65mm,trim={0.5cm 0.2cm 0.5cm 0.5cm},clip]{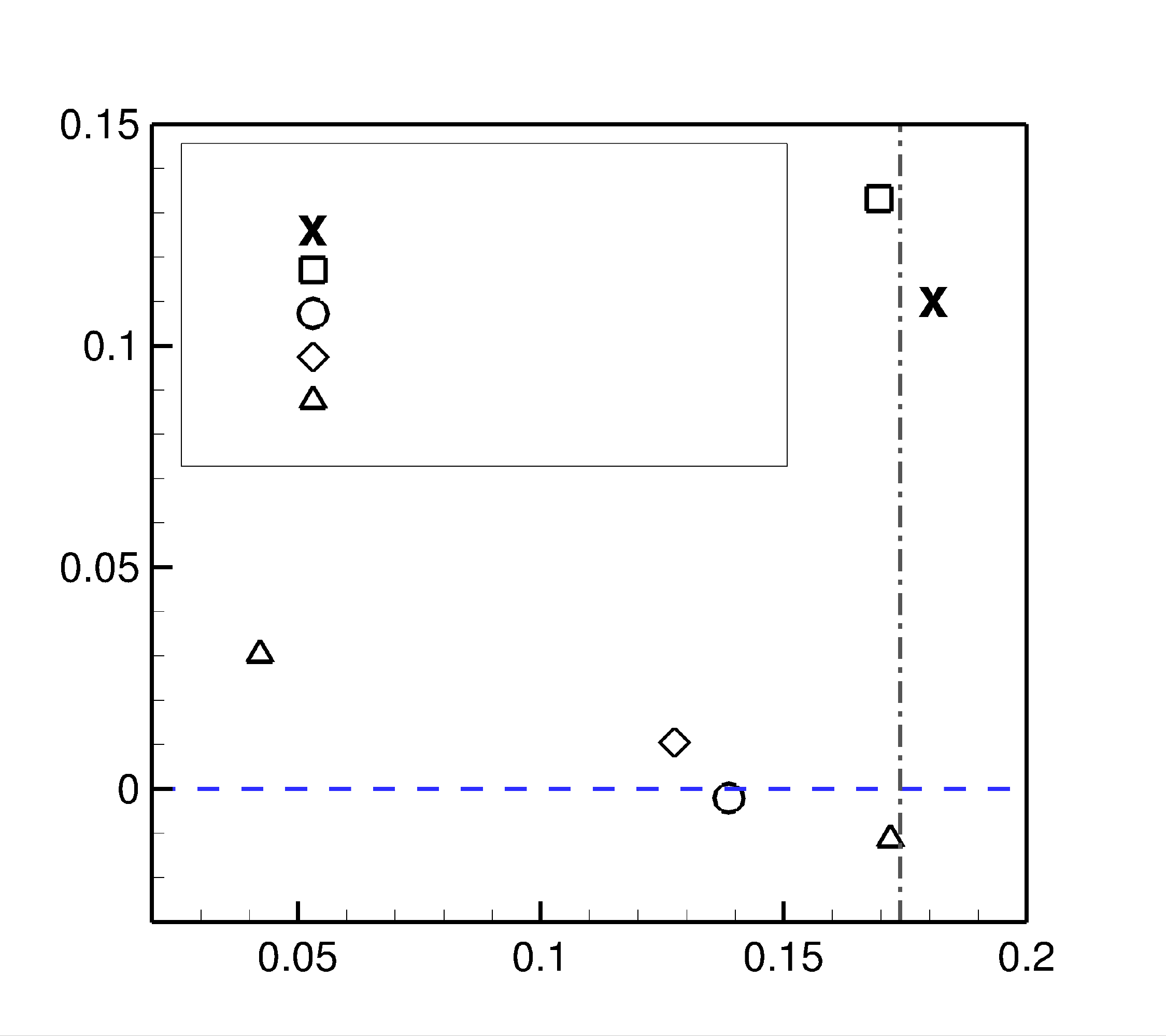}
\put(-185,80){\rotatebox{90}{$\sigma$}}
\put(-90,0){$St$}
\put(-130,131){\tiny{Isolated,$Re_h=600$}}
\put(-130,124){\tiny{$(5h,5h)$,$Re_h=600$}}
\put(-130,117){\tiny{Isolated,$Re_h=475$}}
\put(-130,110){\tiny{$(5h,5h)$,$Re_h=475$}}
\put(-130,103){\tiny{$(10h,5h)$,$Re_h=600$}}
% \put(-50,133){\scriptsize{$Re_h=600$}}
% \put(-180,160){$(a)$}
\caption{Leading eigenvalues for Cases $(5h, 5h)$ and $(10h, 5h)$ at different $Re_h$, with a comparison to the isolated roughness case. The vertical dash-dot line denotes the Strouhal number $St=\omega_a h/(2\pi u_h)$ of the primary hairpin vortices, where $u_h$ is the Blasius velocity at roughness tips.}
\label{fig:eigenspectra}
\end{figure}

\begin{figure}
\includegraphics[width=70mm,trim={0.9cm 1.2cm 0.5cm 2cm},clip]{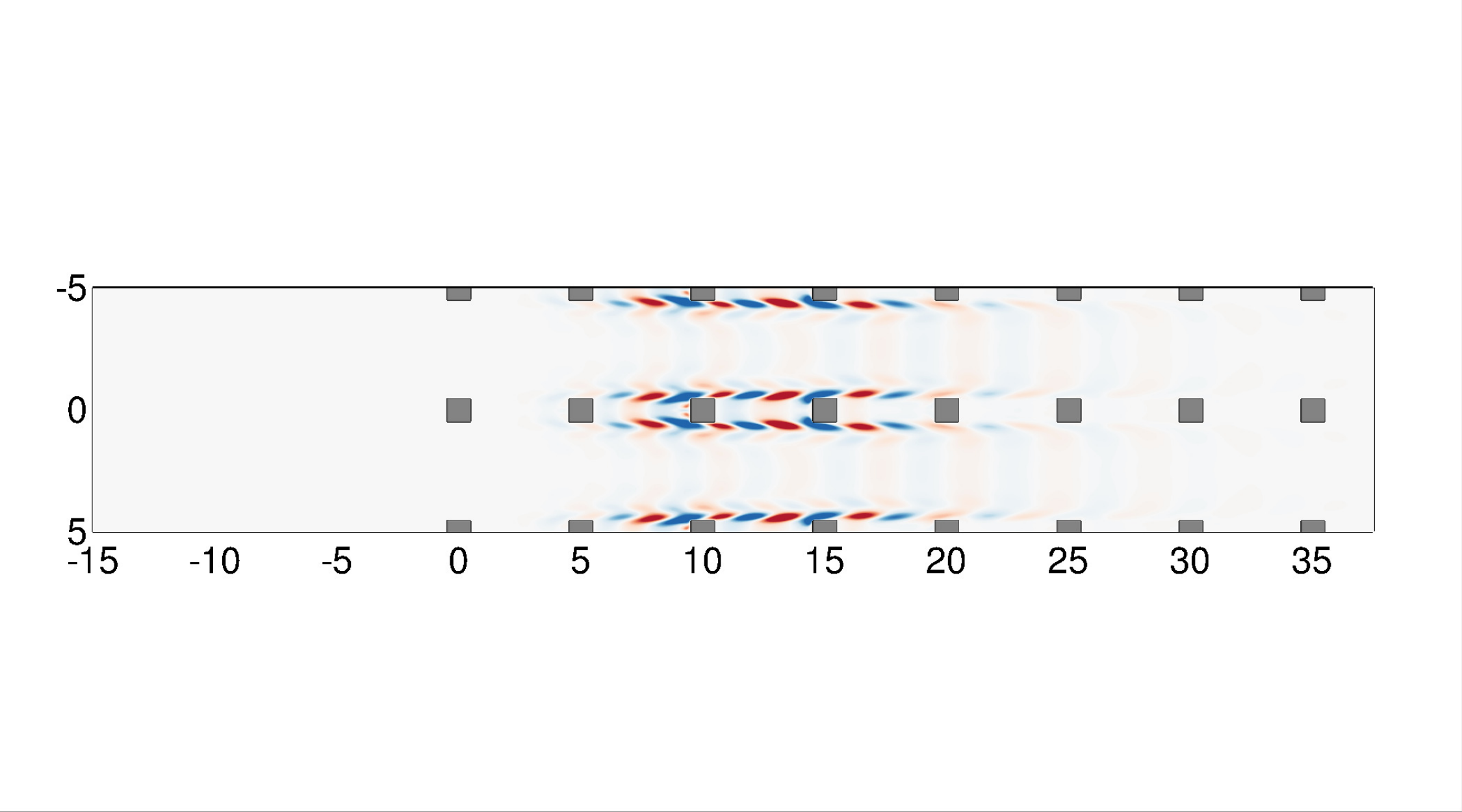}
% \put(-200,80){$(a)$}
% \put(-180,80){$\eta=1,Re_h=475$}
\put(-210,38){\rotatebox{90}{$z/h$}}
\put(-106,15){$x/h$}
\includegraphics[width=70mm,trim={0.2cm 0.2cm 0.5cm 0cm},clip]{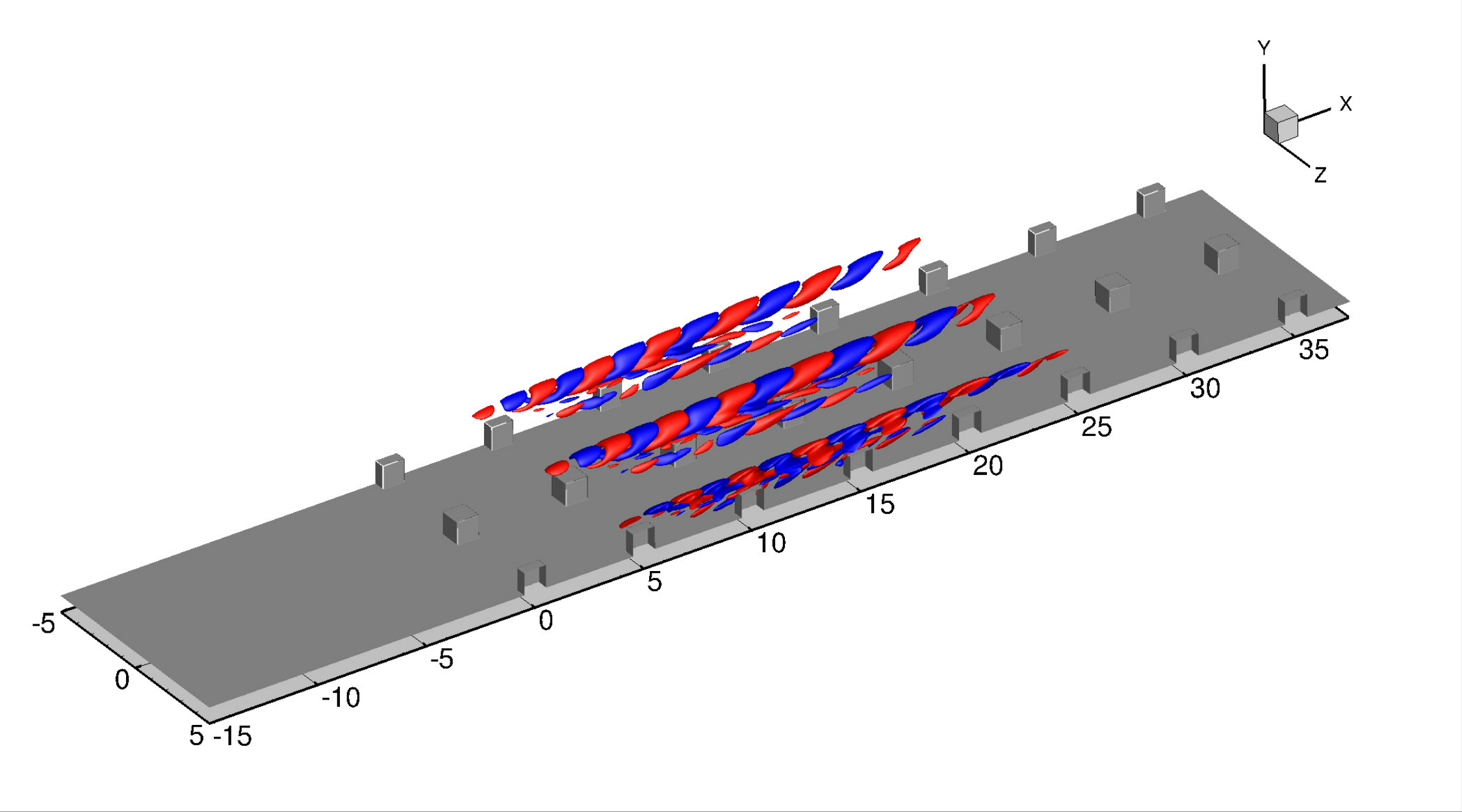}
% \put(-200,90){$(b)$}
\put(-195,3){\rotatebox{90}{$z/h$}}
\put(-90,27){$x/h$}
 \caption{Contour plots in the plane of $y=0.5h$ (left) and isosurfaces (right) of the streamwise velocity component of the leading unstable mode for Case $(5h, 5h)$ at $Re_h=600$. The contour levels depict $\pm 10 \%$ of the mode's maximum streamwise velocity. } 
\label{fig:eigenmode_5h}
\end{figure}

\begin{figure}
% \includegraphics[width=72mm,trim={2cm 1.2cm 0.5cm 2cm},clip]{images/Iy_Lx5_x10.pdf}
%  \put(-200,75){$(a)$}
% \put(-180,75){$x=10$}
% \put(-213,35){\rotatebox{90}{$y/h$}}
% \put(-112,0){$z/h$}
% \includegraphics[width=72mm,trim={2cm 1.2cm 0.5cm 2cm},clip]{images/Iz_Lx5_x10.pdf}
% % \put(-200,90){$(b)$}
% \put(-213,35){\rotatebox{90}{$y/h$}}
% \put(-112,0){$z/h$}
% \hspace{3mm}
\includegraphics[width=72mm,trim={2cm 1.2cm 0.5cm 2cm},clip]{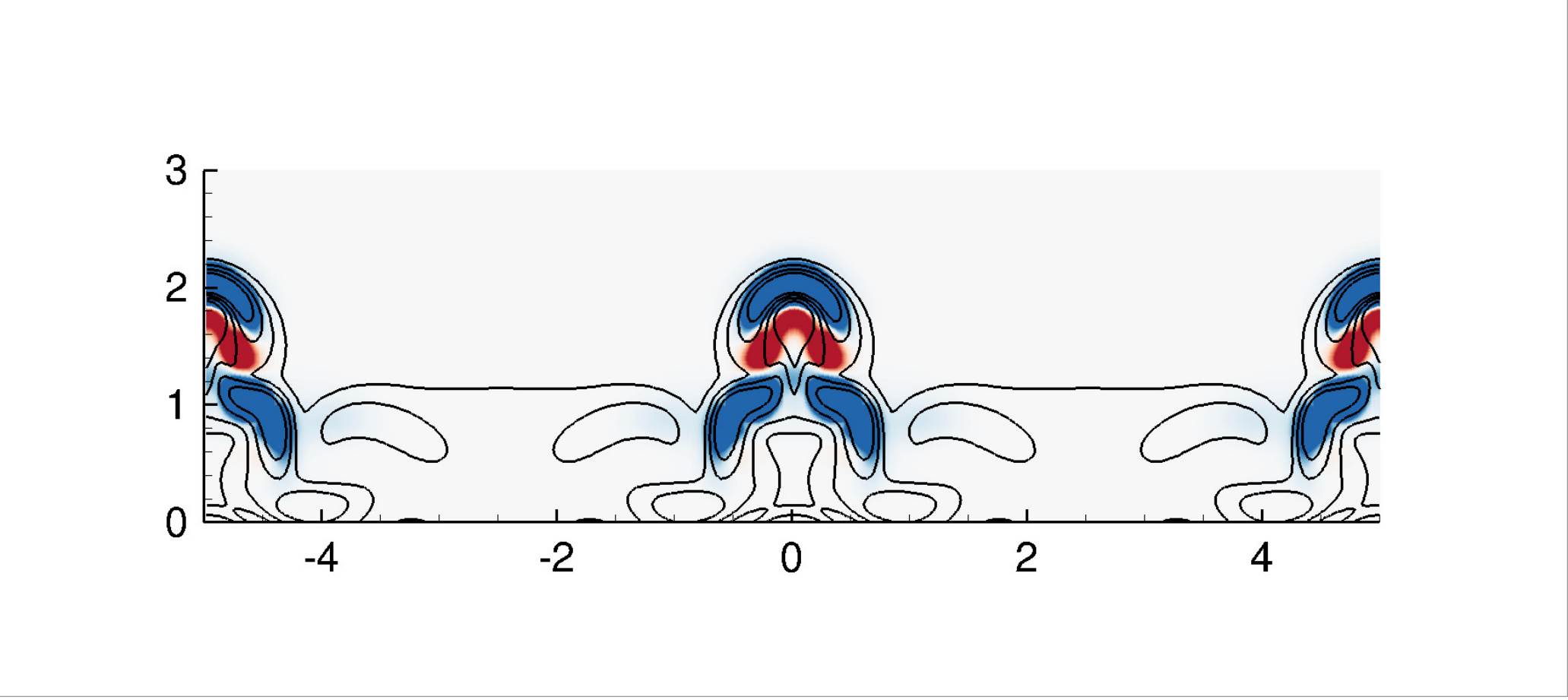}
% \put(-200,75){$(b)$}
% \put(-180,75){$x=12.5$}
\put(-213,35){\rotatebox{90}{$y/h$}}
\put(-112,0){$z/h$}
\includegraphics[width=72mm,trim={2cm 1.2cm 0.5cm 2cm},clip]{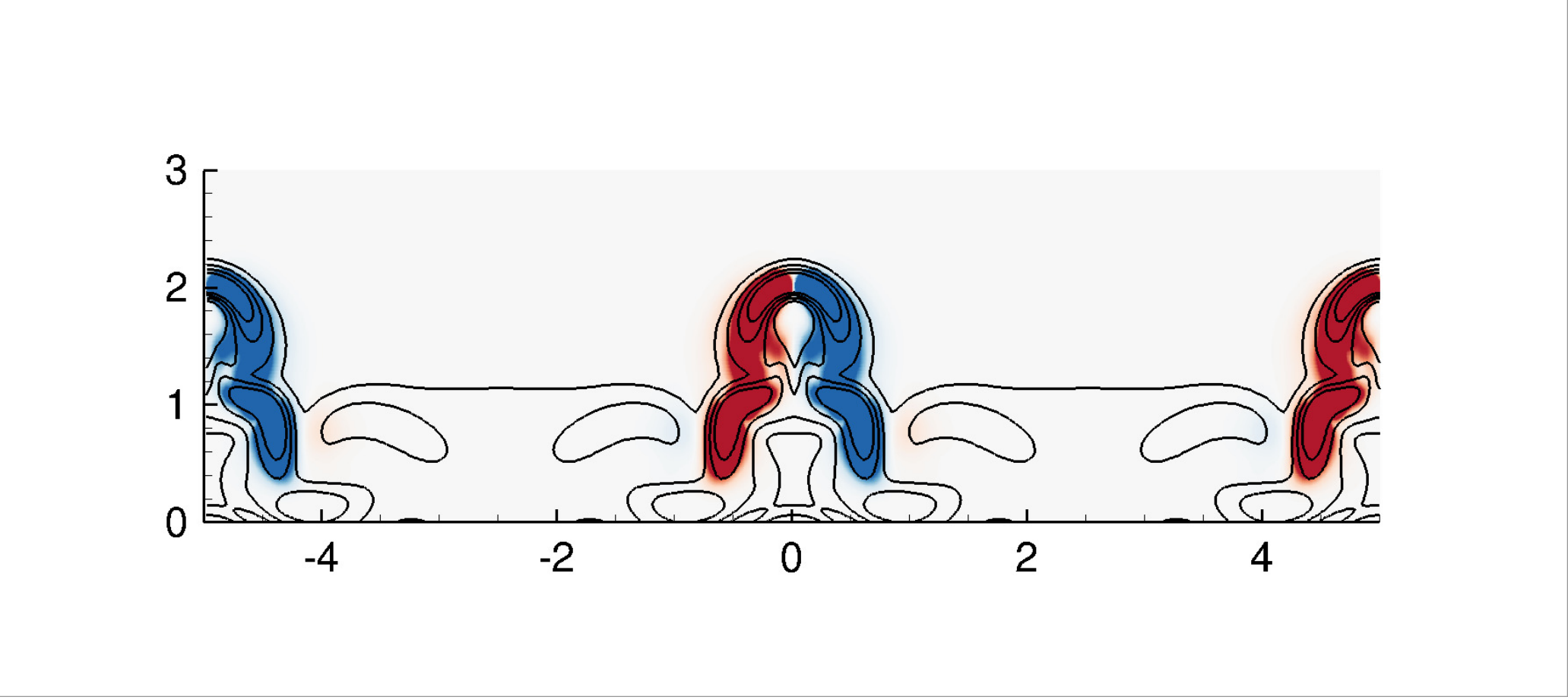}
% \put(-200,90){$(b)$}
\put(-213,35){\rotatebox{90}{$y/h$}}
\put(-112,0){$z/h$}
 \caption{Contour plots of $P_y=-|\hat{u}||\hat{v}|\frac{\partial{\overline{u}}}{\partial y}$ (left) and $P_z=-|\hat{u}||\hat{w}|\frac{\partial{\overline{u}}}{\partial z}$ (right) in the cross-flow plane of %$(a)$ $x=10$ and $(b)$
 $x=12.5h$ for Case $(5h, 5h)$ at $Re_h=600$. %The contour levels are within the range from $-1.0e^{-7}$ (blue) to $1.0e^{-7}$ (red)
 The contour levels depict $\pm 10 \%$ of the maximum $P_y$ and $P_z$. The localized shear is depicted by the solid lines of $u_s=((\partial \overline{u}/\partial y)^2+(\partial \overline{u}/\partial z)^2)^{1/2}$ from $0$ to $2$.} 
\label{fig:prod_5h}
\end{figure}

The leading eigenvalues for Cases $(5h, 5h)$ and $(10h, 5h)$ at different $Re_h$ are plotted in figure \ref{fig:eigenspectra}. For Case $(5h,5h)$, compared to the isolated case at the same $Re_h$, the growth rate is larger and the temporal frequency is lower. This indicates that the distributed roughness elements lead to lower critical $Re_h$ for linear instability to occur compared to the isolated roughness. % Case $(5h, 5h)$ at $Re_h=600$ shows the similar difference from the isolated case as $Re_h=475$. 
The associated eigenmode for Case $(5h, 5h)$ is examined in figure \ref{fig:eigenmode_5h}. The varicose mode is observed along the central low-speed streak, similar to the varicose mode observed in the isolated case \citep{ma2022global}. The dominant production terms of disturbance kinetic energy $P_y$ and $P_z$ are examined in figure \ref{fig:prod_5h}. It shows that the unstable mode extracts energy from both the wall-normal and spanwise shear of the central low-speed streak as well as the localized shear layer induced by the cubes.

\begin{figure}
\includegraphics[width=120mm,trim={0.1cm 0.1cm 0.1cm 0.1cm},clip]{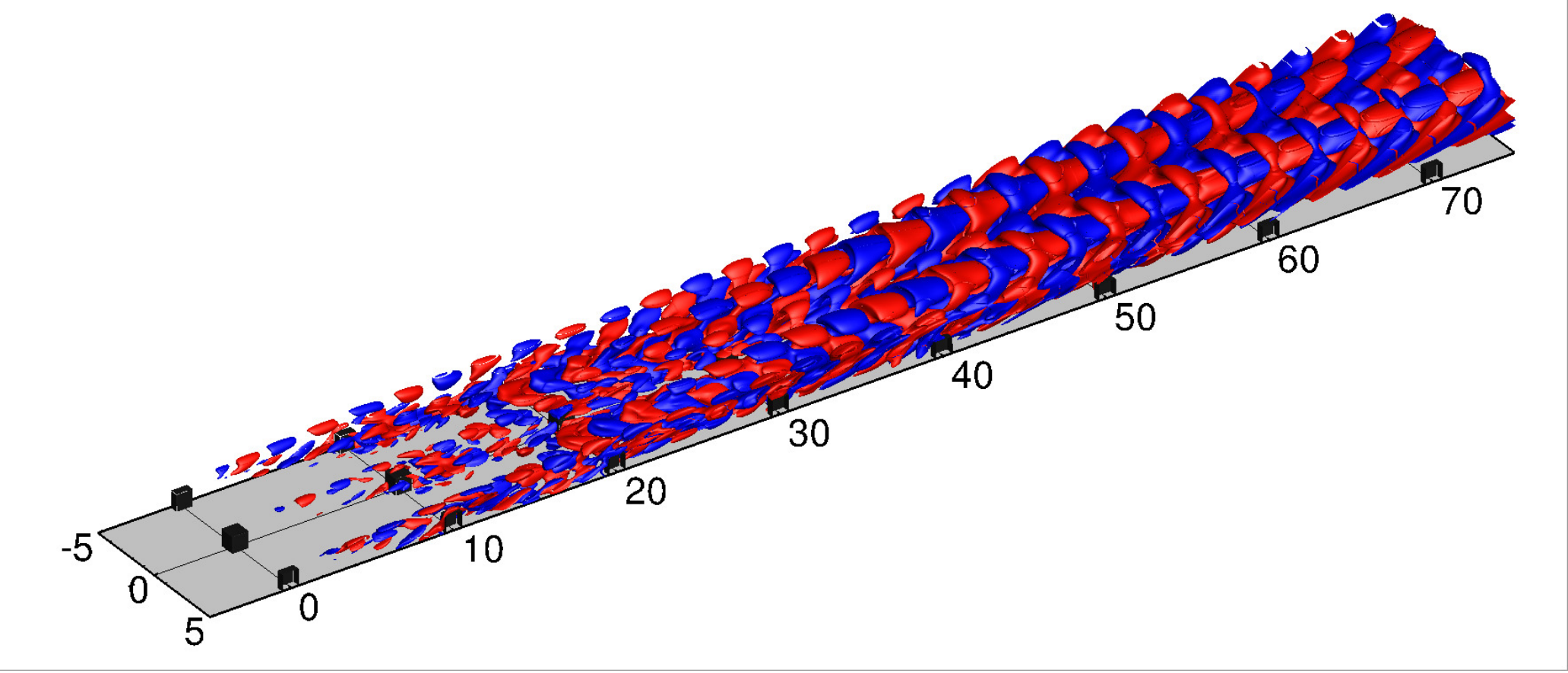}
%\put(-380,60){$(a)$}
% \put(-180,80){$\eta=1,Re_h=475$}
\put(-150,42){$x/h$}
\put(-338,8){$z/h$}
\hspace{3mm}
\includegraphics[width=130mm,trim={0.2cm 1.0cm 0.5cm 0.5cm},clip]{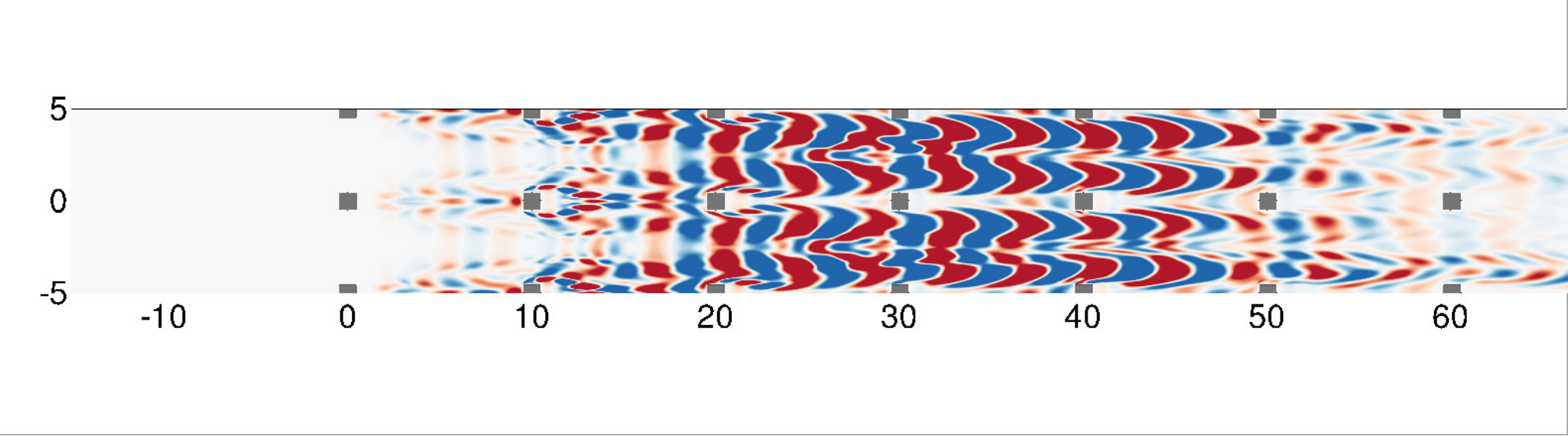}
%\put(-380,60){$(a)$}
% \put(-180,80){$\eta=1,Re_h=475$}
\put(-380,32){\rotatebox{90}{$z/h$}}
\put(-180,0){$x/h$}
\put(-340,50){$y=0.5h$}
% \hspace{3mm}
% \includegraphics[width=130mm,trim={0.2cm 0.5cm 0.5cm 2.5cm},clip]{images/eig1_Lx10_z0.pdf}
% % \put(-200,90){$(b)$}
% \put(-380,25){\rotatebox{90}{$y/h$}}
% \put(-340,40){$z=0$}
% \hspace{3mm}
% \includegraphics[width=130mm,trim={0.2cm 0.5cm 0.5cm 2.5cm},clip]{images/eig1_Lx10_z125.pdf}
% % \put(-200,90){$(b)$}
% \put(-380,25){\rotatebox{90}{$y/h$}}
% \put(-190,0){$x/h$}
\hspace{3mm}
 \caption{Isosurfaces (top) and contour plots at $y=0.5h$ (bottom) of the streamwise velocity component of the leading varicose mode for Case $(10h, 5h)$ at $Re_h=600$. %The contour levels depict $\pm 10 \%$ of the mode's maximum streamwise velocity.
 } 
\label{fig:eigenmode1_10h}
\end{figure}

\begin{figure}
\includegraphics[width=72mm,trim={2cm 1.2cm 0.5cm 2cm},clip]{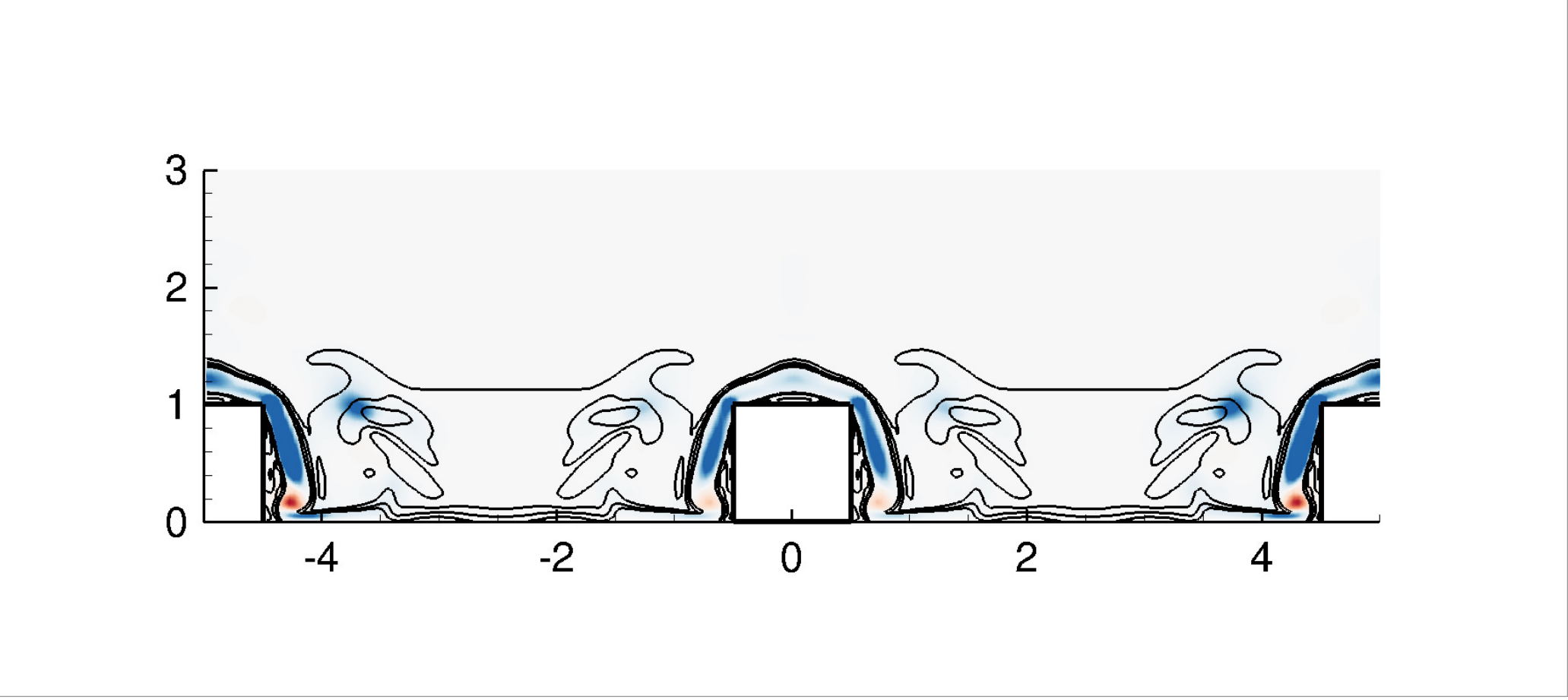}
 \put(-200,75){$(a)$}
\put(-180,75){$x=10h$}
\put(-213,35){\rotatebox{90}{$y/h$}}
\put(-112,0){$z/h$}
\includegraphics[width=72mm,trim={2cm 1.2cm 0.5cm 2cm},clip]{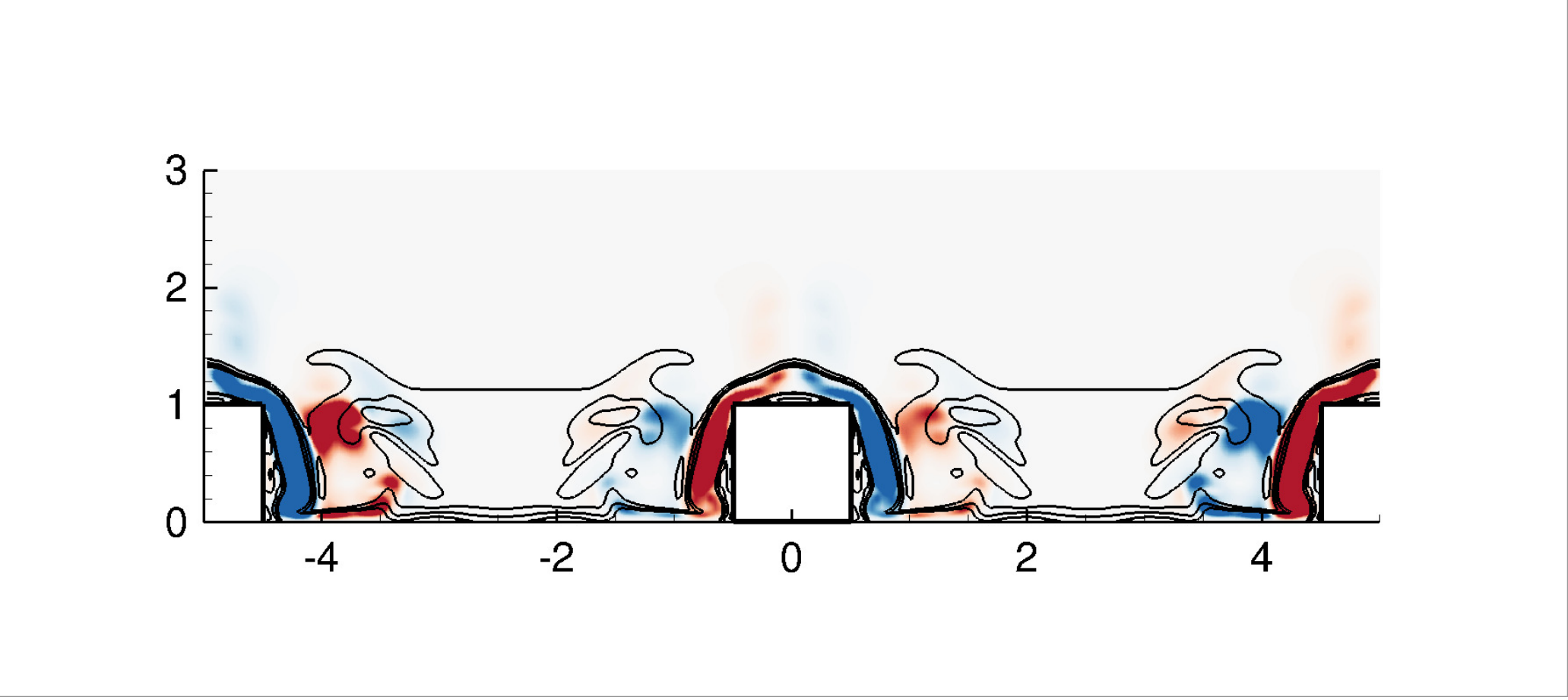}
% \put(-200,90){$(b)$}
\put(-213,35){\rotatebox{90}{$y/h$}}
\put(-112,0){$z/h$}
\hspace{3mm}
\includegraphics[width=72mm,trim={2cm 1.2cm 0.5cm 2cm},clip]{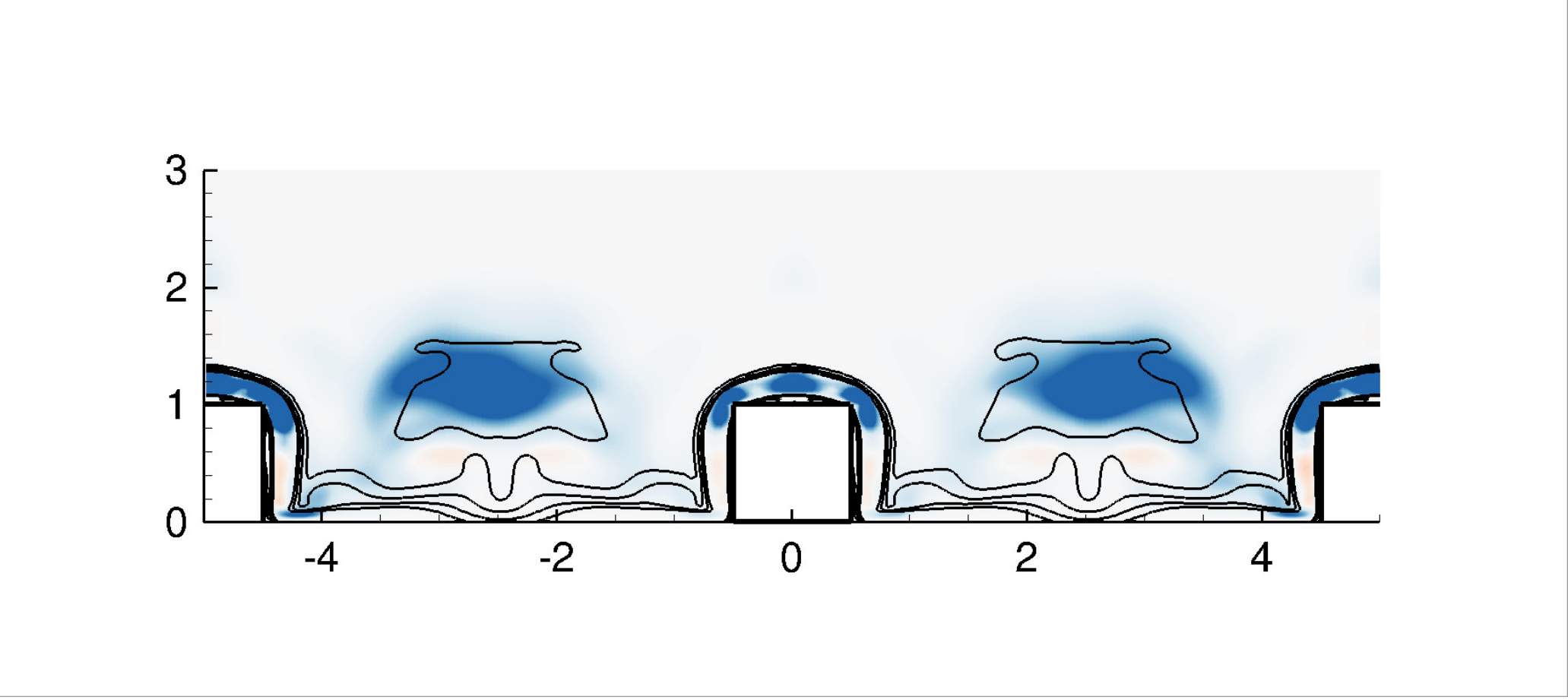}
\put(-200,75){$(b)$}
\put(-180,75){$x=20h$}
\put(-213,35){\rotatebox{90}{$y/h$}}
\put(-112,0){$z/h$}
\includegraphics[width=72mm,trim={2cm 1.2cm 0.5cm 2cm},clip]{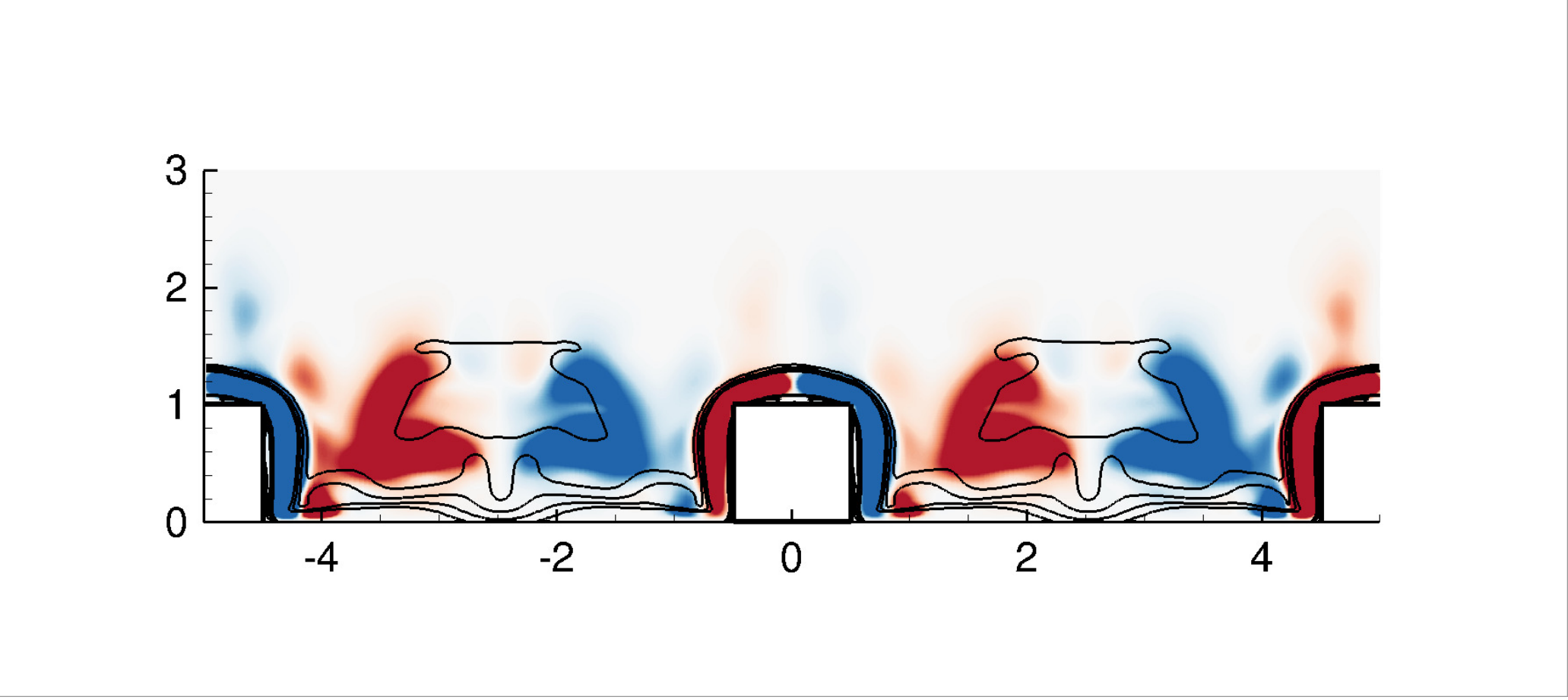}
% \put(-200,90){$(b)$}
\put(-213,35){\rotatebox{90}{$y/h$}}
\put(-112,0){$z/h$}
 \caption{Contour plots of $P_y$ (left) and $P_z$ (right) in the cross-flow planes at $(a)$ $x=10h$ and $(b)$ $x=20h$ for the leading varicose mode of Case $(10h, 5h)$ at $Re_h=600$. The contour levels are the same as figure \ref{fig:prod_5h}.} 
\label{fig:prod_10h}
\end{figure}

\begin{figure}
\includegraphics[width=120mm,trim={0.2cm 0.2cm 0.2cm 0.5cm},clip]{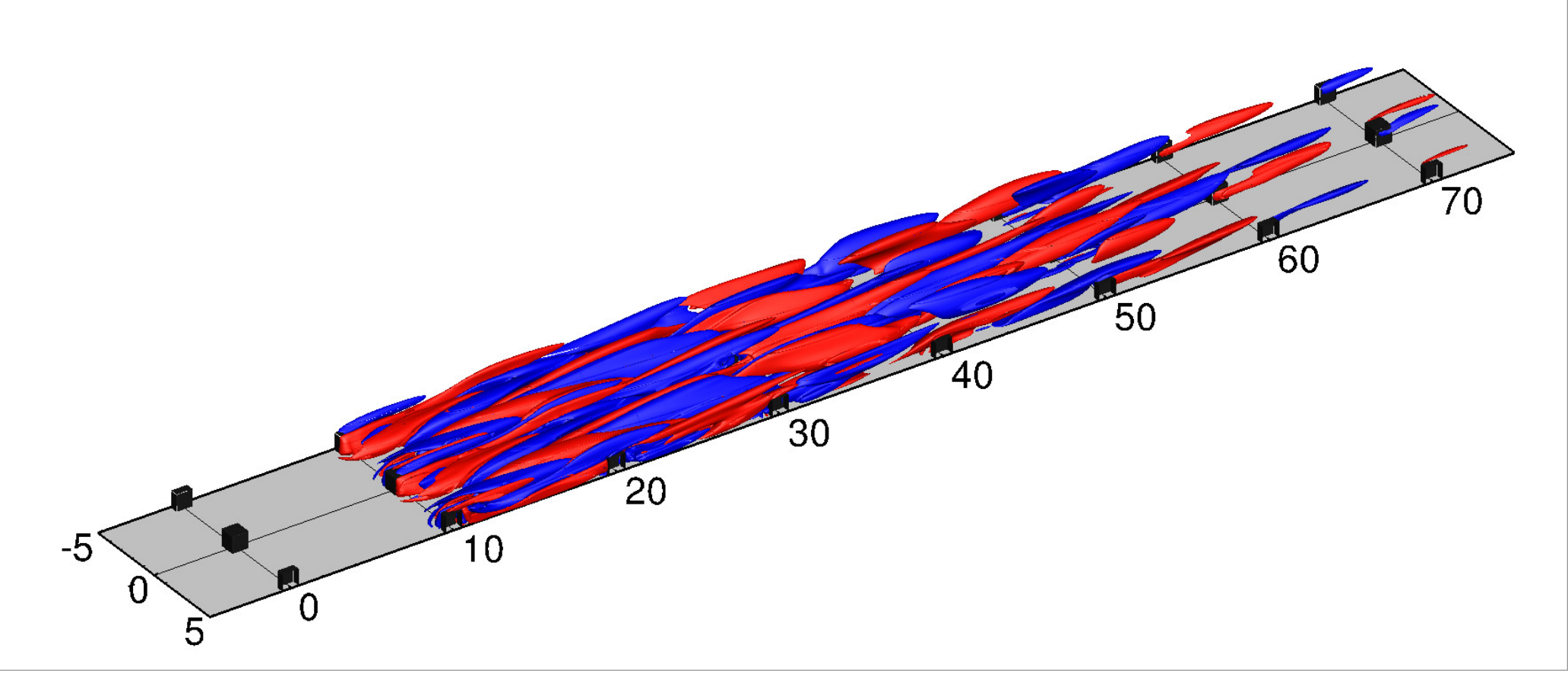}
%\put(-380,60){$(a)$}
% \put(-180,80){$\eta=1,Re_h=475$}
% \put(-380,32){\rotatebox{90}{$z/h$}}
%\put(-106,15){$x/h$}
\put(-150,42){$x/h$}
\put(-338,8){$z/h$}
\hspace{3mm}
\includegraphics[width=130mm,trim={0.2cm 1.0cm 0.5cm 0.5cm},clip]{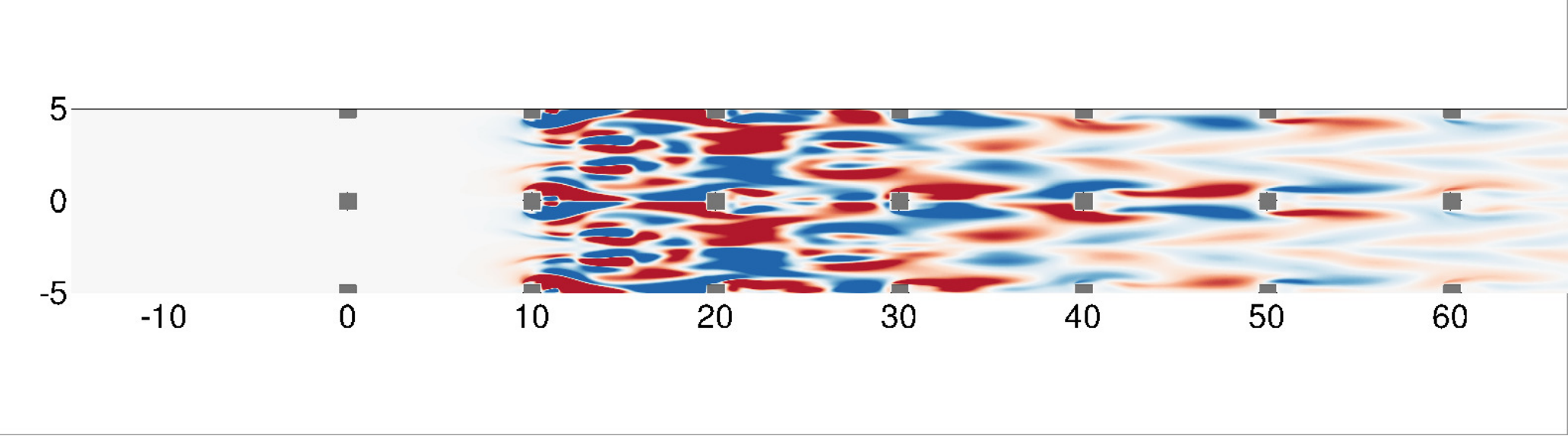}
%\put(-380,60){$(b)$}
% \put(-180,80){$\eta=1,Re_h=475$}
\put(-380,32){\rotatebox{90}{$z/h$}}
\put(-180,0){$x/h$}
\put(-340,50){$y=0.5h$}
% \hspace{3mm}
% \includegraphics[width=130mm,trim={0.2cm 0.5cm 0.5cm 2.5cm},clip]{images/eig5_Lx10_z125.pdf}
% % \put(-200,90){$(b)$}
% \put(-380,25){\rotatebox{90}{$y/h$}}
% \put(-340,40){$z=1.25h$}
 \caption{Isosurfaces (top) and contour plots at $y=0.5h$ (bottom) of the streamwise velocity component of the leading sinuous mode for Case $(10h, 5h)$ at $Re_h=600$. %The contour levels depict $\pm 10 \%$ of the mode's maximum streamwise velocity.
 } 
\label{fig:eigenmode2_10h}
\end{figure}

\begin{figure}
\includegraphics[width=72mm,trim={2cm 1.2cm 0.5cm 2cm},clip]{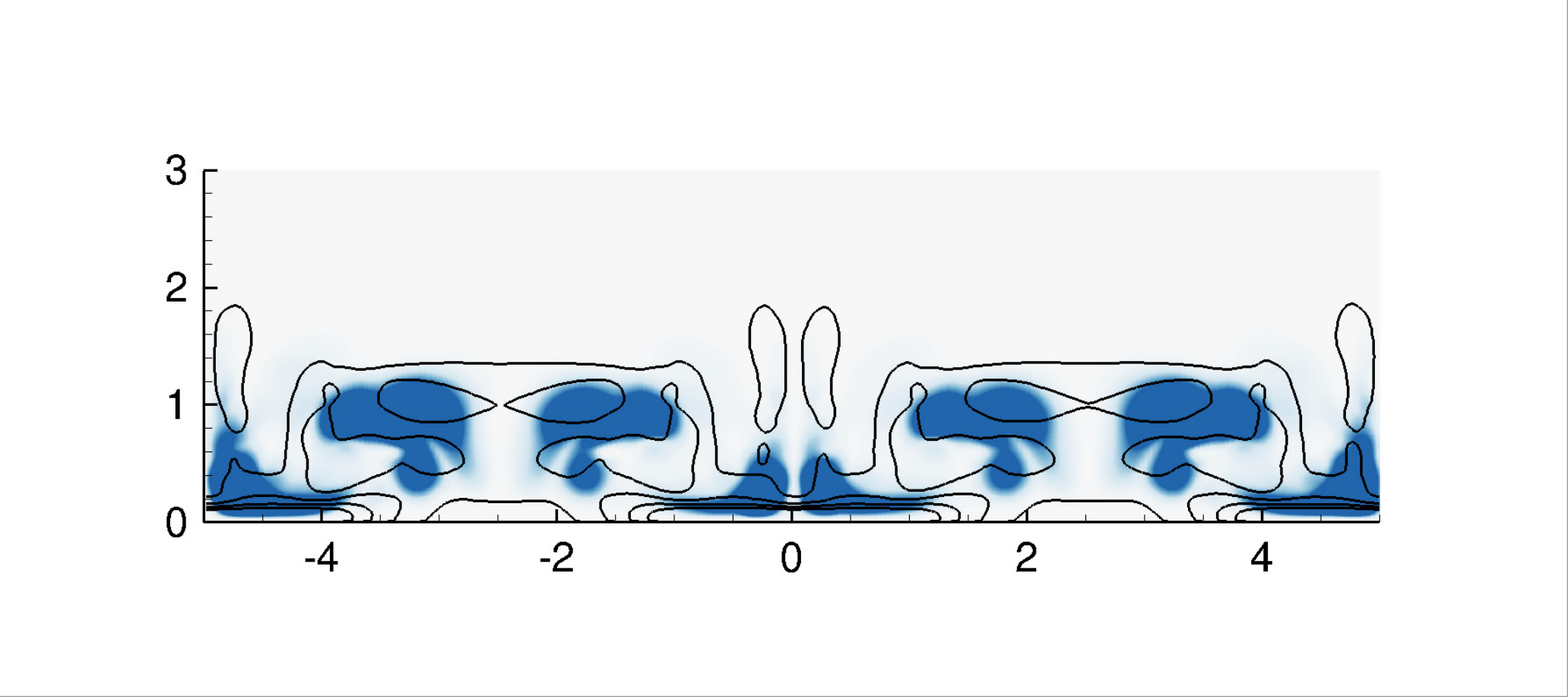}
 \put(-200,75){$(a)$}
\put(-180,75){$x=15h$}
\put(-213,35){\rotatebox{90}{$y/h$}}
\put(-112,0){$z/h$}
\includegraphics[width=72mm,trim={2cm 1.2cm 0.5cm 2cm},clip]{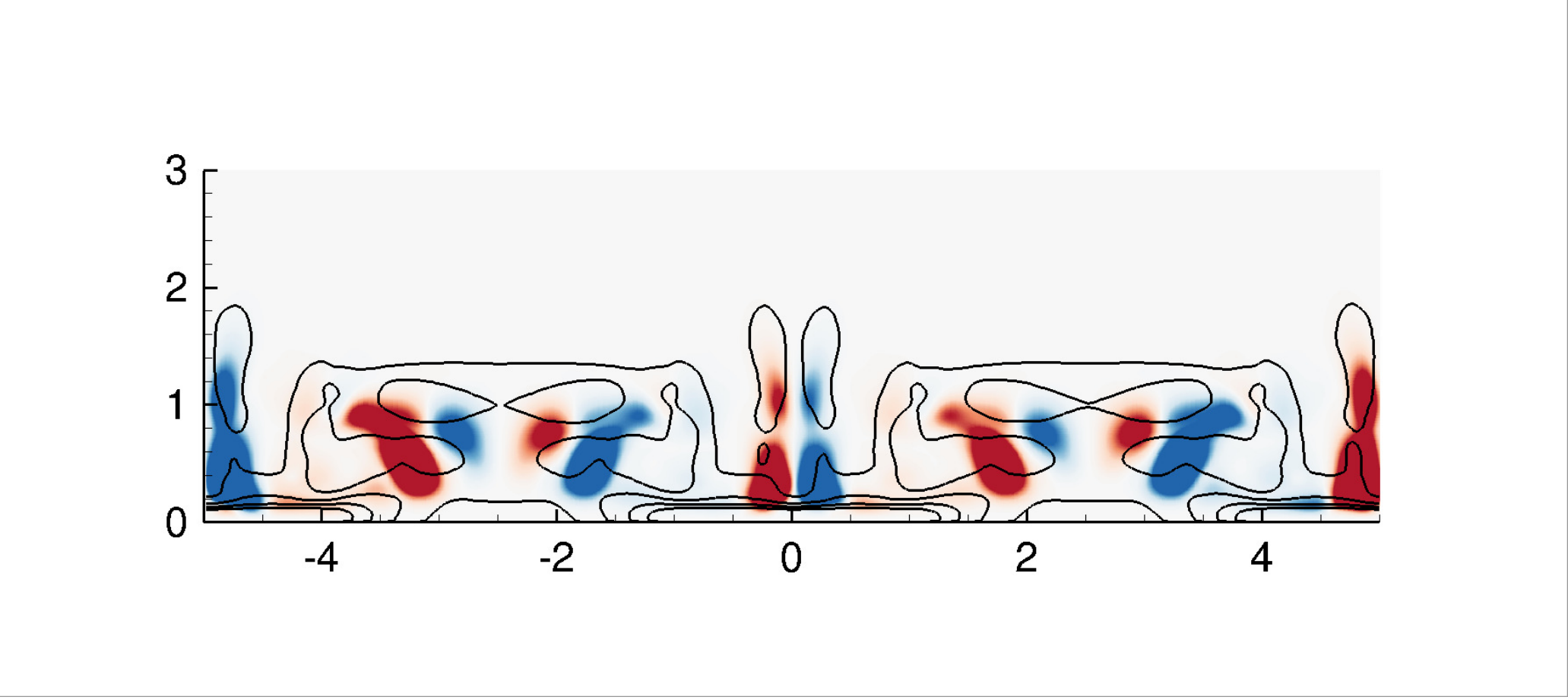}
% \put(-200,90){$(b)$}
\put(-213,35){\rotatebox{90}{$y/h$}}
\put(-112,0){$z/h$}
\hspace{3mm}
\includegraphics[width=72mm,trim={2cm 1.2cm 0.5cm 2cm},clip]{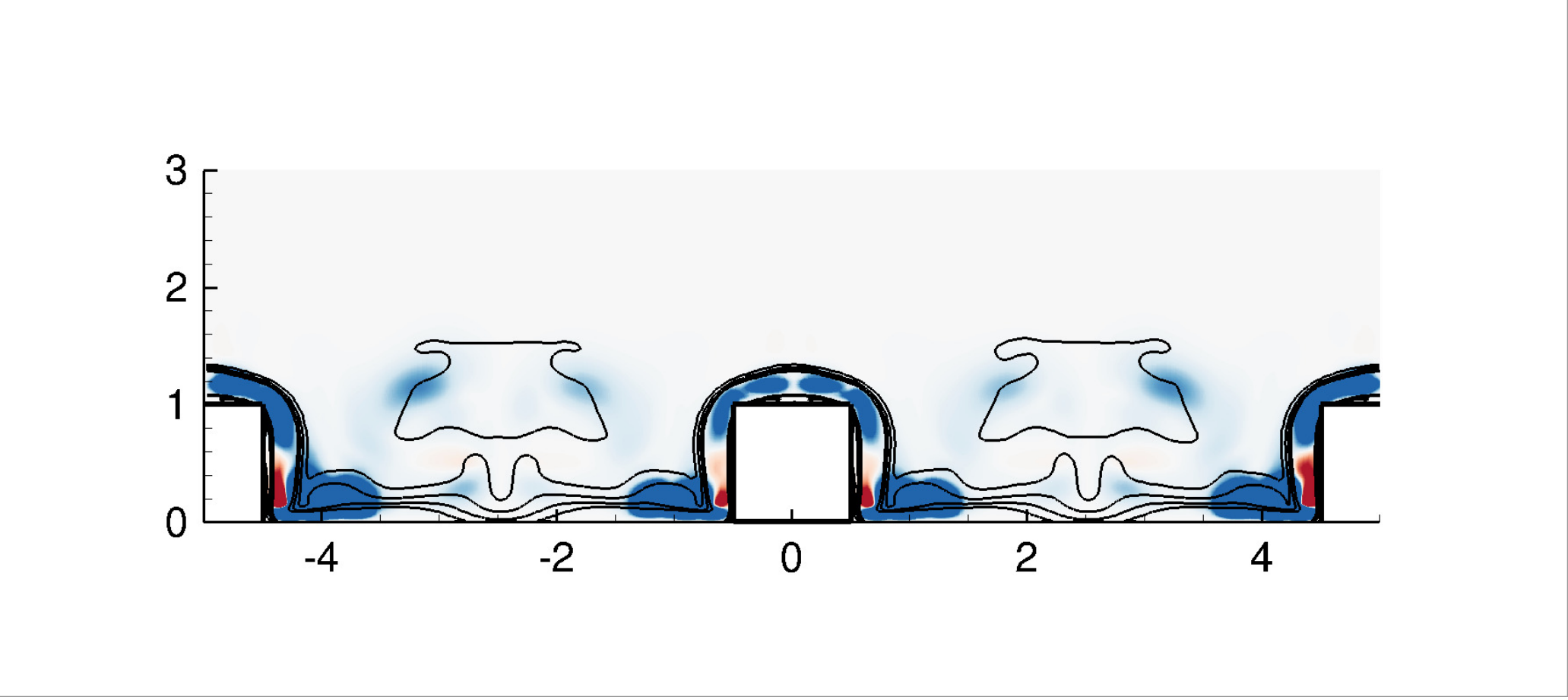}
\put(-200,75){$(b)$}
\put(-180,75){$x=20h$}
\put(-213,35){\rotatebox{90}{$y/h$}}
\put(-112,0){$z/h$}
\includegraphics[width=72mm,trim={2cm 1.2cm 0.5cm 2cm},clip]{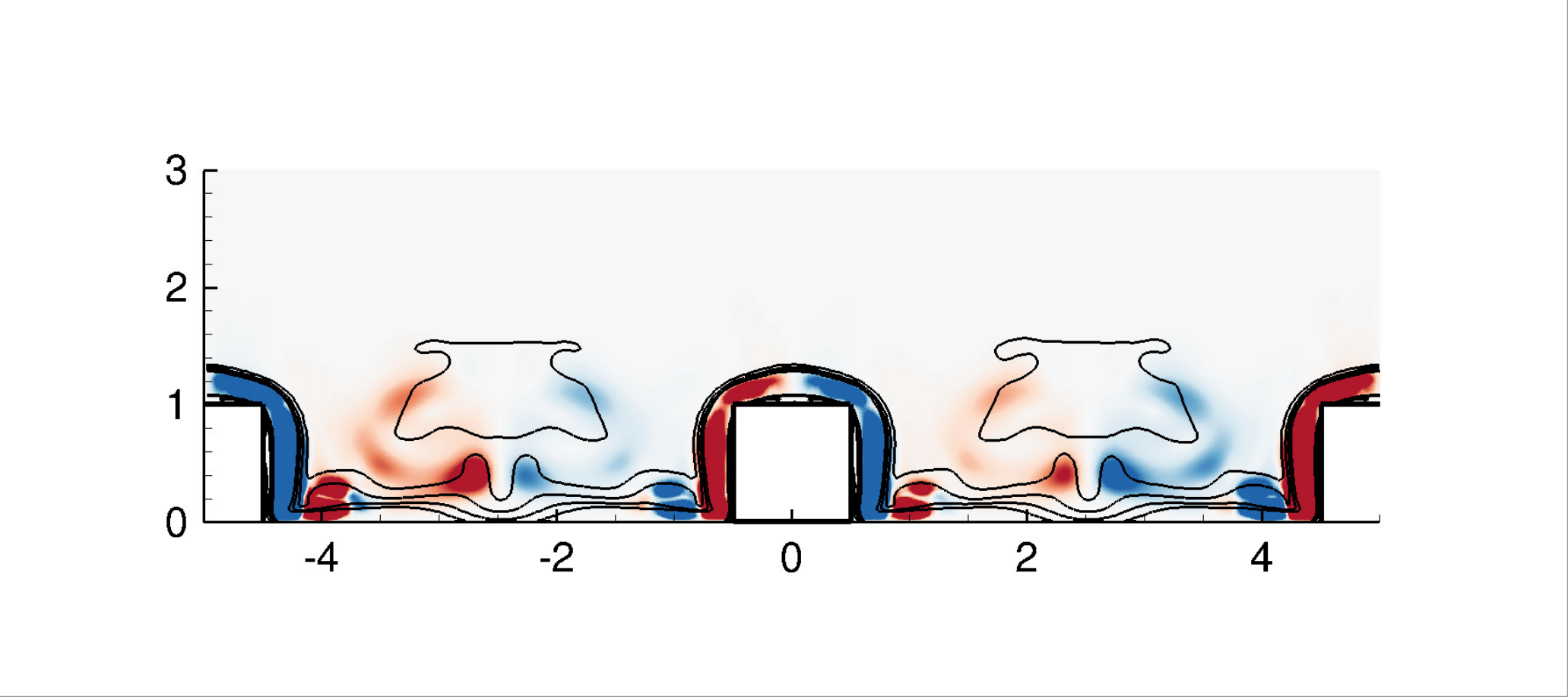}
% \put(-200,90){$(b)$}
\put(-213,35){\rotatebox{90}{$y/h$}}
\put(-112,0){$z/h$}
 \caption{Contour plots of $P_y$ (left) and $P_z$ (right) in the cross-flow planes at $(a)$ $x=15h$ and $(b)$ $x=20h$ for the leading sinuous mode of Case $(10h, 5h)$ at $Re_h=600$. The contour levels are the same as figure \ref{fig:prod_5h}.} 
\label{fig:prod_10h_mode5}
\end{figure}

For Case $(10h, 5h)$, two leading eigenvalues are captured in figure \ref{fig:eigenspectra}. One is the leading eigenvalue whose temporal frequency is close to that of the isolated case and Case $(5h, 5h)$. This leading eigenvalue corresponds to the primary hairpin vortex shedding and is marginally stable, consistent with the state of marginal stability of mean flow for the isolated roughness suggested by \cite{ma2022global}. The corresponding eigenmode in figure \ref{fig:eigenmode1_10h} is associated with the high-speed streaks in the longitudinal grooves as well as the entire shear layer formed above the roughness tips. The production results in figure \ref{fig:prod_10h} indicate that the mode extracts most of energy from the localized shear caused by obstacles and the high-speed streaks in the grooves farther downstream. In contrast to the isolated case and Case $(5h, 5h)$, the mode hardly extracts the energy from the central low-speed streak since the central streak is diminished for larger streamwise spacing. The other unstable leading eigenvalue with a lower frequency is also obtained for Case $(10h, 5h)$ in figure \ref{fig:eigenspectra}. Figure \ref{fig:eigenmode2_10h} shows that the associated eigenmode displays sinuous symmetry, it is induced by the second row of roughness elements and fades away within short downstream distance. This indicates that when the streamwise spacing is larger than the region of flow separation, an additional unstable mode is generated as the wake flow from the first-row roughness impinges on the following roughness. The production results in figure \ref{fig:prod_10h_mode5} demonstrate that this sinuous mode mainly extracts its energy from the spanwise shear induced between the high- and low-speed streaks in the grooves.

% \begin{figure}
% \includegraphics[width=140mm,trim={0.5cm 0.5cm 0.5cm 0.5cm},clip]{images/inst_u_Lx5.pdf}
% \put(-396,75){$(a)$}
% \put(-407,35){\rotatebox{90}{$z/h$}}
% \put(-200,0){$x/h$}
% %\put(-22,50){\scriptsize{$u_d/U_e$}}
% \hspace{3mm}
% \includegraphics[width=140mm,trim={0.5cm 0.5cm 0.5cm 0.5cm},clip]{images/inst_u_Lx10.pdf}
% \put(-396,75){$(b)$}
% \put(-407,35){\rotatebox{90}{$z/h$}}
% \put(-200,0){$x/h$}
% %\put(-22,50){\scriptsize{$u_d/U_e$}}
% \caption{Contour plots of instantaneous streamwise velocity in the plane of $y=0.5h$ from the DNS at $Re_h=600$ for $(a)$ Case $(\lambda_x,\lambda_z)=(5h, 5h)$ and $(c)$ Case $(\lambda_x,\lambda_z)=(10h, 5h)$. } 
% \label{fig:contour_zslice}
% \end{figure}
% %/project/maxx0242/distributed_roughness

% \subsubsection{Boundary layer integral parameters and skin-friction coefficient}
% local variation of integral parameters and spanwise-averaged integral parameters (shape factors)
% skin friction and form drag; mean flow-recirculation

\section{Summary}\label{sec:summary}
The effects of roughness spacing on boundary layer transition due to distributed surface roughness are investigated. Both streamwise and spanwise proximities of roughness elements are considered. The transitional flow behavior and the primary vortical structures due to distributed surface roughness are examined by performing direct numerical simulations. Global stability analysis is performed to study the stability properties of the flow field modified by the distributed roughness, with a comparison to an isolated roughness element with the same geometry \citep{ma2022global}.

When the spanwise spacing is small ($\lambda_z=2.5h$), the OSP vortices in the groove enhance the upward fluid motion and weaken the centrifugal forces, inhibiting the formation of SP vortices. With the absence of CVP, the hairpin vortices are not generated and the downstream shear layer remains steady at $Re_h=600$. The flow might be unstable at higher $Re_h$ due to Kelvin-Helmholtz instability.
When the spanwise spacing increases to $\lambda_z=5h$, the wake flow becomes unsteady and the effects of streamwise spacing on transition are investigated for the fixed $\lambda_z=5h$. The steeper and larger increase in the streamwise variations of boundary layer thickness indicates that the breakdown of boundary layer is more significant for distributed surface roughness. The shape factor profiles suggest that transition occurs at similar downstream locations as the isolated roughness.

When the streamwise spacing is comparable to the length scale of flow separation ($\lambda_x=5h$), the high-momentum fluid above the roughness layer barely penetrates into the cavities, and the primary hairpin vortices with shorter legs shed at the similar frequency as the isolated roughness case. The global stability analysis indicates that the leading unstable eigenvalue is close to that of the isolated case, while the distributed roughness results in a lower critical $Re_h$ for instability to occur. The unstable eigenmode presents varicose symmetry, and extracts the energy from the central low-speed streaks and the localized shear induced by the cubes. 

When the streamwise spacing increases to $\lambda_x=10h$, the high-momentum fluid transfers downward into the cavities. %, and the wake flow of the first-row cubes run into the following cubes. 
The CVP and hairpin vortices induced by the first row of roughness break down into small vortical structures when they run onto the second-row roughness. %The SFD method fails to damp the unsteady oscillations, the time-averaged mean flow is therefore used as the base state to perform global stability analysis. 
Two distinct eigenvalues are obtained from global stability analysis. One %marginally stable leading eigenvalue 
corresponds to the primary hairpin vortex shedding induced by the first-row cubes, %and it is also associated with the high-speed streaks developed in the longitudinal grooves farther downstream. 
whose shedding periodicity is independent on different streamwise spacing.  %The eigenmode extracts the main energy from the high-speed streaks and the entire shear layer. 
It is also associated with the high-speed streaks developed in the longitudinal grooves farther downstream.
The other leading unstable eigenmode with lower frequency presents sinuous symmetry. It is induced as the wake flow of the first row of roughness impinges onto the second row of roughness, and extracts its energy from the spanwise shear between the high- and low-speed streaks. 

\section*{Acknowledgements}
This work was supported by the United States Office of Naval Research (ONR) Grant N00014-17-1-2308 and N00014-20-1-2717 managed by Dr. P. Chang. The authors acknowledge the Minnesota Supercomputing Institute (MSI) and Extreme Science and Engineering Discovery Environment (XSEDE) for providing computing resources that have contributed to the research results reported in this paper.

\section*{Declaration of Interests}
The authors report no conflict of interest.

\bibliographystyle{jfm}
\bibliography{jfm-instructions}

\begin{thebibliography}{35}
\expandafter\ifx\csname natexlab\endcsname\relax\def\natexlab#1{#1}\fi
\def\au#1{#1} \def\ed#1{#1} \def\yr#1{#1}\def\at#1{#1}\def\jt#1{\textit{#1}}
  \def\bt#1{#1}\def\bvol#1{\textbf{#1}} \def\vol#1{#1} \def\pg#1{#1}
  \def\publ#1{#1}\def\arxiv#1{#1}\def\org#1{#1}\def\st#1{\textit{#1}}

\bibitem[{\AA}kervik {\em et~al.\/}(2006){\AA}kervik, Brandt, Henningson,
  H{\oe}pffner, Marxen \& Schlatter]{aakervik2006steady}
{\sc \au{{\AA}kervik, E.}, \au{Brandt, L.}, \au{Henningson, D.~S.},
  \au{H{\oe}pffner, J.}, \au{Marxen, O.} \& \au{Schlatter, P.}} \yr{2006}
  \at{Steady solutions of the navier-stokes equations by selective frequency
  damping}.  \jt{Phys. Fluids}  \bvol{18}~(6),  \pg{068102}.

\bibitem[Alam{\'e} \& Mahesh(2019)]{alame2019wall}
{\sc \au{Alam{\'e}, K.} \& \au{Mahesh, K.}} \yr{2019}  \at{Wall-bounded flow
  over a realistically rough superhydrophobic surface}.  \jt{J. Fluid Mech.}
  \bvol{873},  \pg{977--1019}.

\bibitem[Baker(1979)]{baker1979laminar}
{\sc \au{Baker, C.~J.}} \yr{1979}  \at{The laminar horseshoe vortex}.  \jt{J.
  Fluid Mech.}  \bvol{95}~(2),  \pg{347--367}.

\bibitem[Bucci {\em et~al.\/}(2021)Bucci, Cherubini, Loiseau \&
  Robinet]{bucci2021influence}
{\sc \au{Bucci, M.~A.}, \au{Cherubini, S.}, \au{Loiseau, J-Ch} \& \au{Robinet,
  J-Ch}} \yr{2021}  \at{Influence of freestream turbulence on the flow over a
  wall roughness}.  \jt{Phys. Rev. Fluids}  \bvol{6}~(6),  \pg{063903}.

\bibitem[Casacuberta {\em et~al.\/}(2018)Casacuberta, Groot, Tol \&
  Hickel]{casacuberta2018effectivity}
{\sc \au{Casacuberta, J.}, \au{Groot, K.~J.}, \au{Tol, H.~J.} \& \au{Hickel,
  S.}} \yr{2018}  \at{Effectivity and efficiency of selective frequency damping
  for the computation of unstable steady-state solutions}.  \jt{J. Comput.
  Phys.}  \bvol{375},  \pg{481--497}.

\bibitem[Choudhari \& Fischer(2006)]{choudhari2006roughness}
{\sc \au{Choudhari, M.} \& \au{Fischer, P.}} \yr{2006} Roughness induced
  transient growth: Nonlinear effects.  \bt{In {\em IUTAM Symposium on
  Laminar-Turbulent Transition\/}},  \pg{pp. 237--242}. Springer.

\bibitem[Choudhari {\em et~al.\/}(2010)Choudhari, Li, Chang, Edwards, Kegerise
  \& King]{choudhari2010llaminar}
{\sc \au{Choudhari, M.}, \au{Li, F.}, \au{Chang, C.}, \au{Edwards, J.},
  \au{Kegerise, M.} \& \au{King, R.}} \yr{2010} Llaminar-turbulent transition
  behind discrete roughness elements in a high-speed boundary layer.  \bt{In
  {\em 48th AIAA aerospace sciences meeting including the new horizons forum
  and aerospace exposition\/}},  \pg{p. 1575}.

\bibitem[Citro {\em et~al.\/}(2015)Citro, Giannetti, Luchini \&
  Auteri]{citro2015global}
{\sc \au{Citro, V.}, \au{Giannetti, F.}, \au{Luchini, P.} \& \au{Auteri, F.}}
  \yr{2015}  \at{Global stability and sensitivity analysis of boundary-layer
  flows past a hemispherical roughness element}.  \jt{Phys. Fluids}
  \bvol{27}~(8),  \pg{084110}.

\bibitem[Coceal {\em et~al.\/}(2006)Coceal, Thomas, Castro \&
  Belcher]{coceal2006mean}
{\sc \au{Coceal, O.}, \au{Thomas, T.~G.}, \au{Castro, I.~P.} \& \au{Belcher,
  S.~E.}} \yr{2006}  \at{Mean flow and turbulence statistics over groups of
  urban-like cubical obstacles}.  \jt{Bound.-Layer Meteorol.}  \bvol{121}~(3),
  \pg{491--519}.

\bibitem[Cohen {\em et~al.\/}(2014)Cohen, Karp \& Mehta]{cohen2014minimal}
{\sc \au{Cohen, J.}, \au{Karp, M.} \& \au{Mehta, V.}} \yr{2014}  \at{A minimal
  flow-elements model for the generation of packets of hairpin vortices in
  shear flows}.  \jt{J. Fluid Mech.}  \bvol{747},  \pg{30--43}.

\bibitem[Corke {\em et~al.\/}(1986)Corke, Bar-Sever \&
  Morkovin]{corke1986experiments}
{\sc \au{Corke, T.~C.}, \au{Bar-Sever, A.} \& \au{Morkovin, M.~V.}} \yr{1986}
  \at{Experiments on transition enhancement by distributed roughness}.
  \jt{Phys. Fluids}  \bvol{29}~(10),  \pg{3199--3213}.

\bibitem[De~Tullio {\em et~al.\/}(2013)De~Tullio, Paredes, Sandham \&
  Theofilis]{de2013laminar}
{\sc \au{De~Tullio, N.}, \au{Paredes, P.}, \au{Sandham, N.~D.} \&
  \au{Theofilis, V.}} \yr{2013}  \at{Laminar--turbulent transition induced by a
  discrete roughness element in a supersonic boundary layer}.  \jt{J. Fluid
  Mech.}  \bvol{735},  \pg{613--646}.

\bibitem[von Deyn {\em et~al.\/}(2020)von Deyn, Forooghi, Frohnapfel,
  Schlatter, Hanifi \& Henningson]{von2020direct}
{\sc \au{von Deyn, L.~H.}, \au{Forooghi, P.}, \au{Frohnapfel, B.},
  \au{Schlatter, P.}, \au{Hanifi, A.} \& \au{Henningson, D.~S.}} \yr{2020}
  \at{Direct numerical simulations of bypass transition over distributed
  roughness}.  \jt{AIAA journal}  \bvol{58}~(2),  \pg{702--711}.

\bibitem[Ergin \& White(2006)]{ergin2006unsteady}
{\sc \au{Ergin, F.~G.} \& \au{White, E.~B.}} \yr{2006}  \at{Unsteady and
  transitional flows behind roughness elements}.  \jt{AIAA journal}
  \bvol{44}~(11),  \pg{2504--2514}.

\bibitem[Fransson {\em et~al.\/}(2004)Fransson, Brandt, Talamelli \&
  Cossu]{fransson2004experimental}
{\sc \au{Fransson, J. H.~M.}, \au{Brandt, L.}, \au{Talamelli, A.} \& \au{Cossu,
  C.}} \yr{2004}  \at{Experimental and theoretical investigation of the
  nonmodal growth of steady streaks in a flat plate boundary layer}.  \jt{Phys.
  Fluids}  \bvol{16}~(10),  \pg{3627--3638}.

\bibitem[Fransson {\em et~al.\/}(2005)Fransson, Brandt, Talamelli \&
  Cossu]{fransson2005experimental}
{\sc \au{Fransson, J. H.~M.}, \au{Brandt, L.}, \au{Talamelli, A.} \& \au{Cossu,
  C.}} \yr{2005}  \at{Experimental study of the stabilization of
  tollmien--schlichting waves by finite amplitude streaks}.  \jt{Phys. Fluids}
  \bvol{17}~(5),  \pg{054110}.

\bibitem[Hunt {\em et~al.\/}(1988)Hunt, Wray \& Moin]{hunt1988eddies}
{\sc \au{Hunt, J. C.~R.}, \au{Wray, A.~A.} \& \au{Moin, P.}} \yr{1988}
  \at{Eddies, streams, and convergence zones in turbulent flows}.  \jt{Studying
  Turbulence Using Numerical Simulation Databases, 2. Proceedings of the 1988
  Summer Program} .

\bibitem[Ikeda \& Durbin(2007)]{ikeda2007direct}
{\sc \au{Ikeda, T.} \& \au{Durbin, P.~A.}} \yr{2007}  \at{Direct simulations of
  a rough-wall channel flow}.  \jt{J. Fluid Mech.}  \bvol{571},  \pg{235--263}.

\bibitem[Iyer \& Mahesh(2013)]{iyer2013high}
{\sc \au{Iyer, P.~S.} \& \au{Mahesh, K.}} \yr{2013}  \at{High-speed
  boundary-layer transition induced by a discrete roughness element}.  \jt{J.
  Fluid Mech.}  \bvol{729},  \pg{524--562}.

\bibitem[Jim{\'e}nez(2004)]{jimenez2004turbulent}
{\sc \au{Jim{\'e}nez, J.}} \yr{2004}  \at{Turbulent flows over rough walls}.
  \jt{Annu. Rev. Fluid Mech.}  \bvol{36},  \pg{173--196}.

\bibitem[Jordi {\em et~al.\/}(2014)Jordi, Cotter \&
  Sherwin]{jordi2014encapsulated}
{\sc \au{Jordi, B.~E.}, \au{Cotter, C.~J.} \& \au{Sherwin, S.~J.}} \yr{2014}
  \at{Encapsulated formulation of the selective frequency damping method}.
  \jt{Phys. Fluids}  \bvol{26}~(3),  \pg{034101}.

\bibitem[Leonardi \& Castro(2010)]{leonardi2010channel}
{\sc \au{Leonardi, S.} \& \au{Castro, I.~P.}} \yr{2010}  \at{Channel flow over
  large cube roughness: a direct numerical simulation study}.  \jt{J. Fluid
  Mech.}  \bvol{651},  \pg{519--539}.

\bibitem[Loiseau {\em et~al.\/}(2014)Loiseau, Robinet, Cherubini \&
  Leriche]{loiseau2014investigation}
{\sc \au{Loiseau, J.}, \au{Robinet, J.}, \au{Cherubini, S.} \& \au{Leriche,
  E.}} \yr{2014}  \at{Investigation of the roughness-induced transition: global
  stability analyses and direct numerical simulations}.  \jt{J. Fluid Mech.}
  \bvol{760},  \pg{175--211}.

\bibitem[Ma {\em et~al.\/}(2021)Ma, Alam{\'e} \& Mahesh]{ma2021direct}
{\sc \au{Ma, R.}, \au{Alam{\'e}, K.} \& \au{Mahesh, K.}} \yr{2021}  \at{Direct
  numerical simulation of turbulent channel flow over random rough surfaces}.
  \jt{J. Fluid Mech.}  \bvol{908},  \pg{A40}.

\bibitem[Ma \& Mahesh(2022)]{ma2022global}
{\sc \au{Ma, R.} \& \au{Mahesh, K.}} \yr{2022}  \at{Global stability analysis
  and direct numerical simulation of boundary layers with an isolated roughness
  element}.  \jt{Journal of Fluid Mechanics}  \bvol{949},  \pg{A12}.

\bibitem[Mahesh {\em et~al.\/}(2004)Mahesh, Constantinescu \&
  Moin]{mahesh2004numerical}
{\sc \au{Mahesh, K.}, \au{Constantinescu, G.} \& \au{Moin, P.}} \yr{2004}
  \at{A numerical method for large-eddy simulation in complex geometries}.
  \jt{J. Comput. Phys.}  \bvol{197}~(1),  \pg{215--240}.

\bibitem[Muppidi \& Mahesh(2012)]{muppidi2012direct}
{\sc \au{Muppidi, S.} \& \au{Mahesh, K.}} \yr{2012}  \at{Direct numerical
  simulations of roughness-induced transition in supersonic boundary layers}.
  \jt{J. Fluid Mech.}  \bvol{693},  \pg{28--56}.

\bibitem[Perry {\em et~al.\/}(1969)Perry, Schofield \& Joubert]{perry1969rough}
{\sc \au{Perry, A.~E.}, \au{Schofield, W.~H.} \& \au{Joubert, P.~N.}} \yr{1969}
   \at{Rough wall turbulent boundary layers}.  \jt{J. Fluid Mech.}
  \bvol{37}~(2),  \pg{383--413}.

\bibitem[Reshotko(2001)]{reshotko2001transient}
{\sc \au{Reshotko, E.}} \yr{2001}  \at{Transient growth: a factor in bypass
  transition}.  \jt{Phys. Fluids}  \bvol{13}~(5),  \pg{1067--1075}.

\bibitem[Tani(1987)]{tani1987turbulent}
{\sc \au{Tani, I.}} \yr{1987}  \at{Turbulent boundary layer development over
  rough surfaces}.  \bt{In {\em Perspectives in turbulence studies\/}},
  \pg{pp. 223--249}.  \publ{Springer}.

\bibitem[Theofilis(2011)]{theofilis2011global}
{\sc \au{Theofilis, V.}} \yr{2011}  \at{Global linear instability}.  \jt{Annu.
  Rev. Fluid Mech.}  \bvol{43},  \pg{319--352}.

\bibitem[Vadlamani {\em et~al.\/}(2018)Vadlamani, Tucker \&
  Durbin]{vadlamani2018distributed}
{\sc \au{Vadlamani, N.~R.}, \au{Tucker, P.~G.} \& \au{Durbin, P.}} \yr{2018}
  \at{Distributed roughness effects on transitional and turbulent boundary
  layers}.  \jt{Flow Turbul. Combust}  \bvol{100}~(3),  \pg{627--649}.

\bibitem[Vanderwel \& Ganapathisubramani(2015)]{vanderwel2015effects}
{\sc \au{Vanderwel, C.} \& \au{Ganapathisubramani, B.}} \yr{2015}  \at{Effects
  of spanwise spacing on large-scale secondary flows in rough-wall turbulent
  boundary layers}.  \jt{J. Fluid Mech.}  \bvol{774}.

\bibitem[Xie \& Fuka(2018)]{xie2018note}
{\sc \au{Xie, Z.} \& \au{Fuka, V.}} \yr{2018}  \at{A note on spatial averaging
  and shear stresses within urban canopies}.  \jt{Bound.-Layer Meteorol.}
  \bvol{167}~(1),  \pg{171--179}.

\bibitem[Xu {\em et~al.\/}(2021)Xu, Altland, Yang \& Kunz]{xu2021flow}
{\sc \au{Xu, H.}, \au{Altland, S.~J.}, \au{Yang, X.} \& \au{Kunz, R.~F.}}
  \yr{2021}  \at{Flow over closely packed cubical roughness}.  \jt{J. Fluid
  Mech.}  \bvol{920}.

\end{thebibliography}

\end{document}